\documentclass[12 pt]{article}

\usepackage{graphicx}
\usepackage[cmex10]{amsmath}
\usepackage{algorithm,algorithmic}
\usepackage{amsfonts}
\usepackage{multirow}
\usepackage{color}
\usepackage{balance}

\newcommand{\eye}{\mbox{I}} 
\newcommand{\EV}[1]{\mathbb{E}\left[#1 \right]} 
\newcommand{\bigparenth}[1]{\left( #1\right)} 
\newcommand{\bigsquare}[1]{\left[ #1 \right]}
\newcommand{\inv}[1]{#1^{-1}} 

\newcommand{\set}[1]{\left\{ #1 \right\}}

\newcommand{\pareqref}[1]{(\ref{eq:#1})}

\newcommand{\pgenericddf}[1]{p^{f}(#1)}
\newcommand{\pexactddf}[1]{p^{f,\smC{E}}(#1)}
\newcommand{\pwepddf}[1]{p^{f,\smC{W}}(#1)}
\newcommand{\GaussianMix}[7]{\displaystyle \sum_{#2=1}^{#3^{#1}}{#4^{#1}_{#2}{\cal N}(#7;#5^{#1}_{#2},#6^{#1}_{#2})}}

\newcommand{\smC}[1]{\mathrm{\textsc{#1}}} 

\newcommand{\parinv}[1]{\bigparenth{#1}^{-1}} 

\newcommand{\omegacurr}[0]{\omega_{\mbox{\tiny curr}}}
\newcommand{\omegaold}[0]{\omega_{\mbox{\tiny old}}}

\begin{document}




\title{Decentralized Gaussian Mixture Fusion \\ through Unified Quotient Approximations}


\author{Nisar R. Ahmed \\ 
        Smead Aerospace Engineering Sciences  \\ 
				University of Colorado Boulder, CO USA}
\date{}
\maketitle

\begin{abstract}
This work examines the problem of using finite Gaussian mixtures (GM) probability density functions in recursive Bayesian peer-to-peer decentralized data fusion (DDF). It is shown that algorithms for both exact and approximate GM DDF lead to the same problem of finding a suitable GM approximation to a posterior fusion pdf resulting from the division of a `naive Bayes' fusion GM (representing direct combination of possibly dependent information sources) by another non-Gaussian pdf (representing removal of either the actual or estimated `common information' between the information sources). The resulting quotient pdf for general GM fusion is naturally a mixture pdf, although the fused mixands are non-Gaussian and are not analytically tractable for recursive Bayesian updates. Parallelizable importance sampling algorithms for both direct local approximation and indirect global approximation of the quotient mixture are developed to find tractable GM approximations to the non-Gaussian `sum of quotients' mixtures.  
Practical application examples for multi-platform static target search and maneuverable range-based target tracking demonstrate the higher fidelity of the resulting approximations compared to existing GM DDF techniques, as well as their favorable computational features.
\end{abstract}







\section{Introduction}
Bayesian Decentralized Data Fusion (DDF) is a well-established framework for state estimation-based information sharing and autonomous perception in sensor networks. The strength of Bayesian DDF resides in its ability to replicate idealized centralized Bayesian data fusion results through parallel distributed computing and asynchronous communication, in which all raw sensor data is sent to a single location for maximum information extraction. However, compared to centralized fusion, DDF achieves far greater computational efficiency, scalability, and robustness to network node failures through peer-to-peer `message passing' algorithms that exchange of local node beliefs \cite{Chong-C3Workshop-1975, Chong-ACC-1982, Chong-ACC-1983, Grime-CEP-1994, Campbell-CSM-2016}.

Many techniques have been developed for implementing DDF with state uncertainties modeled by Gaussian or other exponential family distributions. However, many real-world sensor network applications, such as robotic mapping \cite{Schoenberg-FUSION-2009, ASC-RSS-2012, Tse-TRO-2015} or dynamic target search and tracking  \cite{Ahmed-TRO-2013, Julier-FUSION-2006}, involve uncertainties described by distributions outside the exponential family. In these domains, uncertainties are inadequately characterized by the first two moments of the statistical processes under consideration, e.g. due to the presence of discrete random variables or multi-modal/heavy-tailed noise distributions. As such, the use of simple Gaussian approximations for DDF runs the risk of losing important information about the true fused state posterior. This motivates the use of alternative models such as finite Gaussian mixtures (GMs) to approximate the required pdfs for DDF as accurately as possible. 

Although DDF can theoretically support complex pdf models like GMs, practical non-Gaussian DDF implementations require balancing important tradeoffs between approximation accuracy and computational efficiency. The main challenge here lies in the fact that GMs (unlike Gaussian or other exponential family pdfs) do not admit recursive closed-form solutions for Bayesian DDF, i.e. the posterior pdf that results from fusing two GMs via Bayesian DDF is generally not a GM. A number of approximations based on semi-parametric and non-parametric density estimation methods have been proposed to address this issue \cite{Ridley-ParzenDDF-2004, Julier-FUSION-2006, Ong-MFI-2006, Ong-IROS-2006, Ong-FUSION-2008, Chang-ICARV-2010}. Unfortunately, these approximations rely on a variety of heuristic assumptions that are narrowly tailored to specific applications, and thus do not generalize well. 

This paper proposes a novel computationally efficient and unified approximation strategy for recursive Bayesian DDF with arbitrary GM models. Specifically, this work derives and exploits the important fact that the true fusion posterior for the general GM DDF problem is exactly equal to a mixture of non-Gaussian pdfs. This makes it possible to obtain naturally parallelizable decompositions of the GM fusion posterior, where the moments of the individual non-Gaussian mixands are computed and used to approximate the overall fusion result as another GM. This insight leads to a set of high-fidelity GM approximations of otherwise analytically intractable GM fusion posteriors, which retain important higher order moment information for GM-based decentralized recursive probabilistic data fusion. Two approximation strategies based on fast Monte Carlo importance sampling methods are considered here.  
In indirect global sampling (IGS), importance samples are drawn with respect to the entire fusion pdf and then probabilistically assigned to non-Gaussian posterior mixands via a novel single shot weighted expectation-maximization (SS-WEM) algorithm. 
In direct local sampling (DLS), the non-Gaussian mixands are separately sampled and approximated by moment-matched Gaussians (or other pdfs). 

IGS and DLS are `unified' approaches since they directly applies to both exact Bayesian DDF methods that explicitly track common information dependencies between platforms (e.g. the channel filter for tree structured networks \cite{Grime-CEP-1994}) and to approximate DDF methods, where common information dependencies are not explicitly known but mitigated via heuristic fusion rules (e.g. the weighted exponential product rule for ad hoc networks \cite{Bailey-FUSION-2012}). In this sense, IGS and DLS represent an important advance over other existing GM DDF approximations, which treat exact and conservative DDF problems separately. The IGS and DLS methods also allow higher order features of the fusion pdf to emerge naturally as the number of importance samples used to form the local mixand approximations increases. This obviates the need for the overly restrictive heuristics used by conventional GM fusion approximations, which lead to significant information loss when assumptions about the global fusion pdf are invalid. 
This work builds significantly on the author's initial work in \cite{Ahmed-MFI-2015}, which presented the DLS method only. 
The current paper develops the new IGS and SS-WEM algorithms for the first time, and presents more thorough simulation results and analysis. 

Section II provides preliminaries for the general GM DDF problem, followed by a derivation of the unified quotient result in Section III. The IGS and DLS approximation techniques are presented in Section IV. Section V provides numerical simulation studies for toy problems and simulated decentralized multi-sensor fusion applications, and Section VI concludes the paper. 
\section{Background} \label{background}
\subsection{Bayesian DDF Problem Formulation}
Let $x$ be a $d$-dimensional vector of random variables monitored by a decentralized network of $n_A$ autonomous Bayesian agents. Assume each agent $i \in \set{1,...,n_A}$ performs local recursive Bayesian updates on a common prior pdf $p_0(x)$ with independent sensor data $y^i_k$ having conditional likelihood $p(y^i_k|x)$ at discrete time step $k\geq0$, so that each agent's local posterior state pdf is given by
\begin{align}
p^i(x|y^i_{1:k}) \propto p^i(x|y^i_{1:k-1}) \cdot p(y^i_k|x),
\label{eq:bayeslocal}
\end{align}
where $p^i(x|y^i_{1:k-1}) = p_0(x)$ for $k=0$. Given some agent-to-agent communication topology at $k$, assume agent $i$ is aware only of its connected neighbors and unaware of the complete network topology. Let $N(i,k)$ denote the set of neighbors $i$ receives information from \textit{at} time $k$, and let $Z^i_k$ denote the information set received by $i$ \textit{up to} time $k$, i.e. $y^i_{1:k}$ plus new external information \textit{previously fused} by $i$ from other agents. At time $k$, just before new information from $N(i,k)$ is fused, eq. \pareqref{bayeslocal} is the same as 
\begin{align}
p^i(x) \equiv p^i(x|Z^i_k),
\end{align}
where $p^i(x)$ hereafter always implies local conditioning on $Z^i_k$. The DDF problem for each agent $i$ is to thus find the pdf representing fusion of $Z^i_k$ and information from $N(i,k)$,
{
\allowdisplaybreaks
\abovedisplayskip = 2pt
\abovedisplayshortskip = 1pt
\belowdisplayskip = 2pt
\belowdisplayshortskip = 1pt
\begin{align}
p^{f, N(i,k)}(x) \equiv p^i(x| Z^i_{k} \bigcup_{j \in N(i,k)} Z^{j}_k).
\label{eq:mainFuse}
\end{align} 
}
The random vector $x$ can be generalized to a dynamic random vector $x_k$ for discrete time $k$, where $p^i(x_{k}|y^i_{1:k-1})$ is computed via the Chapman-Kolmogorov equation 
\begin{align}
p^i(x_{k}|y^i_{1:k-1}) = \int{ p(x_{k}|x_{k-1}) p^i(x_{k-1}|y^i_{1:k-1}) dx_{k-1}}
\end{align}
for some process transition pdf $p(x_{k}|x_{k-1})$. 
%
\subsection{Exact DDF: Distributed Bayesian Inference}
If $i$ recursively computes (\ref{eq:mainFuse}) via a FIFO queue for each $j \in N(i,k)$, then it is easy to show that, for any $i$ and $j$,
\begin{align}
p^i(x_k) \propto p(x_k|Z^i_{k} \cap Z^j_k) p(x_k|Z^{i\slash j}_{k}) = p^{c,ij}(x_k) p^{i\slash j}(x_k),
\label{eq:pXZdecomp}
\end{align}
where $p^{i \slash j}(x_k) \equiv p(x_k | Z^{i\slash j}_{k})$ is $i$'s exclusive information relative to $j$, and $p^{c,ij}(x_k) \equiv p(x_k|Z^i_{k} \cap Z^j_k)$ is the common information shared by $i$ and $j$. Refs. \cite{Chong-C3Workshop-1975, Chong-ACC-1982, Chong-ACC-1983, Grime-CEP-1994} use eq. \pareqref{pXZdecomp} to show that agent $i$ can recover the desired joint fusion posterior pdf \textit{exactly} by applying a distributed variant of Bayes' rule,
\begin{align}
p^i(x_k | Z^i_{k} \cup Z^j_k) &\propto  p(x_k|Z^i_{k} \cap Z^j_k) p(x_k|Z^{i\slash j}_{k}) p(x_k|Z^{j\slash i}_{k}) \nonumber \\
& =\frac{p^{f,ij'}(x_k) p^j(x_k)}{p^{c,ij}(x_k)},
\label{eq:exactDef}
\end{align}
where $p^{f,ij'}(x_k)$ denotes the fusion posterior obtained by $i$ for all previous $j'<j \in N(i,k)$ in the FIFO recursion. 
Note that this update rule exploits the fact that the posterior pdfs $p^i(x_k)$ and $p^j(x_k)$ compactly summarize all knowledge received by $i$ and $j$ from local sensor data and other network neighbors at time $k$. Importantly, to avoid double-counting of common information, \pareqref{exactDef} requires explicit knowledge of $p^{c,ij}(x_k)$, which arises due to: (i) use of shared state transition models $p(x_{k}|x_{k-1})$ by all $n_A$ agents; and (ii) existence of multiple communication pathways between $i$ and $j$ at any time $k$. Failure to properly account for $p^{c,ij}(x_k)$ eventually produces overconfident and incorrect posterior beliefs across the network, thus leading to `rumor propagation' or data incest. 

\textit{Exact DDF algorithms} employ special data structures such as channel filters \cite{Grime-CEP-1994, Ong-FUSION-2008} to explicitly track $p^{c,ij}(x_k)$ across all fusion instances. These approaches are theoretically `optimal' in the sense that each agent can recover the idealized centralized fusion posterior pdf \footnote{up to a time delay proportional to the maximum time required to receive a message from any other agent, and assuming messages are sent every time step}  
and are straightforward to implement in networks with either tree-connected or fully-connected bilateral communication topologies. 
However, exact DDF methods are cumbersome and computationally expensive for more general network topologies. In particular, eq. \pareqref{exactDef} requires tracking the pedigree of each new piece of information sent or received by $i$ for \textit{all} fusion instances prior to time $k$ to properly account for $p^{c,ij}(x_k)$ \cite{Martin05}. 
\subsection{Conservative WEP DDF for Ad Hoc Networks}
Suboptimal \textit{conservative approximations} to Bayesian DDF can be used to guarantee that the common information is never double-counted and thus never has to be explicitly tracked. The \textit{weighted exponential product (WEP) rule} provides one way to guarantee consistent fusion when $p^{c,ij}(x_k)$ is unknown \cite{Chong-ACC-1983, Bailey-FUSION-2012}, 
\begin{align}
\pwepddf{x_k;\omega} \propto [p^i(x_k)]^{\omega} [p^j(x_k)]^{1-\omega},
\ \omega \in [0,1].
\label{eq:wepDef}
\end{align}
The WEP fusion parameter $\omega$ trades off the amount of new information fused from $p^i(x_k)$ and $p^j(x_k)$, while always counting $p^{c,ij}(x_k)$ exactly once for any $\omega$; this can be readily seen upon substitution of \pareqref{pXZdecomp} into \pareqref{wepDef}
\begin{align}
\pwepddf{x_k} &\propto [p^{c,ij}(x_k)p_{i\slash j}(x_k)]^{\omega} [p^{c,ij}(x_k)p_{j\slash i}(x_k)]^{1-\omega}  \nonumber \\
&= p^{c,ij}(x_k)[p^{i\slash j}(x_k)]^{\omega}[p^{j\slash i}(x_k)]^{1-\omega}. \nonumber
\end{align}

WEP fusion requires application of a fusion rule to select the parameter $\omega$ in eq. \pareqref{wepDef}. 
The fusion rule is typically specified in the form of a variational optimization problem, so that $\omega$ minimizes some predetermined functional $f_{ij}(\omega)$ on $p_i(x_k)$ and $p_j(x_k)$, 
\begin{align}
\omega^* &= \arg \min_{\omega \in [0,1]} f_{ij}(\omega). \label{eq:genWEPFunctional} 
\end{align}
For instance, various information-theoretic strategies could be used to define $f_{ij}(\omega)$. 
The widely recognized \textit{Chernoff rule} sets $\omega$ to the argument corresponding to the Chernoff information between $p_i$ and $p_j$ \cite{Hurley-FUSION-2002, Julier-FUSION-2006, Farrell-FUSION-2009},  
\begin{align}
\omega^* &= \arg \min_{\omega \in [0,1]} -\ln \int_{-\infty}^{\infty}{[p_i(x_k)]^{\omega} [p_j(x_k)]^{1-\omega} dx_k}, \nonumber \\
&= \arg \min_{\omega \in [0,1]} \int_{-\infty}^{\infty}{[p_i(x_k)]^{\omega} [p_j(x_k)]^{1-\omega} dx_k}.
\label{eq:simpleChernoffDef}
\end{align}
It is easily shown that this minimization problem is convex, and that the necessary and sufficient condition for $\omega^*$ yields
\begin{align}
D_{\smC{KL}}[p_f^{\smC{W}}(x_k; \omega^*)|| p_{i}(x_k) ] = D_{\smC{KL}}[ p_f^{\smC{W}}(x_k ; \omega^*) || p_j(x_k)].
\label{eq:kldEquiv}
\end{align}
where $D_{\smC{KL}}$ denotes the KLD.  
This provides the oft-cited basis for the Chernoff fusion rule, since $p_i(x_k)$ and $p_j(x_k)$ become `equidistant' from the fused pdf (\ref{eq:wepDef}) in the KLD sense, such that $i$ and $j$ furnish each other with the same amount of new information. 
However, as shown in \cite{ASC-RSS-2012}, the inherent lossiness of WEP fusion means, in practice, that any significant new information `gains' made by $j$ according to the Chernoff rule may in fact come at the expense of losing equally significant amounts of new exclusive information acquired by $i$ prior to fusion.  
That is, the Chernoff fusion rule does not account for the possibility (for instance) that $j$ simply switches off its sensors while $i$ collects vast amounts of new exclusive data prior to fusion. 
To guard against such lop-sided `information washout' scenarios, ref. \cite{ASC-RSS-2012} proposes an alternative \textit{minimax information loss fusion rule}, which minimizes an upper bound on the maximum possible information loss between the WEP fusion posterior in eq. \pareqref{wepDef} and the exact Bayesian fusion pdf in eq. \pareqref{exactDef}. 
This information loss upper bound is given by the KLD between the exact Bayes fusion pdf, which assumes no common information dependence between $p_i(x_k)$ and $p_j(x_k)$ (i.e. the Naive Bayes pdf $p_{\smC{NB}}(x_k) \propto p_i(x_k)p_j(x_k)$), and the WEP fusion pdf (which depends on $\omega$), so that
\begin{align}
\omega^* &= \arg \min_{\omega \in [0,1]} D_{\smC{KL}}[p_{\smC{NB}(x_k)} || p_f^{\smC{W}}(x_k; \omega^*)]. \label{eq:minimaxInfoLossDef}
\end{align}
It is easily shown that this minimization problem is also convex. 

As discussed in \cite{ASC-RSS-2012}, and more recently in \cite{Taylor-InfoFusion-2019}, a host of other alternative WEP functionals $f_{ij}(\omega)$ can also be used. Alternatives to WEP for conservative fusion of pdfs, such as those based on ellipsoidal intersection \cite{Sijs-ACC-2010} or Schur/Lorentz dominance \cite{Rendas-MFI-2010}, could also be considered. However, these alternatives typically deal with fusion problems where the first and second state pdf moments are of primary interest, and thus have not yet been adapted to more complex pdfs such as those represented by Gaussian mixture models. So, attention here is restricted to WEP techniques. 

\subsection{DDF with Gaussian Mixtures}
Eqs. \pareqref{exactDef} and \pareqref{wepDef} lead to recursive updates for the sufficient statistics of exponential family distributions (e.g. Gaussian, Bernoulli, etc.). 
The well-known covariance intersection (CI) algorithm \cite{Julier-ACC-1997, Chen-FUSION-2002} is a special case of eq. \pareqref{wepDef} that deals with fusion of local pdf means and covariances only \cite{Hurley-FUSION-2002} \footnote{this is sufficient for MMSE state estimation and does not require the underlying pdfs to be Gaussian}. Unfortunately, neither eq. \pareqref{exactDef}  nor \pareqref{wepDef} can be evaluated in closed-form for more complex distributions such as Gaussian mixtures (GMs), 
{
\allowdisplaybreaks
\abovedisplayskip = 1pt
\abovedisplayshortskip = 1pt
\belowdisplayskip = 1pt
\belowdisplayshortskip = 0pt
\begin{align}
p^i(x_k) = \GaussianMix{i}{q}{M}{w}{\mu}{\Sigma}{x_k}
\label{eq:GMdef}
\end{align}
}
\noindent where $M^i$ is the number of mixands, $\mu^i_q$ and $\Sigma^i_q$ are the $q^{\mbox{th}}$ mixand's mean vector and covariance matrix, and $w^i_q \in[0,1]$ is the $q^{\mbox{th}}$ mixand's weight, where $\sum_{q=1}^{M^i}{w^i_q}=1$. GMs arise in many applications such as multi-target tracking \cite{Kaupp-JFR-2007}, target search \cite{Ahmed-TRO-2013}, robotic terrain mapping \cite{Schoenberg-FUSION-2009, Tse-MFI-2012, Tse-TRO-2015}, robotic navigation and planning \cite{Brunskill2010, Porta-IJCAI-2011, Burks-CDC-2017, Burks-FUSION-2018}, hybrid control systems \cite{lesser2017approximate}, and image processing \cite{wainwright2000scale, portilla2003image, Goldberger08, lagrange2017large}, to name a few. GMs are especially useful in contexts where the posterior distribution over multiple hypotheses and/or other highly non-Gaussian uncertainties must be maintained beyond the first two moments (e.g. such that MMSE point estimates are insufficient for describing state uncertainties for subsequent decision making).  

A basic strategy for implementing exact and WEP DDF with GMs is to closely approximate the desired fusion pdfs by GMs, so that the recursive forms of \pareqref{exactDef} and \pareqref{wepDef} can be (approximately) maintained. 
In cases where $p^{c,ij}(x_k)$ is given by a GM pdf, Chang and Sun \cite{Chang-ICARV-2010} derived a closed-form GM approximation to \pareqref{exactDef} which replaces $p^{c,ij}(x_k)$ with a single moment-matched Gaussian, i.e. 
\begin{align}
p^{c,ij}(x_k) \approx {\cal N}(\bar{\mu},\bar{\Sigma}), \label{eq:ChangApprox}
\end{align}
where $\bar{\mu}$ and $\bar{\Sigma}$ are the mixture mean and mixture covariance of $p^{c,ij}(x_k)$, respectively. 
Substitution of \pareqref{ChangApprox} into \pareqref{exactDef} leads to the product of two GMs in the numerator, divided by a Gaussian pdf in the denominator, which can be resolved into a GM. This is  referred to as the moment-matched Gaussian denominator (MMGD) approximation. 

Ref. \cite{Julier-FUSION-2006} proposed a GM approximation to $\pwepddf{x_k}$ for Chernoff fusion that is based on a pair-wise CI rule for the Gaussian mixture component pdfs for $i$ and $j$, 
\begin{align}
&\pwepddf{x_k;\omega} \approx \GaussianMix{f}{m}{M}{w}{\mu}{\Sigma}{x_k}, \label{eq:FOCIApprox} \\
&\Sigma^f_m = \parinv{ \omega \parinv{ \Sigma^{i}_{q} } + (1-\omega)\parinv{\Sigma^{j}_{r}} } \\
&\mu^f_m = \parinv{\omega \parinv{\Sigma^{i}_{q}}\mu^{i}_{q} + (1-\omega)\parinv{\Sigma^{j}_{r}} \mu^{j}_{r} }  \\
&w^f_m = \frac{ (w^i_q)^{\omega} (w^j_r)^{1-\omega}  }{ \sum_{q',r'} (w^i_{q'})^{\omega} (w^j_{r'})^{1-\omega} }
\end{align}
where $M^f = M^i M^j$ and each component index $m \in \set{1,\cdots,M^f}$ corresponds to a pair of component indices $q \in \set{1,\cdots,M^i}$ and $r \in \set{1,\cdots,M^j}$. 
This is referred to as the first order covariance intersection (FOCI) approximation. 
Ref. \cite{Upcroft-FUSION-2005} develops a related GM-based approximation that applies a separate CI operation and corresponding $\omega$ weight to every possible pair of mixands formed by the product of GMs $p^i(x_k)$ and $p^j(x_k)$; this is referred to as Pairwise Component CI, or PCCI. 
Ref. \cite{Julier-FUSION-2006} uses a highly non-Gaussian multi-platform target tracking scenario to show that PCCI provides inferior results compared to FOCI (which uses the same $\omega$ for all fused mixture terms). 
However, to obtain $\omega$ values, both PCCI and FOCI attempt to minimize the size of the component or overall GM covariance instead of the actual Chernoff information, which is not easy to compute for GMs. 

Although fast and convenient, the MMGD, FOCI, and PCCI approximations all rely on strong heuristic assumptions that lead to poor approximations of \pareqref{exactDef} and \pareqref{wepDef} whenever $p^i(x_k), p^j(x_k),$ and/or $p^{c,ij}(x_k)$ are highly non-Gaussian. 
Alternative techniques for Bayesian DDF with particle-based pdf approximations have also been developed, which are closely related to GM fusion. In these methods, weighted samples (particles) can be smoothed through the use of nonparametric Gaussian density kernels (Parzen smoothing), and then can be subsequently `compressed' into GM pdfs via batch learning methods (e.g. the expectation-maximization or EM algorithm) or sequential condensation methods \cite{West-JRSSB-1993}. 

In \cite{Ridley-ParzenDDF-2004}, particle pdf approximations are smoothed by Parzen kernels to enable exact DDF via channel filtering. 
Specifically, Ridley, et al. approximate the division of Parzen-smoothed particle sets via weighted sums of kernel functions. This bears some resemblance to the approach developed here using GMs, except that the present work considers division of GMs with full $d-$dimensional covariance matrices rather than isotropic Parzen kernel functions (which in general must also be tuned through computationally expensive bandwidth optimization procedures). GMs generally require fewer weighted mixture parameters to accurately represent non-Gaussian pdfs in high-dimensional settings compared to the number of weighted particles typically required for a particle pdf approximation. Furthermore, the GM approach developed here readily generalizes and extends to WEP DDF, whereas the approach in \cite{Ridley-ParzenDDF-2004} has no such obvious extension, especially when $\omega$ is not known a priori. 

In \cite{Ong-FUSION-2008}, the exact DDF channel filter update for particles is initially approximated by drawing importance samples from $p^{c,ij}(x_k)$ to produce a weighted sample approximation of \pareqref{exactDef}. In each iteration of the channel filter, nonparametric Parzen kernels are placed around each sample point in the resulting $p^{c,ij}(x_k)$, $p^i(x_k)$ and $p^j(x_k)$ particle sets, so that the multiplication and division operations between particle sets in the RHS of \pareqref{exactDef} is well defined for an importance sampling approximation. 
This approach also bears some resemblance to the technique developed here using GMs, but again suffers from the same drawbacks as in \cite{Ridley-ParzenDDF-2004}. Moreover, the approach from \cite{Ong-FUSION-2008} does not generalize well, since the use of $p^{c,ij}(x_k)$ as an importance sampling proposal density can lead to particle depletion when the exact fusion posterior is significantly different from the common information pdf. 

Refs. \cite{Ong-MFI-2006, Ong-IROS-2006} developed a GM-based approach for WEP fusion of particle sets. This involves first converting particle sets for $p^i(x_k)$ and $p^j(x_k)$ into GMs using Parzen kernels and a condensation algorithm. PCCI is then used for GM fusion. Finally, the fused GM produced by PCCI is resampled to obtain new particles. This approach is computationally expensive due to the kernel tuning and condensation steps. It also inherits the ad hoc/heuristic nature of the PCCI algorithm by selecting $\omega$ to minimize the size of the PCCI mixture covariance, rather than minimizing a more suitable functional for non-Gaussian DDF. 

Refs. \cite{ASC-RSS-2012, Ahmed-TSP-2012} developed a generalizable importance sampling approximation to address these limitations for GM-based WEP DDF. The key idea behind this approach is that, regardless of which WEP functional $f_{ij}(\omega)$ is used, finding $\omega^*$ practically requires \emph{simultaneous} approximation of \emph{both} $p_f^{\smC{W}}(x_k; \omega)$ and 
$f_{ij}(\omega)$ (e.g. the RHS of \pareqref{simpleChernoffDef} and \pareqref{minimaxInfoLossDef}), since neither is closed-form for $\omega \in [0,1]$ when $p_i(x_k)$ and $p_j(x_k)$ are distinct GM pdfs. 
This leads to a general two-step approximation process for WEP fusion of GMs: 
\begin{enumerate}
	\item stochastic minimization of the desired WEP functional $f_{ij}(\omega)$ with respect to $\omega$ using a fixed importance sampling-based particle set representing $p_f^{\smC{W}}(x_k; \omega)$; this particle set can be easily reweighted as a function of $\omega$ given a suitable choice of importance sampling density $q(x_k)$;  
	\item condensation of the final optimally weighted particle set produced by step 1 (which is at some optimal $\omega$) into a GM using the weighted EM algorithm \cite{Goldberger08}.  
\end{enumerate}

\begin{algorithm}[t] 
\caption{IS Optimization for GM WEP Fusion}
{
\begin{algorithmic} \label{alg:ISOptAlg}
\STATE \textbf{Input}: GM pdfs $p_i(x_k)$ and $p_j(x_k)$; number of samples $N_s$; initial guess $\omega_0$; IS pdf exponent $\bar{\omega}$; 1D convex minimization rule $R_{\smC{1D}}[\hat{f}_{ij}(\omegacurr),\omegacurr,\omegaold]$; 
\STATE \textbf{Output}: $\hat{\omega}^* \in [0,1]$; samples $\set{x^s_k}^{N_s}_{s=1}$, unnormalized weights $\set{\theta_s}^{N_s}_{s=1}$ 
\STATE 1. Initialize  $\omegacurr \leftarrow \omega_0$ and $\omegaold$ according to $R_{\smC{1D}}$
\STATE 2. construct GM IS pdf $q(x_k)$ via eq. \pareqref{FOCIApprox} with $\omega = \bar{\omega}$
\STATE 3. draw $N_s$ samples $\set{x^s_k}_{s=1}^{N_s} \sim q(x_k)$
\STATE 4. store pdf values $p_i(x^s_k)$, $p_j(x^s_k)$, $q(x^s_k)$ for $\set{x^s_k}_{s=1}^{N_s}$
\WHILE{$\omegacurr$ not converged}
\STATE 5. compute $\theta_s(x^s_k;\omegacurr) = \frac{[p_i(x^s_k)]^{\omegacurr}[p_j(x^s_k)]^{1-\omegacurr}}{q(x^s_k)}$
\STATE 6. compute WEP cost estimate $\hat{f}_{ij}(\omegacurr)$ 
\STATE 7. modify $\omegaold$ and $\omegacurr$ via $R_{\smC{1D}}[\hat{f}(\omegacurr),\omegacurr,\omegaold]$
\ENDWHILE
\end{algorithmic}
}
\end{algorithm}

\begin{algorithm}[t] 
\caption{GM Learning by Weighted EM}
{
\begin{algorithmic} \label{alg:WEMLearning}
\STATE \textbf{Input}: samples $\set{x^s_k}^{N_s}_{s=1}$, unnormalized weights $\set{\theta(x^s_k;\hat{\omega}^*)}^{N_s}_{s=1}$, number of components $M^{f}$, maximum number of steps $N_{\smC{max}}$
\STATE \textbf{Output}: GM approximation of $\set{x^s_k,\theta(x^s_k;\hat{\omega}^*)}^{N_s}_{s=1}$
\STATE 1. enter initial guess of GM parameters $\set{\mu_z, \Sigma_z, w_z}_{z=1}^{M^f}$
\STATE 2. normalize $\set{\theta(x^s_k;\hat{\omega}^*)}_{s=1}^{N_s}$ s.t. $\sum_{s=1}^{N_s}{\theta(x^s_k;\hat{\omega}^*)}=1$
\STATE 3. set counter $k=0$
\WHILE{($\set{\mu_z, \Sigma_z, w_z}_{z=1}^{M^f}$ not converged \AND $k\leq N_{\smC{max}}$)}
\STATE 4. \textit{E-step:} for $s\in \set{1,...,N_s}$ and $z\in \set{1,...,M^f}$, compute weighted component responsibilities and normalizers:
\begin{align}
\gamma_s^z = \frac{\theta(x^s_k;\hat{\omega}^*) \cdot w_z \cdot {\cal N}(x^s_k; \mu_z, \Sigma_z)}{\sum_{y=1}^{M^f}{w_y \cdot {\cal N}(x^s_k; \mu_y, \Sigma_y)}}, \ \ \ 
\bar{N}^z = \sum_{s=1}^{N_s}{\gamma_s^z} \nonumber
\end{align}  		
\STATE 5. \textit{M-step:} for $z\in \set{1,...,M^f}$, compute GM parameters
\begin{align}
w_z = {\bar{N}^z}, \ \ \ \mu_z = \frac{1}{\bar{N}^z} \sum_{s=1}^{N_s}{\gamma_s^z \cdot x^s_k}, \ \ \ \nonumber \\
\Sigma_z = \frac{1}{\bar{N}^z} \sum_{s=1}^{N_s}{ \gamma_s^z \cdot ( x^s_k x^{s,T}_k - \mu_z\mu_z^T) } \nonumber
\end{align}
\STATE 6. update $k = k + 1$;
\ENDWHILE
\end{algorithmic}
}
\end{algorithm}

These two steps are detailed in Algorithms \ref{alg:ISOptAlg} and \ref{alg:WEMLearning}, respectively. 
In the first step (Algorithm \ref{alg:ISOptAlg}), $N_s$ samples $x^s_k$ in $x_k$ space are drawn only once according to the importance density $q(x_k)$ and used to estimate $f_{ij}(\omega)$ as a function of $\omega$. The importance weights 
\begin{align}
{\theta_s}(x^s_k;\omega) = \frac{[p_i(x^s_k)]^{\omega} [p_i(x^s_k)]^{1-\omega}}{q(x^s_k)} \label{eq:ISwtswep}
\end{align}
are adjusted as a function of new $\omega$ values during a search over $\omega \in [0,1]$ to minimize $f_{ij}(\omega)$, which can be implemented using a fast zeroth-order 1D optimization algorithm, e.g. golden section or bisection search. This does not require gradients or higher order derivative information for $f_{ij}(\omega)$, but only point evaluations of $f_{ij}(\omega)$ via the importance sampling estimator (such that $q(x_k)$ can remain fixed throughout). 
For the Chernoff fusion rule, $f_{ij}(\omega)$ is thus approximated as
\begin{align}
&f_{ij}(\omega) = \int_{-\infty}^{\infty}{[p_i(x_k)]^{\omega} [p_j(x_k)]^{1-\omega} dx_k} \nonumber \\
\approx &\hat{f}_{ij}(\omega) = \sum_{s=1}^{N_s}{\theta_s}(x^s_k;\omega) \label{eq:ISchernoff}.
\end{align}
Likewise, for the minimax information loss fusion rule, $f_{ij}(\omega)$ is (up to an additive constant independent of $\omega$) approximated as 
\begin{align}
&f_{ij}(\omega) = \mbox{const.} \ + \omega \cdot \kappa + \log \int_{-\infty}^{\infty}{[p_i(x_k)]^{\omega} [p_j(x_k)]^{1-\omega} dx_k} \nonumber \\
\approx &\hat{f}_{ij}(\omega) = \mbox{const.} + \omega \cdot \kappa  + \log \left( \sum_{s=1}^{N_s}{\theta_s}(x^s_k;\omega) \right) \label{eq:ISminimax}, \\
&\kappa = \int_{-\infty}^{\infty}{p_i(x_k) p_j(x_k) \log \left( \frac{p_j(x_k)}{p_i(x_k)} \right) dx_k}, 
\end{align}
where $\kappa$ can be pre-computed and stored prior to optimization of $\omega$ in the case of GMs, e.g. via sigma point approximation \cite{Goldberger08} or averaging of upper-/lower-bounds for logarithms of GMs \cite{Huber-MFI-2008}. 
The importance sampling density $q(x_k)$ can be chosen freely, but in refs. \cite{Ahmed-TSP-2012, ASC-RSS-2012} it was empirically found that setting $q(x_k)$ to the FOCI approximation in \pareqref{FOCIApprox} with a constant $\bar{\omega}=0.5$ generally offers good performance for a wide range of input pdfs, when using either the Chernoff or minimax information loss fusion rules. 
It is also interesting to note that, for any convex $f_{ij}(\omega)$, the number of approximate cost function evaluations and search iterations required to converge on $\omega^*$ within a desired tolerance can be pre-determined via the golden section search method \cite{Vanderplaats}. 

The second step (Algorithm \ref{alg:WEMLearning}) uses the weighted EM algorithm to condense the weighted particles produced by the first step into a GM. The weighted EM algorithm generalizes the classical EM algorithm for maximum likelihood estimation by accounting for the relative influence of individual data points in the log-likelihood function. By weighting individual data points relative to one another via scalar weights (in this case, the importance weights $\theta(x^s_k;\omega)$ produced from the first step), the log-likelihood function can be interpreted more generally as a `free-energy' cost function to be minimized with respect to the unknown parameters of the GM representing the WEP fusion result \cite{Goldberger08}. As with classical EM, the weighted EM free-energy cost results in a non-convex optimization problem, for which convergence to a local minimum can be assured via iterative minimization of convex lower bounds to the free-energy cost. This results in an iterative coordinate-descent strategy akin to classical EM, with an `E-step' that re-computes weighted expectations for latent variables (posterior mixand association probabilities) for fixed GM parameter values, and an `M-step' that re-computes model parameters (mixand weights, means and covariances) from the weighted data with fixed data-to-mixand assignment probabilities. Ref. \cite{Goldberger08} derives the weighted EM algorithm for GM pdf estimation. Note that classical EM can also be used in place of Algorithm \ref{alg:WEMLearning}, if the samples produced by Algorithm \ref{alg:ISOptAlg} are first resampled according to their importance weights in a particle resampling step, such that the resulting samples can then be reweighted to have uniform importance weights. 

As demonstrated in refs. \cite{ASC-RSS-2012, Ahmed-TSP-2012}, the stochastic optimization approach of Algorithm \ref{alg:ISOptAlg} is not only computationally cheap and fast, but also generally provides accurate estimates of $\omega^*  = \arg \min f_{ij}(\omega)$ along with an efficient set of weighted particles that accurately represent $\pwepddf{x_k;\omega}$, even in high-dimensional settings. However, this technique only applies to approximate WEP DDF, and as of yet has no obvious analog for exact DDF. 
The weighted EM particle to GM condensation approach in Algorithm \ref{alg:WEMLearning} also requires at least several iterations through non-trivial E-step and M-step calculations to achieve convergence, and also requires multiple GM parameter initializations/restarts to avoid getting trapped by poor local maxima in the non-convex weighted log-likelihood function landscape. 
As such, there is no guarantee that the weighted EM GM condensation step will produce an accurate approximation of the GM WEP fusion pdf, even if $\omega^*$ is reliably identified and samples are reliably drawn from the corresponding $\pwepddf{x_k;\omega}$ following Algorithm \ref{alg:ISOptAlg}. 

It is interesting to note that all of the aforementioned methods for GM-based DDF seek to simultaneously approximate all parts of the GM fusion pdf at once, and yet use relatively little information from the full fusion pdf itself. The MMGD and FOCI approximations, for example, rely only on the `inputs' to the fusion problem, i.e. the resulting GM approximations are not guided by comparison to the RHS of \pareqref{exactDef} or \pareqref{wepDef} (aside from the choice of $\omega$ for FOCI). 
On the other hand, existing importance sampling methods extract `local' information from the RHS of \pareqref{exactDef} or \pareqref{wepDef}, but only in the vicinity of a given particle. 
The approximations derived next exploit the fact that the true GM fusion pdf (for both exact and WEP fusion) is a mixture of non-Gaussian component pdfs, which can be naturally approximated component-wise by a GM. This structural insight leads to the development of alternative `divide and conquer' strategies for GM DDF approximations that allow various parts of the global GM fusion pdf to be obtained via simpler and more accurate parallel update operations on a per mixand basis. Furthermore, these techniques provide a more unified picture of GM-based DDF for both exact and WEP-based implementations. 
\section{Mixture Posteriors for General GM DDF }
Replacing $p^i(x_k)$, $p^j(x_k)$, and $p^{c,ij}(x_k)$ with GM pdfs in eqs. \pareqref{exactDef} and \pareqref{wepDef} yields, respectively,
\begin{align}
\pexactddf{x_k} \propto 
\frac{\bigparenth{ \GaussianMix{i}{v}{M}{w}{\mu}{\Sigma}{x_k}} \bigparenth{ \GaussianMix{j}{r}{M}{w}{\mu}{\Sigma}{x_k} }}{\GaussianMix{c,ij}{z}{M}{w}{\mu}{\Sigma}{x_k}}, \label{eq:exactGM} \\
\pwepddf{x_k} \propto 
\bigsquare{\GaussianMix{i}{v}{M}{w}{\mu}{\Sigma}{x_k}}^{\omega}\bigsquare{\GaussianMix{j}{r}{M}{w}{\mu}{\Sigma}{x_k}}^{1-\omega} \label{eq:wepGM},
\end{align}
for which we seek recursive GM approximations. Note, however, that \pareqref{wepDef} can also be rewritten as
\begin{align}
\pwepddf{x_k} &\propto \frac{p^i(x_k)p^j(x_k)}{[p^i(x_k)]^{1-\omega} [p^j(x_k)]^{\omega}} 
\propto \frac{p^i(x_k)p^j(x_k)}{\hat{p}^{c,ij}(x_k)} \\
&\propto \frac{\bigparenth{ \GaussianMix{i}{v}{M}{w}{\mu}{\Sigma}{x_k}} \bigparenth{ \GaussianMix{j}{r}{M}{w}{\mu}{\Sigma}{x_k} }}{\bigsquare{\GaussianMix{i}{v}{M}{w}{\mu}{\Sigma}{x_k}}^{1-\omega}\bigsquare{\GaussianMix{j}{r}{M}{w}{\mu}{\Sigma}{x_k}}^{\omega}},
\end{align}
where (in view of \pareqref{exactDef}) $\hat{p}^{c,ij}(x_k; \omega) \equiv \frac{1}{\tau}[p^i(x_k)]^{1-\omega}[p^j(x_k)]^{\omega}$ (with normalizing constant $\tau$) can be interpreted as an `estimated common information pdf' for $i$ and $j$ \cite{Ahmed-MFI-2014}.  
Since $\omega$ can be obtained for arbitrary WEP fusion cost functions even when $\pwepddf{x_k}$ is not available in closed-form \cite{ASC-RSS-2012, Ahmed-TSP-2012}, this implies that the problems of approximating the LHS of eqs. \pareqref{exactGM} and \pareqref{wepGM} by GMs are essentially equivalent once $\omega$ is given. In particular, we must find a recursive GM approximation to the generic fusion pdf $p^{f}(x_k)$ given by the quotient
\begin{align}
p^{f}(x_k) 
&\propto \frac{ \bigparenth{ \GaussianMix{i}{v}{M}{w}{\mu}{\Sigma}{x_k}} \bigparenth{ \GaussianMix{j}{r}{M}{w}{\mu}{\Sigma}{x_k} } }{ u(x_k)}, \label{eq:unifiedGM}
\end{align}
where $u(x_k)$ is an arbitrary non-Gaussian pdf that is given by either the exact common information pdf $p^{c,ij}(x_k)$ (which can be approximated as a GM for recursive implementation) 
or estimated common information pdf $\hat{p}^{c,ij}(x_k;\omega)$ for some known $\omega$. 

The numerator of the `unified' GM fusion expression in \pareqref{unifiedGM} can be rearranged to give  
\begin{align}
\pgenericddf{x_k} \propto \sum_{v=1}^{M^i}{ \sum_{r=1}^{M^j}{ w^i_v w^j_r \frac{ {\cal N}(x_k; \mu^i_v, \Sigma^i_v) {\cal N}(x_k; \mu^j_r, \Sigma^j_r) }{ u(x_k) } } }. \label{eq:doublesum}
\end{align}
Using the fact that the product of two Gaussian pdfs results in another unnormalized Gaussian pdf, this can be further simplified to
\begin{align}
\pgenericddf{x_k} &\propto \sum_{v=1}^{M^i}{ \sum_{r=1}^{M^j}{ w^{ij}_{vr}  \frac{ \bar{z}^{ij}_{vr} {\cal N}(x_k; \mu^{ij}_{vr}, \Sigma^{ij}_{vr})}{ u(x_k)} } } \nonumber \\
&= \sum_{v=1}^{M^i}{ \sum_{r=1}^{M^j}{ \tilde{w}^{ij}_{vr} m_{vr}(x_k) } },
\label{eq:ddfGMsum}
\end{align}
where 
\begin{align}
m_{vr}(x_k) &= \frac{{\cal N}(x_k; \mu^{ij}_{vr}, \Sigma^{ij}_{vr})}{u(x_k)}, \\
\Sigma^{ij}_{vr} &= \inv{ \bigsquare{ \inv{\bigparenth{\Sigma^{i}_{v}}} + \inv{\bigparenth{ \Sigma^{j}_{r}}} } }, \\
\mu^{ij}_{vr} &= \Sigma^{ij}_{vr} \bigsquare{ \inv{\bigparenth{\Sigma^{i}_{v}}}\mu^{i}_{v} + \inv{\bigparenth{ \Sigma^{j}_{r}}}\mu^{j}_{r} },  \\
\tilde{w}^{ij}_{vr} &= w^i_v w^j_r \bar{z}^{ij}_{vr}, \label{eq:wtildeTrue} \\
\bar{z}^{ij}_{vr} &= {\cal N}(\mu^{i}_{v}; \mu^{j}_{r}, \bigparenth{\Sigma^{i}_{v} + \Sigma^{j}_{r} }).
\end{align}
Eq. \pareqref{ddfGMsum} thus shows that \pareqref{unifiedGM} is naturally a \emph{mixture of non-Gaussian pdfs}, where each mixand $m_{vr}(x_k)$ is the ratio of a Gaussian pdf ${\cal N}(x_k; \mu^{ij}_{vr}, \Sigma^{ij}_{vr})$ (resulting from component-wise `naive Bayes' fusion of $p^i(x_k)$ and $p^j(x_k)$) and a generally non-Gaussian pdf $u(x_k)$. Note that, by virtue of \pareqref{ddfGMsum}, $u(k)$ is generally a (non-Gaussian) mixture model even in the case of WEP fusion, since $\hat{p}^{c,ij}(x_k)$ can be expanded the same way as $\pwepddf{x_k}$ with the exponents reversed. Also note that this expansion closely resembles the MMGD approximation for exact GM-based DDF described earlier. Namely, MMGD is a special case of the above mixture expansion for exact DDF, where the common information term $u(x_k)$ is approximated by a moment-matched Gaussian pdf. 
\section{GMs from Moment-Matching Approximations} \label{methods}

The mixands $m_{vr}(x_k)$ in \pareqref{ddfGMsum} cannot be integrated analytically to obtain normalization constants. Furthermore, they do not possess sufficient statistics for scalable DDF recursions. However, a GM approximation of the mixture fusion pdf $\pgenericddf{x_k}$ could be obtained if each non-Gaussian mixand $m_{vr}(x_k)$ were replaced with a moment-matched Gaussian pdf, so that  
\begin{align}
\pgenericddf{x_k} &\approx \frac{1}{\eta} \sum_{v=1}^{M^i}{ \sum_{r=1}^{M^j}{ \tilde{w}^{*}_{vr} {\cal N}(x_k; \mu^{*}_{vr}, \Sigma^{*}_{vr})} },
\label{eq:ddfgmapprox} \\
\mbox{where \ } \ \tilde{w}^{*}_{vr} &= w^i_v w^j_r \bar{z}^{ij}_{vr} \cdot \EV{ 1 }_{\tilde{m}_{vr}(x_k)}, \label{eq:wstar} \\
\mu^{*}_{vr} &= \EV{ x_k } _{\tilde{m}_{vr}(x_k)}, \label{eq:mustar} \\
\Sigma^{*}_{vr} &= \EV{ x_k x^T_k }_{\tilde{m}_{vr}(x_k)} - \mu^{*}_{vr} (\mu^{*}_{vr})^T, \label{eq:sigstar} \\
\eta &= \sum_{v=1}^{M^i}{ \sum_{r=1}^{M^j}{ \tilde{w}^{*}_{vr} }}, \ \ \
\tilde{m}_{vr}(x_k) \propto \frac{ {\cal N}(x_k; \mu^{ij}_{vr}, \Sigma^{ij}_{vr})}{ u(k)}, \nonumber
\end{align}
and where $\tilde{m}_{vr}(x_k)$ is the $(v,r)^{th}$ normalized mixand term from \pareqref{ddfGMsum}. 
This component-wise approximation could also be augmented with additional Gaussian mixture terms to capture higher order moments for each non-Gaussian $\tilde{m}_{vr}$ mixand. In the remainder of this paper, it will be assumed that only the zeroth, first and second moments are of interest for each $\tilde{m}_{vr}$, since these are sufficient for constructing a Gaussian mixture approximation to $\pgenericddf{x_k}$. 
In any case, the GM approximation to the fusion posterior pdf allows (both exact and WEP) GM DDF to proceed recursively.   
However, since the required moments cannot be found analytically, they must also somehow be approximated.
  
Numerical quadrature methods offer the most straightforward pathway to approximating the required mixand moments. 
Monte Carlo importance sampling (IS) \cite{LiuBook} is one such approach, which exploits the identity
\begin{align}
\EV{f(x_k)}_{b(x_k)} &= \EV{ \frac{b(x_k)}{q(x_k)} f(x_k) }_{q(x_k)} \nonumber \\
 &= \EV{ \theta(x_k) f(x_k) }_{q(x_k)} \label{eq:ISdef} 
\end{align} 
where $f(x_k)$ is a given moment function and $q(x_k)$ is a proposal pdf (which is easy to directly sample from) for the target distribution $b(x_k)$ (which is difficult to directly sample). Ideally $q$ has a shape `close' to $b(x_k)$, and has support on $x_k$ such that $b(x_k)>0 \Rightarrow q(x_k)>0$. 
Both $b(x_k)$ and $q(x_k)$ need only be known up to normalizing constants for point-wise evaluation at any given $x_k$. 
Given a set of $N_s$ samples ${\cal X} = \set{x^s_k}_{s=1}^{N_s} \sim q(x_k)$, \pareqref{ISdef} has the sampling estimate
\begin{align}
\EV{f(x_k)}_{b(x_k)} \approx \sum_{s=1}^{N_s}{ \theta(x^s_k) f(x^s_k) } \label{eq:MonteCarloIS}, 
\end{align}
where $\theta(x^s_k) \propto \frac{b(x_k)}{q(x_k)}$ is the importance weight for sample $s$. 
Informally, $\theta(x^s_k)$ indicates `how much' the sample $s$ contributes to the estimate. 
The \emph{effective sample size (ESS)} is a useful figure of merit for assessing the efficiency of the weighted samples ${\cal X}$ for estimating \pareqref{MonteCarloIS} \cite{LiuBook}, 
\begin{align}
ESS = \frac{N_s}{1+\mbox{cv}^2(\theta)}, \ \ \mbox{for \ }
\mbox{cv}^2(\theta) = \frac{ \sum_{s=1}^{N_s} (\theta^s - \bar{\theta})^2 }{ \sum_{s=1}^{N_s}(N_s - 1)\bar{\theta}^2 },
\label{eq:ESS}
\end{align}
where $\mbox{cv}^2$ is the coefficient of variation for the unnormalized importance weights and $\bar{\theta}$ is the sample mean of the importance weights. 
As $ESS \rightarrow N_s$, the weighted ${\cal X}$ from $q(x_k)$ provide a better representation of the target distribution $b(x_k)$ and thus lead to more consistent moment estimates. 

Therefore, the problem of accurately approximating the moments of each $\tilde{m}_{vr}$ in \pareqref{ddfgmapprox} can be addressed through selection of a suitable IS pdf $q$ and number of samples $N_s$. 
IS is easily parallelized and permits many strategies for selecting/tailoring the proposal function $q(x_k)$, so a few different possibilities can be considered. 
Since either the full posterior fusion pdf or any posterior fusion mixand in \pareqref{ddfGMsum} can be evaluated up to a normalizing constant as the target distribution $b(x_k)$ of interest, two sets of IS techniques are developed, which are each applicable to both exact and WEP DDF: 
\begin{enumerate}
	\item \emph{indirect global sampling (IGS)} wherein $b(x_k) \propto p^{f}(x_k)$ or $\pwepddf{x_k}$, such that a single set of weighted ${\cal X}$ samples is generated only once and used to estimate the moments of all $\tilde{m}_{vr}$ terms simultaneously; 
	\item \emph{direct local sampling (DLS)} wherein $b(x_k) \propto \tilde{m}_{vr}$, such that a different set of weighted samples ${\cal X}$ are generated once for each $\tilde{m}_{vr}$ and used to estimate $\tilde{m}_{vr}$'s moments only.  
\end{enumerate}

\subsection{Indirect global sampling}
\begin{figure}[t]
\centering
\newcommand{\figsize}{9.5cm}
\includegraphics[width=\figsize]{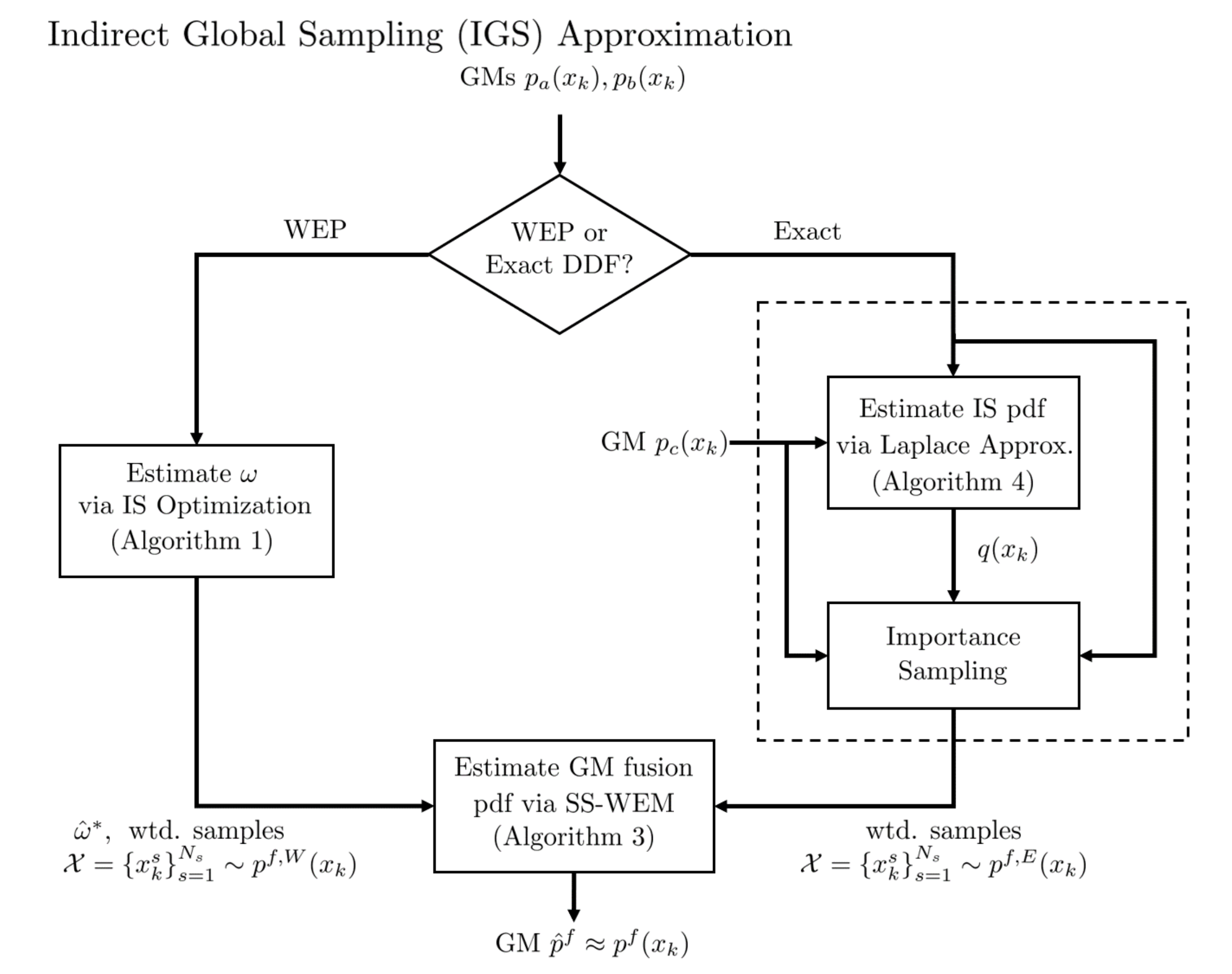}
\caption{Block diagram of IGS for GM-based DDF.}
\label{fig:IGSApproxBlockDiag}
\end{figure}

This subsection describes the indirect global sampling (IGS) approximation for WEP and exact DDF with GMs, respectively. 
The IGS approximation generally consists of a two-step optimization process for either form of DDF. 
The first step generates importance samples over the global posterior fusion mixture in eq. \pareqref{unifiedGM}.  
The second step probabilistically associates these importance samples to the various posterior fusion pdf mixands to facilitate weighted maximum likelihood estimation of their individual zeroth, first and second moments for the final GM approximation in \pareqref{ddfgmapprox}. 
The first step uses Algorithm \ref{alg:ISOptAlg} for WEP DDF; a newly developed IS approximation based on Laplace's method is used instead for exact DDF. 
The second step for both WEP and exact DDF is outlined in the newly developed \textit{single-shot weighted EM (SS-WEM)} procedure Algorithm \ref{alg:SSWEMLearning}, which is far more efficient and stable than the WEM approximation outlined in Algorithm \ref{alg:WEMLearning} and used in previous work. Figure \ref{fig:IGSApproxBlockDiag} shows the main steps for the IGS approximation in WEP and exact GM DDF; these are each explained in greater detail next. 

\subsubsection{IGS approximation for WEP GM DDF}

The IGS implementation for WEP GM DDF modifies the two-stage IS-based optimization process developed originally in refs. \cite{ASC-RSS-2012, Ahmed-TSP-2012}.  
The initial IS-based optimization of $\omega$ in Algorithm \ref{alg:ISOptAlg} remains the same, since this simultaneously produces reliable estimates of the optimal $\omega$ and weighted samples that closely approximate the WEP fusion posterior pdf \pareqref{unifiedGM} at the optimal $\omega$ value. 
However, the final iterative weighted expectation-maximization step in Algorithm \ref{alg:WEMLearning} is replaced by single shot weighted expectation-maximization (SS-WEM) to recover posterior fusion pdf. As shown in Algorithm \ref{alg:SSWEMLearning}, the SS-WEM algorithm computes the E-step \emph{only once} via evaluations of the true fusion pdf mixands for $p^f$ in \pareqref{ddfgmapprox} (up to a normalization constant) for each sample point. The subsequent M-step is also computed only once to estimate the required GM parameters of $\hat{p}^f$. In this way, the SS-WEM algorithm addresses the WEM algorithm's non-trivial computational overhead and inability to reliably converge to a GM $\hat{p}^f$ which closely approximates $p^f$. 

\begin{algorithm}[t] 
\caption{GM Learning by Single-Shot Weighted EM}
{
\begin{algorithmic} \label{alg:SSWEMLearning}
\STATE \textbf{Input}: samples $\set{x^s_k}^{N_s}_{s=1}$, unnormalized weights $\set{\theta(x^s_k;\hat{\omega}^*)}^{N_s}_{s=1}$, number of components $M^{f}=M_i \cdot M_j$, true fusion pdf Gaussian parameters $\set{\tilde{w}^{ij}_{z}, \mu^{ij}_{z}, \Sigma^{ij}_{z}}_{z=1}^{M^f}$ for platforms $i$ and $j$ as defined in \pareqref{ddfGMsum}. 
\STATE \textbf{Output}: GM approximation of $\set{x^s_k,\theta(x^s_k;\hat{\omega}^*)}^{N_s}_{s=1}$
\STATE 1. \textit{Constrained E-step:} for $s\in \set{1,...,N_s}$ and $z\in \set{1,...,M^f}$, compute weighted component responsibilities and normalizers using true fusion pdf:
\begin{align}
\gamma_s^z = \frac{\theta(x^s_k;\hat{\omega}^*) \cdot \tilde{w}^{ij}_{z} \cdot {\cal N}(x^s_k; \mu^{ij}_{z}, \Sigma^{ij}_{z})}{\sum_{z'=1}^{M^f}{\tilde{w}^{ij}_{z'} \cdot {\cal N}(x^s_k; \mu^{ij}_{z'}, \Sigma^{ij}_{z'})}}, \ \ \ 
\bar{N}^z = \sum_{s=1}^{N_s}{\gamma_s^z} \nonumber
\end{align}  		
\STATE 2. \textit{M-step:} for $z\in \set{1,...,M^f}$, compute GM parameters
\begin{align}
w_z = {\bar{N}^z}, \ \ \ \mu_z = \frac{1}{\bar{N}^z} \sum_{s=1}^{N_s}{\gamma_s^z \cdot x^s_k}, \ \ \ \nonumber \\
\Sigma_z = \frac{1}{\bar{N}^z} \sum_{s=1}^{N_s}{ \gamma_s^z \cdot ( x^s_k x^{s,T}_k - \mu_z\mu_z^T) } \nonumber
\end{align}
\end{algorithmic}
}
\end{algorithm}

The SS-WEM can be understood in more detail as follows, building on derivation of the WEM algorithm for GM learning provided in \cite{Goldberger08}. Given sample data ${\cal X} = \set{x^s_k}_{k=1}^{s}$ with non-negative weights $\theta^s_k$, WEM seeks to fit parameters $\Theta^{*} = \set{\mu^{*}_{vr}, \Sigma^{*}_{vr} , \tilde{w}^{*}_{vr}}_{(v,r)}$ for each of the components indexed by pair $(v,r) \mapsto z$ in eq. \pareqref{ddfgmapprox} by maximizing the weighted log-likelihood function, which in turn is the same as double-maximization of the associated `free-energy' function,
\begin{align}
\Theta^* = \arg \max \sum_{s=1}^{N_s}{\theta(x^s_k) \log \hat{p}^{f}(x^s_k; \Theta) } = \arg \max_{\Theta} \max_{b({\cal Y})} FE(b,\Theta), \label{eq:wtdLogLikelihood}
\end{align}
where ${\cal Y} = \set{y^s_k}_{k=1}^s$ is the set of latent mixand labels for ${\cal X}$ (whose realizations are one-hot vectors), $b({\cal Y})$ is a probability distribution over ${\cal Y}$, and $FE(b,\Theta)$ is the free-energy, 
\begin{align}
FE(b,\Theta) = \sum_{s=1}^{N_s}{\theta^s_k} \sum_{y^s_k}{b(y^s_k)} \log \hat{p}^{f}(y^s_k,x^s_k; \Theta) + \sum_{s=1}^{N_s}{\theta^s_k {\cal H}[b(y^s_k)]}, \label{eq:freeEnergy}
\end{align}
where ${\cal H}[b(y^s_k)]$ is the entropy of $b(y^s_k)$ and $\hat{p}^{f}(y^s_k,x^s_k; \Theta^{*}) = [\tilde{w}^{*}_{vr} \cdot {\cal N}(x^s_k; \mu^{*}_{vr} ,\Sigma^{*}_{vr}) ]^{y^s_k}$ is the joint distribution for the unknown parameters, latent mixand labels, and observed samples under the \emph{approximate} fusion GM pdf in eq. \pareqref{ddfgmapprox}. 
Double-maximization of $FE(b,\Theta)$ is achieved by performing alternating E-step and M-step updates to maximize $FE(b,\Theta)$ with respect to $b({\cal Y})$ (holding $\Theta$ fixed) and $\Theta$ (holding $b({\cal Y})$ fixed) until convergence, as shown in Algorithm \ref{alg:WEMLearning}. Recall that the E-step finds the weighted posterior component responsibilities for each datum $x^s_k$ with respect to the approximate fusion GM pdf \pareqref{ddfgmapprox} using the current $\Theta^{*}$ estimate,
\begin{align} 
b(y^s_k=z) = \gamma^z_s= \theta(x^s_k)\cdot \hat{P}(y^s_k=z|x^s_k, \Theta^{*}) = \theta(x^s_k) \frac{\tilde{w}^{*}_{z} \cdot {\cal N}(x^s_k; \mu^{*}_{z} ,\Sigma^{*}_{z}) }{ \sum_{z'=1}^{M^f} \tilde{w}^{*}_{z'} \cdot {\cal N}(x^s_k; \mu^{*}_{z'} ,\Sigma^{*}_{z'}) }, 
\label{eq:probEStep}
\end{align}
which reduces to $\hat{P}(y^s_k=z|x^s_k, \Theta^{*})$ in the classical unweighted EM algorithm. 

Since \pareqref{wtdLogLikelihood} is non-concave, WEM updates can converge to any one of a large number of poor local maxima without careful initialization of $\Theta$. 
The existence of these local maxima stem from three factors: (i) identifiability issues, i.e. aliasing of mixture labels; (ii) the fact that E-step iterations rely on \emph{estimated} joint probability distributions for the latent mixture labels and observed data; and (iii) inherent mismatches between the true mixture fusion pdf \pareqref{unifiedGM} (which truly describes the samples ${\cal X}$) and the GM approximation \pareqref{ddfgmapprox} being estimated. 
Factors (ii) and (iii) in particular pose problems for selecting appropriately sized Monte Carlo IS sample sets. A large $N_s$ helps control the variance of estimated GM parameters for $\hat{p}^{f}$ and better explore $p^f$, at greater computational expense. On the other hand, a small $N_s$ tends to `simplify' the weighted log-likelihood function landscape, at the risk of potentially overfitting parameters in $\hat{p}^{f}$ and losing features of $p^f$. 

The SS-WEM algorithm reduces sensitivity to all three factors by enforcing a constraint in the E-step that removes the need to continuously re-estimate $b({\cal Y})$ and $\Theta^*$. 
Namely, the weighted posterior mixand association probabilities $b(y^s_k=z)$ in \pareqref{probEStep} are modified to use the \emph{true} fusion pdf mixture $p^f$ in \pareqref{ddfGMsum}, instead of the approximate GM fusion pdf $\hat{p}^f$. 
If the joint distribution for $y^s_k$ and $x^s_k$ via \pareqref{ddfGMsum} is $p(y^s_k=z,x^s_k) = \tilde{w}_{z} \cdot m_{z}(x_k)$, then the constrained E-step update then becomes 
\begin{align}
b(y^s_k=z) = \theta(x^s_k)\cdot P(y^s_k=z|x^s_k) = \theta(x^s_k) \frac{ \tilde{w}^{ij}_{z} \cdot m^{ij}_{z}(x_k) }{ \sum_{z'=1}^{M^f} \tilde{w}^{ij}_{z'} \cdot m^{ij}_{z'}(x_k) }, 
\label{eq:modEStep}
\end{align}
where the index $z$ maps onto the $M^f = M^i \cdot M^j$ component index realizations $(v,r)$ from the product of the platform $i$ and $j$ GM pdfs in \pareqref{ddfGMsum}.  
From the definition of $m^{ij}_{z}$, it follows that 
\begin{align}
P(y^s_k=z|x^s_k) &= \frac{ \tilde{w}^{ij}_{z} \cdot m^{ij}_{z}(x^s_k) }{ \sum_{z'=1}^{M^f} \tilde{w}^{ij}_{z'} \cdot m^{ij}_{z'}(x^s_k)} \nonumber \\
&= \frac{ \tilde{w}^{ij}_{z} \cdot \frac{{\cal N}(x^s_k; \mu^{ij}_{z}, \Sigma^{ij}_{z})}{u(x^s_k)} }{ \sum_{z'=1}^{M^f} \tilde{w}^{ij}_{z'} \cdot \frac{{\cal N}(x^s_k; \mu^{ij}_{z'}, \Sigma^{ij}_{z'})}{u(x^s_k)} } \nonumber \\
&= \frac{ \tilde{w}^{ij}_{z} \cdot {\cal N}(x^s_k; \mu^{ij}_{z}, \Sigma^{ij}_{z}) }{ \sum_{z'=1}^{M^f} \tilde{w}^{ij}_{z'} \cdot {\cal N}(x^s_k; \mu^{ij}_{z'}, \Sigma^{ij}_{z'}) }.
\end{align} 
That is, $P(y^s_k=z|x^s_k)$ depends on neither the common information pdf value $u(x^s_k)$ nor the GM parameters $\Theta$, but rather depends only on the `Naive Bayes' Gaussian component terms that define the numerator of each $m^{ij}_z(x^s_k)$ mixand of $p^f$. 
The constrained E-step is thus computationally attractive, since the numerator Gaussians of each $m^{ij}_z(x^s_k)$ mixand are readily available and easily evaluated at each $x^s_k$ sample (e.g. these values can be stored as additional outputs for Algorithm \ref{alg:ISOptAlg}). Moreover, since $P(y^s_k=z|x^s_k)$ is constant, the constrained E-step and subsequent M-step only need to be computed once. Finally, the label aliasing problem is bypassed, since the mixand labels of \pareqref{ddfGMsum} are naturally fixed. 
\footnote{One minor caveat is that the number of mixand terms in $\hat{p}^f$ and $p^f$ must be the same. 
If this is not the case, then ad hoc merging/splitting of mixands in $\hat{p}^f$ and/or $p^f$ could be used, for instance, to ensure the same support for $\hat{P}(y^s_k|x^s_k, \Theta^{*})$ and $P(y^s_k|x^s_k)$.  }

This modified single-shot E-step is also theoretically justified from an optimization standpoint: since the underlying structure of $p^f$ can be inferred from the weighted sample set ${\cal X}$ and point-wise evaluation of \pareqref{ddfGMsum}, both sources of information should be combined to improve the GM approximation $\hat{p}^f$. 
The true data source pdf is typically unavailable for point-wise evaluation in GM pdf estimation.  
Yet, the free-energy view of the E-step as maximization of $FE(b,\Theta)$ with respect to $b({\cal Y})$ advantageously allows for direct exploitation of constraints on latent variable probabilities \cite{Neal-EMView-1998}. 
In this case, if $p^f$ truly is well-approximated by some GM $\hat{p}^f$, then $\hat{P}(y^s_k=z|x^s_k, \Theta^{*}) \rightarrow P(y^s_k=z|x^s_k)$ is expected upon convergence to $\Theta^*$. 
Hence, by fixing the E-step with respect to the `expected asymptotically optimal' $b({\cal Y})$ (which is independent of $\Theta^{*}$ via $P(y^s_k=z|x^s_k)$), the GM parameter estimates $\Theta^{*}$ from the M-step become local maxima of an \emph{equality-constrained} free-energy function.

\subsubsection{IGS approximation for exact GM DDF}

IGS for exact GM DDF is similar to the WEP DDF case, but assumes the availability of a GM common information pdf. Since it is not necessary to estimate an optimal $\omega$ value in this case, Algorithm \ref{alg:ISOptAlg} is not used. Instead, a different first stage optimization procedure is used to generate a suitable IS pdf $q$, whose weighted samples ${\cal X}$ approximate the true exact fusion pdf \pareqref{ddfGMsum}. This IS pdf is obtained by adapting a Laplace approximation \cite{azevedo1994laplace} to the fusion mixture \pareqref{ddfGMsum}, as shown in  Algorithm \ref{alg:LaplaceGMAlg}. Weighted importance samples from the Laplace approximation mixture model are then used to recover the GM approximation \pareqref{ddfgmapprox} via the second stage SS-WEM in Algorithm \ref{alg:SSWEMLearning}, as before. 

The construction of the IS pdf via the Laplace approximation can be understood in detail as follows; a more in depth review of the Laplace approximation can be found in ref. \cite{azevedo1994laplace}. 
In general, the Laplace approximation applies to $d$-dimensional integrals of the form 
\begin{align}
I_n = \int_{c^{d}_1}^{c^{d}_2} \cdots \int_{c^{1}_1}^{c^{1}_2} f(x) \exp(-n g(x)) dx^1 \cdots dx^d,
\end{align}
where $n$ is a large positive number, $g(x)$ is continuous, unimodal and twice differentiable with minimum at $\hat{x}$ inside the region of integration, and $f(x)$ is continuous, differentiable and nonzero at $\hat{x}$. For a sufficiently large $n$, the bulk of the contribution to the value of $I_n$ is from the region close to the minimum $\hat{x}$ of $g(x)$. Using a first-order Taylor series expansion of $f(x)$ and a second-order Taylor series expansion of $g(x)$ around $\hat{x}$ (where the gradient $\nabla_x g(x)=0$ at $\hat{x}$), it can be shown for sufficiently large $n$ that 
\begin{align}
I_n \approx 
f(\hat{x}) e^{-n g(\hat{x})} \int_{c^{d}_1}^{c^{d}_2} \cdots \int_{c^{1}_1}^{c^{1}_2} \exp(-\frac{n}{2} [x-\hat{x}]^{T} \Sigma_{\hat{x}} [x-\hat{x}])  dx^1 \cdots dx^d,
\end{align}
where $\Sigma_{\hat{x}}$ is the positive-definite Hessian of $g(x)$ at $\hat{x}$. Upon recognizing the unnormalized multivariate Gaussian pdf ${\cal N}_{x}(\hat{x},\Sigma^{-1}_{\hat{x}})$ in the integral, 
\begin{align}
I_n \approx f(\hat{x})e^{-n g(\hat{x})} \int_{c^{d}_1}^{c^{d}_2} \cdots \int_{c^{1}_1}^{c^{1}_2} \bigparenth{\frac{2\pi}{n}}^{\frac{d}{2}} |\Sigma^{-1}_{\hat{x}}|^{\frac{1}{2}} {\cal N}_{x}(\hat{x},\Sigma^{-1}_{\hat{x}}) dx^1 \cdots dx^d,
\end{align}
so that taking the limits for $c^{1}_1 \rightarrow -\infty, \cdots, c^{d}_1 \rightarrow -\infty,$ and $c^{1}_2 \rightarrow \infty, \cdots, c^{d}_2 \rightarrow \infty$ yields
\begin{align}
I_n \approx f(\hat{x})e^{-n g(\hat{x})}\bigparenth{\frac{2\pi}{n}}^{\frac{d}{2}} |\Sigma^{-1}_{\hat{x}}|^{\frac{1}{2}} \label{eq:LaplaceIntegralApprox}. 
\end{align}
From here it is possible to estimate the zeroth, first and second moments of the individual non-Gaussian quotient mixands $m_{vr}(x_k)$ of \pareqref{ddfGMsum}, by setting $n=1$ and $g(x_k)= -\log m_{vr}(x_k)$ with $f(\hat{x}_k)=1$ for zeroth moment, $f(\hat{x}_k)=\hat{x}_k$ for the first moment and $f(\hat{x}_k)=\hat{x}_k \hat{x}^T_k$ for the second moment. 
Note that this involves finding the minimum $\hat{x}_{vr,k}$ of $g(x_k)$ via numerical optimization. This can be done efficiently using Newton-Raphson or quasi-Newton methods, which can make use of the gradient and Hessian of $g(x_k)$. 
Upon convergence to $\hat{x}_{vr,k}$, $\tilde{m}_{vr}(x_k)$ can be approximated as 
\begin{align}
\tilde{m}_{vr}(x_k) \approx {\cal N}_{x_k,vr}(\hat{x}_{vr,k},\Sigma^{-1}_{\hat{x}_{vr,k}}). \label{eq:LaplaceGaussian}
\end{align}
Collecting the RHS along with the zeroth moments for each $vr$ term leads to following the mixture approximation of \pareqref{ddfGMsum} (with normalizing constant $c$)
\begin{align}
p^f \approx c \cdot \sum_{v=1}^{M^i}{ \sum_{r=1}^{M^j}{ \tilde{w}_{vr} {\cal N}_{x_{vr,k}}(\hat{x}_{vr,k},\Sigma^{-1}_{\hat{x}_{vr,k}}) } }, \label{eq:LaplaceGMApprox}.
\end{align} 
Since the Laplace approximation assumes that the probability mass for $\tilde{m}_{vr}(x_k)$ is concentrated and distributed symmetrically near $\hat{x}_k$, the estimated moments for each $m_{vr}(x_k)$ can be biased and thus accumulate errors for the overall Laplace mixture approximation of $p^f$. 
However, the RHS of \pareqref{LaplaceGMApprox} still gives useful information about the overall shape of \pareqref{ddfGMsum}, especially in regions of highest probability for each $\tilde{m}_{vr}(x_k)$. 
Hence, instead of using the RHS of \pareqref{LaplaceGMApprox} to directly approximate \pareqref{ddfGMsum}, it used to define an IS pdf $q(x_k)$. As with WEP DDF, the weighted samples from this $q(x_k)$ can then be compressed via the SS-WEM algorithm into a GM $\hat{p}^{f,E}$. This mitigates potential biases for the $m_{vr}(x_k)$ mixand moments in the Laplace approximation, and thus leads to an overall more accurate approximate global approximation of $p^{f,E}$. It is interesting to note that the idea of combining Laplace approximations with IS to mitigate biases has also been explored previously in the statistics literature for parameter estimation \cite{Kuk-JSCS-1999, Diciccio-JASA-1997}. 

\begin{algorithm}[t] 
\caption{Laplace GM IS pdf Computation}
{
\begin{algorithmic} \label{alg:LaplaceGMAlg}
\STATE \textbf{Input}: GM pdfs $p^i(x_k)$ and $p^j(x_k)$; common information GM pdf $p^c(x_k)$;  
\STATE \textbf{Output}: GM IS pdf $q(x_k) \propto \sum_{v=1}^{M^i}\sum_{r=1}^{M^j}{ \tilde{w}_{vr} {\cal N}_{x_{k},vr}(\hat{x}_{vr,k},\Sigma^{-1}_{\hat{x}_{vr,k}})}$
\FOR{$v=1:M^i$}
\FOR{$r=1:M^j$}
\STATE 1. initialize $\hat{x}_{vr,k}$;    
\STATE 2. find $\hat{x}_{vr,k} =\arg\min \ g(x_k) = \arg\min \ -\log \frac{{\cal N}(x_k;\mu^{ij}_{vr},\Sigma^{ij}_{vr})}{p^c(x_k)}$ (via quasi-Newton/Newton-Raphson);
\STATE 3. compute $\Sigma^{-1}_{\hat{x}_{vr,k}}$ from Hessian of $g(x_k)$ at $\hat{x}_{vr,k}$;  
\STATE 4. compute $\tilde{w}_{vr}$ using RHS of \pareqref{wtildeTrue}; 
\ENDFOR
\ENDFOR
\end{algorithmic}
}
\end{algorithm}
 
In practical terms, numerical optimization for each $m_{vr}(x_k)$ mixand term will converge to a local minimum for $\hat{x}_{vr,k}$ since the corresponding $g(x_k)$ is non-convex. Good initialization is therefore required for reliable results. 
One routine strategy that has been observed to work well in practice can be found from a `naive' Laplace approximation (\cite{azevedo1994laplace}, eq. 5) with $g(x_k)= - \log {\cal N}(x_k; \mu_{vr},\Sigma_{vr})$ (the negative log of $m_{vr}(x_k)$'s numerator, i.e. the Naive Bayes' pdf), giving the initial guess $\hat{x}_{vr,k} = \mu_{vr}$. 
Furthermore, while the Laplace approximation GM has been observed to work well as an IS pdf in practice, the sampling efficiency for IS can theoretically be improved. Similar to \cite{Kuk-JSCS-1999}, one strategy is to modify the Gaussians terms on the RHS of \pareqref{LaplaceGMApprox} to obtain wider or heavier-tailed mixand pdfs for $q(x_k)$. For example, the estimated covariance terms $\Sigma^{-1}_{\hat{x}_{vr,k}}$ could be inflated via heuristic scaling factors. Alternatively, each Gaussian could be replaced with a multivariate Student's t, Laplace, etc., or a unimodal heavy-tailed GM pdf whose components all share mean $\hat{x}_{vr,k}$ but each used different scaled versions of $\Sigma^{-1}_{\hat{x}_{vr,k}}$ to define covariances. As presented next, these and other similar efficiency-boosting strategies could also be adapted from/for DLS approximation.

\subsection{Direct local sampling}
\begin{figure}[t]
\centering
\newcommand{\figsize}{9.5cm}
\includegraphics[width=\figsize]{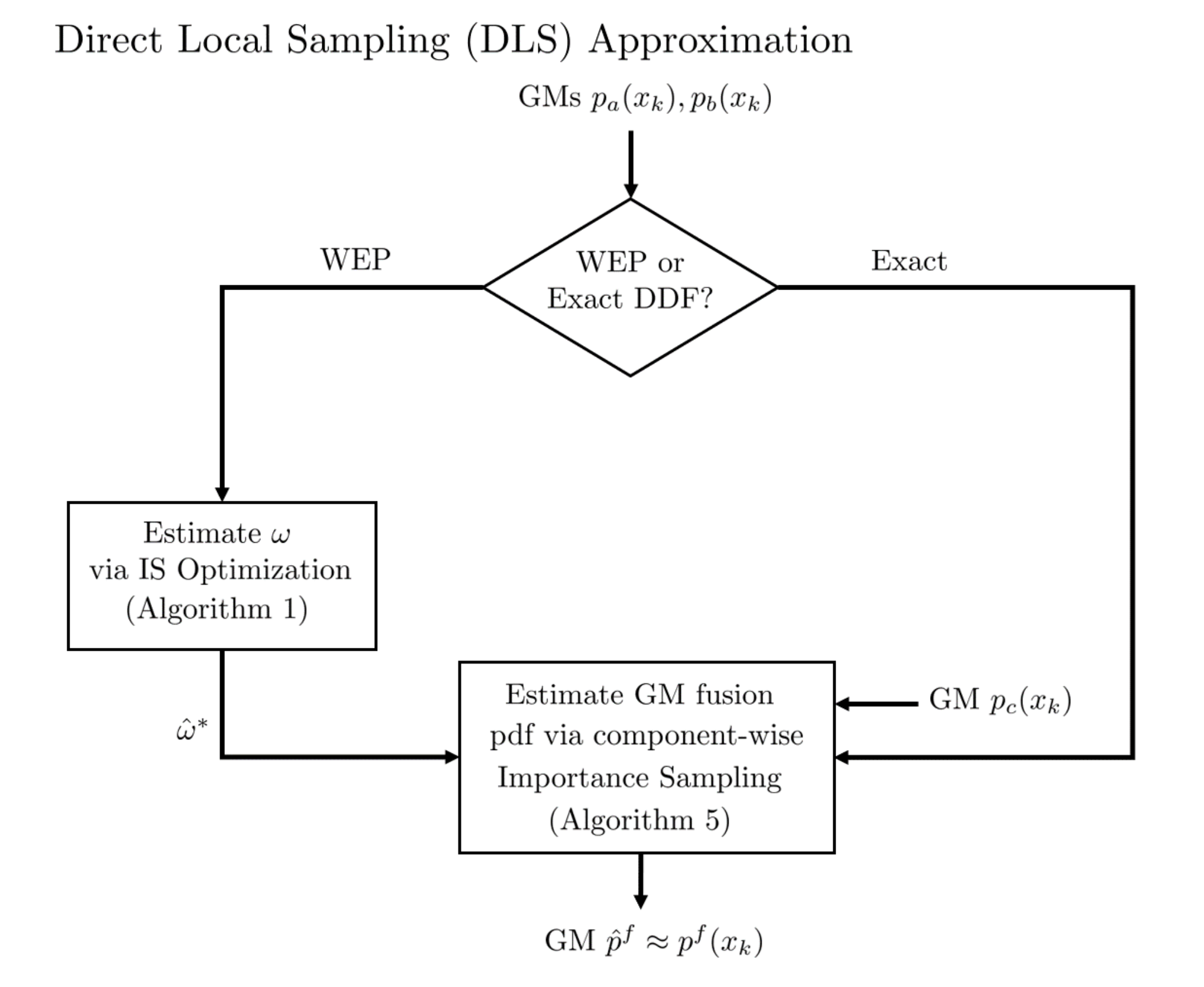}
\caption{Block diagram of DLS for GM-based DDF.}
\label{fig:DLSApproxBlockDiag}
\end{figure}

This subsection describes the direct local sampling (DLS) approximation for WEP and exact DDF with GMs. 
Figure \ref{fig:DLSApproxBlockDiag} and Algorithm \ref{alg:DLSGMAlg} show the main steps for the DLS approximation, which uses the target pdf $b(x_k) \propto \tilde{m}_{vr}(x_k)$ for IS and thus operates on a per mixand basis to directly estimate the zeroth, first, and second moments of each non-Gaussian mixand term in the RHS of \pareqref{ddfGMsum}. The procedure is essentially identical for WEP and exact DDF; the main difference is that Algorithm \ref{alg:ISOptAlg} must be run first for WEP DDF so that $u(x_k)$ can be computed, whereas $u(x_k)=p_c(x_k)$ is a known GM for exact DDF. Unlike IGS, the DLS approximation does not require a first stage optimization for exact DDF. However, DLS still requires a first stage optimization to find $\omega$ for WEP DDF, as in IGS. But unlike IGS, DLS for WEP fusion does not use the weighted samples obtained from Algorithm \ref{alg:ISOptAlg} to estimate $\hat{p}^{f,W}$. 

\begin{algorithm}[t] 
\caption{Direct Local Sampling Approximation for GM Fusion }
{
\begin{algorithmic} \label{alg:DLSGMAlg}
\STATE \textbf{Input}: GM pdfs $p^i(x_k)$ and $p^j(x_k)$; $u(x_k)=p^c(x_k)$ (exact) or $u(x_k)=[p^i(x_k)]^{1-\omega} [p^j(x_k)]^{\omega}$ (WEP);  
\STATE \textbf{Output}: GM approximation to $p^{f,E}(x_k)$ or $p^{f,W}(x_k)$
\FOR{$v=1:M^i$}
\FOR{$r=1:M^j$}
\STATE 1. Construct IS pdf $q_{vr}(x_k)$, e.g. using INGIS \pareqref{naiveGaussianIS}, LAGIS \pareqref{LaplaceGIS}, or heavy-tail mixture \pareqref{scaleLAGIS};   
\STATE 2. Draw $N_s$ samples ${\cal X} = \left\{x^s_k \right\}_{s=1}^{N_s}$ from $q_{vr}(x_k)$;
\STATE 3. Compute importance weights $\theta(x^s_k) \propto \frac{m_{vr}(x^s_k)}{q_{vr}(x^s_k)}$;
\STATE 4. Estimate $\tilde{w}^{*}_{vr}$, $\mu^{*}_{vr}$ and $\Sigma^{*}_{vr}$ in \pareqref{ddfgmapprox} via IS approximation of \pareqref{wstar}-\pareqref{sigstar}.
\ENDFOR
\ENDFOR
\STATE 5. Normalize weights such that $\sum_{v=1}^{M^i} \sum_{r=1}^{M^j} \tilde{w}^{*}_{vr}=1$.

\end{algorithmic}
}
\end{algorithm}

Overall, DLS is more computationally intensive than the IGS approximation, since IS sampling must now be carried out separately for each mixand. However, the IS sampling steps can be easily parallelized across the mixands of \pareqref{ddfGMsum} as well as across the samples generated for each mixand, making it possible to speed up implementation. 
Assuming $u(x_k)$ has been suitably identified for exact or WEP DDF, the IS pdf $q_{vr}(x_k)$ for each $\tilde{m}_{vr}(x_k)$ could be defined in a number of ways for DLS. A few strategies are considered next. 

\subsubsection{Inflated Naive Gaussian approximation}
A particularly simple and convenient (though suboptimal) strategy is to set 
\begin{align}
q_{vr}(x_k) = {\cal N}(x_k; \mu_{vr}, \Sigma^{\smC{samp}}_{vr}) \label{eq:naiveGaussianIS}
\end{align} 
for some suitable covariance matrix $\Sigma^{\smC{samp}}_{vr}$. This approach, dubbed here as \emph{ Inflated Naive Gaussian IS (INGIS)}, is generally effective for low-dimensional $x_k$ as long as $(\Sigma^{\smC{samp}}_{vr} - \Sigma^{*}_{vr})$ is positive semi-definite and the mode(s) of $m_{vr}(x_k)$ are not far from $\mu_{vr}$. One possible rule of thumb is to choose
\begin{align}
\Sigma^{\smC{samp}}_{vr} =  \arg \max(|\Sigma_v|, |\Sigma_r|, |\Sigma^{\smC{def}}|), \nonumber
\end{align}
where $\Sigma^{\smC{def}} = \alpha \cdot \eye$ and tuning parameter $\alpha$ represents an upper bound on the expected variance for any posterior mixand in any dimension (see Sec \ref{impdetails}). 
However, INGIS can perform poorly for high-dimensional $x_k$, since IS is generally much more sensitive to discrepancies between $\tilde{m}_{vr}(x_k)$ and $q(x_k)$ in such cases. This proposal distribution can also lead to inefficient sampling if the mean of $\tilde{m}_{vr}(x_k)$ is far from $\mu_{vr}$. 

\subsubsection{Laplace Approximation} 
The Laplace approximation can also be used to approximate the mixand pdf $\tilde{m}_{vr}(x_k)$ as a Gaussian in the neighborhood of its (dominant) mode. This leads to the importance pdf, 
\begin{align}
q_{vr}(x_k) = {\cal N}(\mu^{+}_{vr},\Sigma^{+}_{vr}), \label{eq:LaplaceGIS}
\end{align}
which defines \emph{Laplace Approximation Gaussian IS} (LAGIS). The mode point $x_{k} = \mu^{+}_{vr}$ of $\tilde{m}_{vr}(x_k)$ can again be found via first/second-order search techniques, and the covariance $\Sigma^{+}_{vr}$ can be calculated as the inverse of the Hessian of $-\log \tilde{m}_{vr}(x_k)$ at $x_k = \mu^{*}_{vr}$. 

Newton-Raphson search can provide fast convergence for relatively low computational cost, especially if GM pre-compression techniques are used to reduce the total number of mixands in $\pgenericddf{x_k}$ (see Sec \ref{impdetails}) and if eqs. \pareqref{ddfgmapprox}-\pareqref{sigstar} are parallelized. In high-dimensional spaces, however, the Hessian can become ill-conditioned or lose positive definiteness. In such cases, quasi-Newton search methods can provide more stable performance with only slightly slower convergence rates. 
If $\tilde{m}_{vr}(x_k)$ has multiple distinct modes that are not close to each other, $q_{vr}(x_k)$ could be replaced by a mixture pdf, as long as the distinct modes can be quickly identified. 
\subsubsection{Heavy tail mixture IS} 
LAGIS implicitly assumes that the covariance matrix $\Sigma^{+}_{vr}$ obtained from the inverse Hessian of $\log \tilde{m}_{vr}(x_k)$ at $x_k = \mu^{+}_{vr}$ provides adequate information for sampling $\tilde{m}_{vr}(x_k)$ via \pareqref{LaplaceGIS}. However, $\tilde{m}_{vr}(x_k)$ can be highly asymmetric and skewed, in which case Gaussian proposal pdfs will lead to inefficient sampling due to shape mismatch and produce unreliable high variance estimates in eqs. \pareqref{ddfgmapprox}-\pareqref{sigstar}. 
This mismatch can be mitigated by replacing the Gaussian proposal pdf in \pareqref{LaplaceGIS} with a heavier tailed distribution that has the same mean/mode and `shape' vis-a-vis the covariance, so that samples can be generated further away from $\mu^{+}_{vr}$ in appropriate directions. One possibility is to use a \emph{scale mixture model} proposal pdf derived from the Laplace approximation, 
\begin{align}
q_{vr}(x_k) = \sum_{c=1}^{M^q}\beta_{c}{\cal N}(\mu^{+}_{vr}, \xi_c \cdot \Sigma^{+}_{vr}), \label{eq:scaleLAGIS}
\end{align}
The scalar terms $\xi_c \geq 1$ allow samples to be drawn at larger distances from $\mu^{+}_{vr}$ and the weights $\beta_{c}$ control the proportion of IS samples drawn at scale $\xi_c$. The parameters $M^q$, $\beta_c$ and $\xi_c$ should be set to maximize sampling efficiency, i.e. to ensure that most samples actually lie inside the areas of high support for $\tilde{m}_{vr}(x_k)$. For example, the proposal mixture parameters can be adapted via the importance weights $\theta_s$ after a sampling pass (although this can be expensive for large $M^q$). 

\subsection{IS Algorithm Summary and Practicalities} \label{impdetails}
For both IGS and DLS, the resulting number of mixands $M^f = M^iM^j$ can grow very large if either $M^i$ or $M^j$ is large. Mixture compression strategies should are therefore needed to ensure computational efficiency for recursive GM fusion updates \cite{Runnalls-AES-2007}. Three general strategies are possible to keep $M^f$ at/below some desired upper bound $M^{max}$: (i) `pre-fusion': compress $p^i(x_k)$ and $p^j(x_k)$ to mixtures $p^{i'}(x_k)$ and $p^{j'}(x_k)$ with sizes $M^{i'}<M^i$ and $M^{j'}<M^j$, respectively; (ii) `mid-fusion': merge/prune $\pgenericddf{x_k}$ on the fly, e.g. by truncating mixands with weights falling below some threshold; (iii) `post-fusion': perform compression only after all $M^iM^j$ components are calculated. Approach (i) can lose too much information prior to fusion, but may still provide acceptable results if the majority of mixands in both $p^i(x_k)$ and $p^j(x_k)$ have significantly small weights, and are in close proximity to each other and/or other mixands with much larger weights. Approach (iii) retains the most information but requires the most computational effort for DDF. As such, it may yield approximate fusion pdfs $p^f$ that are too cumbersome for conventional `one mixand at a time' GM compression methods (which typically scale as $O([M^f]^2)$ or $O([M^f]^3)$). 
Approach (ii) offers the best balance of speed and accuracy, as long as candidates for mixand merging/pruning can be identified prior to IS approximation. 

In exact DDF, for instance, it is easy to show that the following bound holds for each unnormalized fusion component pdf and common information GM pdf mixand $t \in \set{1,...,M^{c,ij}} \ \forall x_k$
\begin{align}
m_{vr}(x_k) &= \frac{ {\cal N}(x_k; \mu^{ij}_{vr}, \Sigma^{ij}_{vr})}{ u(k)}  
\leq \frac{{\cal N}(x_k; \mu^{ij}_{vr}, \Sigma^{ij}_{vr})}{w^{c,ij}_t {\cal N}(x_k; \mu^{c,ij}_{t}, \Sigma^{c,ij}_{t})}= \frac{1}{w^{c,ij}_t } {\cal N}(x_k; \mu^{t \#}_{vr}, \Sigma^{t \#}_{vr}),  \label{eq:upperboundDef} 
\end{align}
so that the unnormalized component weights \pareqref{wstar} also obey
\begin{align}
\tilde{w}^{t\#}_{qr} &\leq w^i_v w^j_r \bar{z}^{ij}_{vr} \cdot \kappa^{t}_{vr},  \label{eq:upperboundWt} \\ 
\kappa^{t}_{vr} &=\frac{1}{w^{c,ij}_t }\int_{-\infty}^{\infty}{\cal N}(x_k; \mu^{t \#}_{vr}, \Sigma^{t \#}_{vr}) dx_k \nonumber
\end{align}
where $\mu^{t\#}_{vr}, \Sigma^{t\#}_{vr},$ and $\kappa^{t\#}_{vr}$ are all easily obtained in closed-form.
Since the bound holds $\forall \ t \in \set{1,...,M^{c,ij}}$, the smallest $\kappa^{\#}_{qr} = \min_{t} \kappa^{t}_{vr}$ for each mixand $vr$ can be used to prioritize updates, such that those mixands for which the RHS of \pareqref{upperboundWt} falls below a certain threshold are either ignored entirely or calculated later. 

\section{Simulation Studies} \label{expts}
This section studies various features of the proposed GM fusion approximations through four sets of numerical examples. 
The first example compares the IGS and DLS methods to other state of the art GM fusion approximations on synthetic problems. 
The second synthetic fusion example compares and contrasts the approximation accuracy and sampling efficiency of IGS and various DLS implementations. 
The third example demonstrates how the DLS approximation behaves in a decentralized static target search application for exact GM-based DDF using multiple mobile search platforms. 
The final example demonstrates how the IGS approximation behaves in a more challenging decentralized dynamic tracking scenario, involving multiple range-only sensing platforms and a highly maneuverable target that yield highly non-Gaussian uncertainties. 


\subsection{Example 1: 2D synthetic problems} 
\begin{figure}[t]
\centering
\begin{tabular}{@{}c@{}c@{}c@{}}
\includegraphics[width=4.5cm]{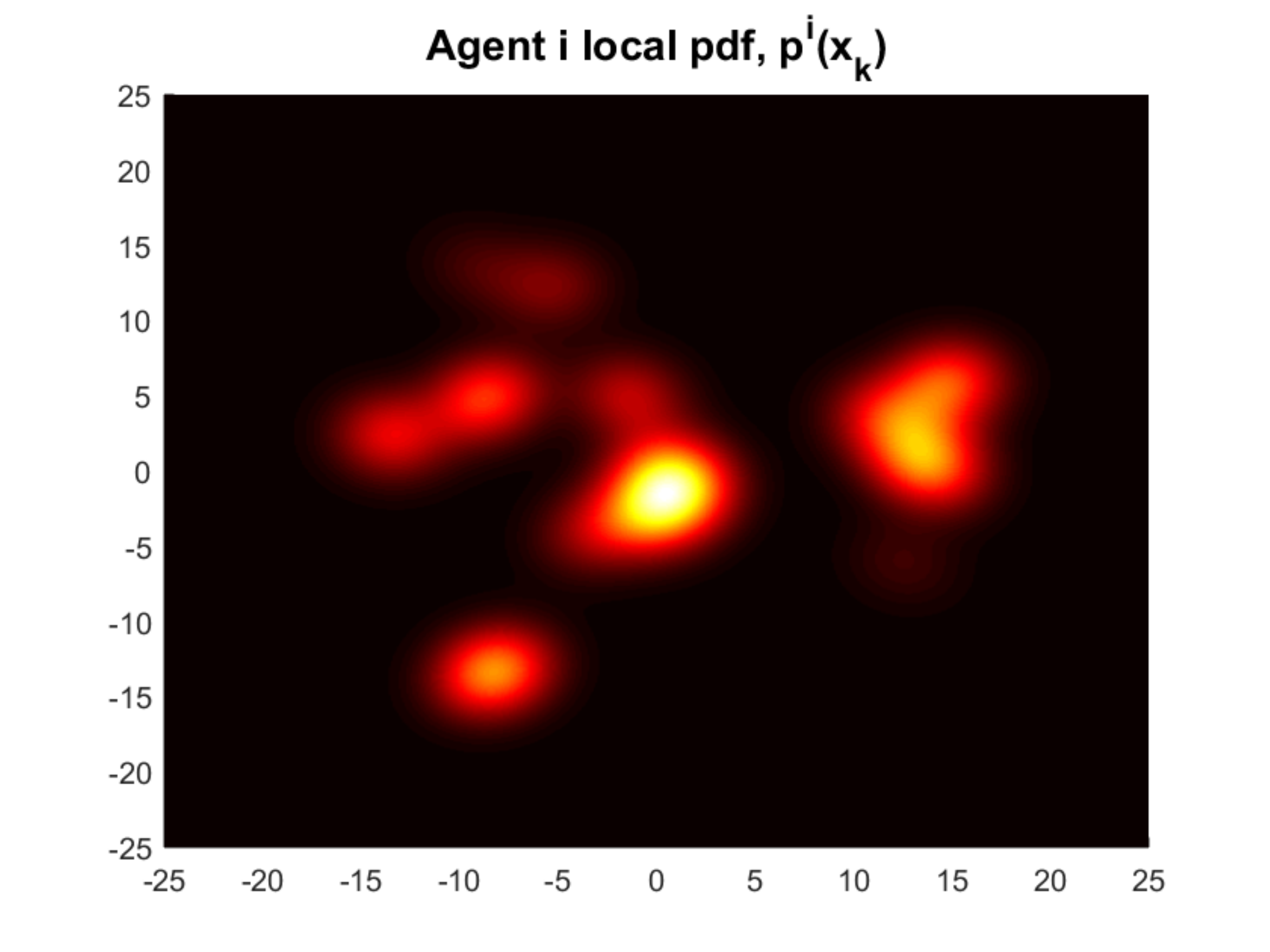} &
\includegraphics[width=4.5cm]{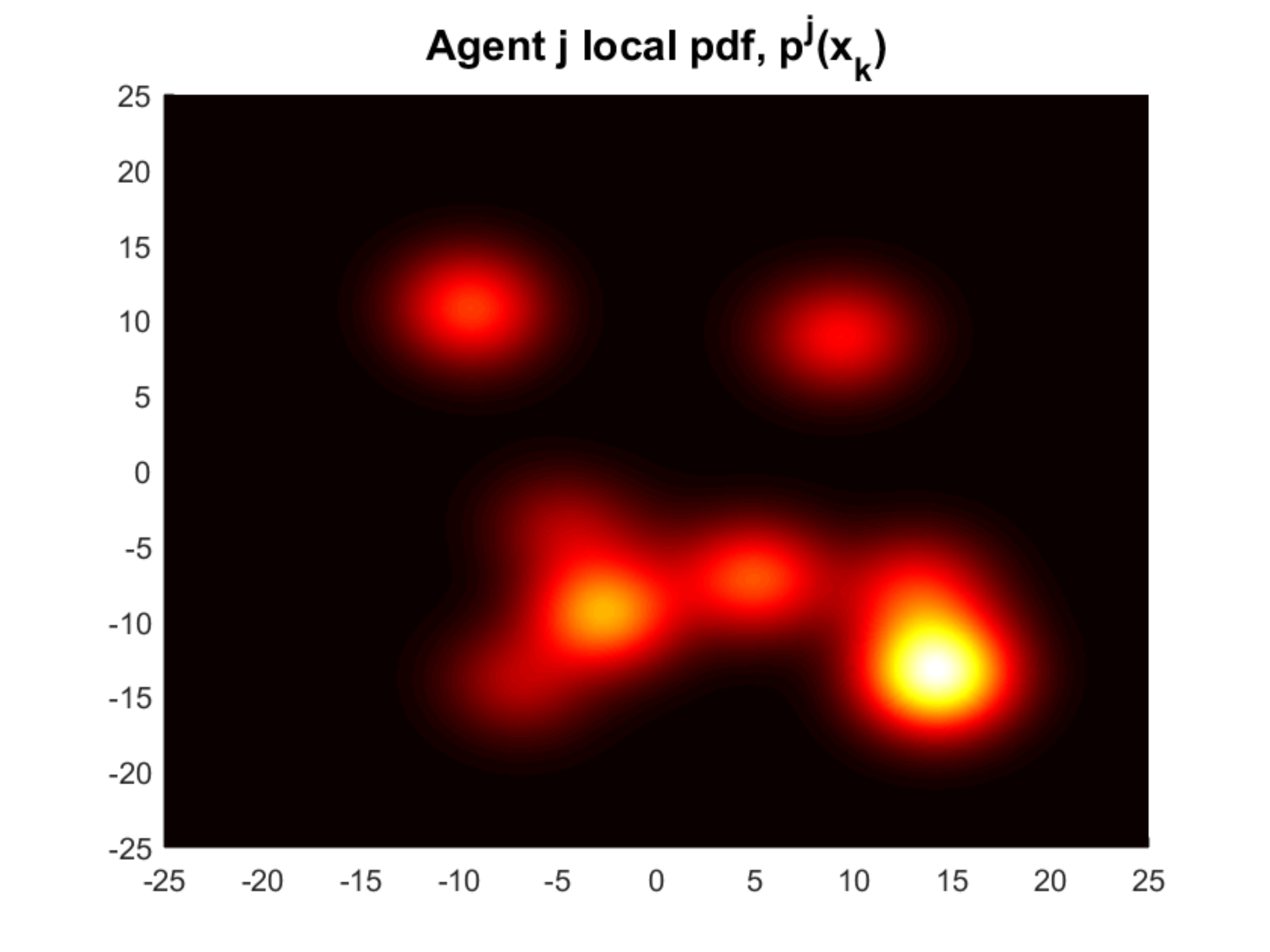} &
\includegraphics[width=4.5cm]{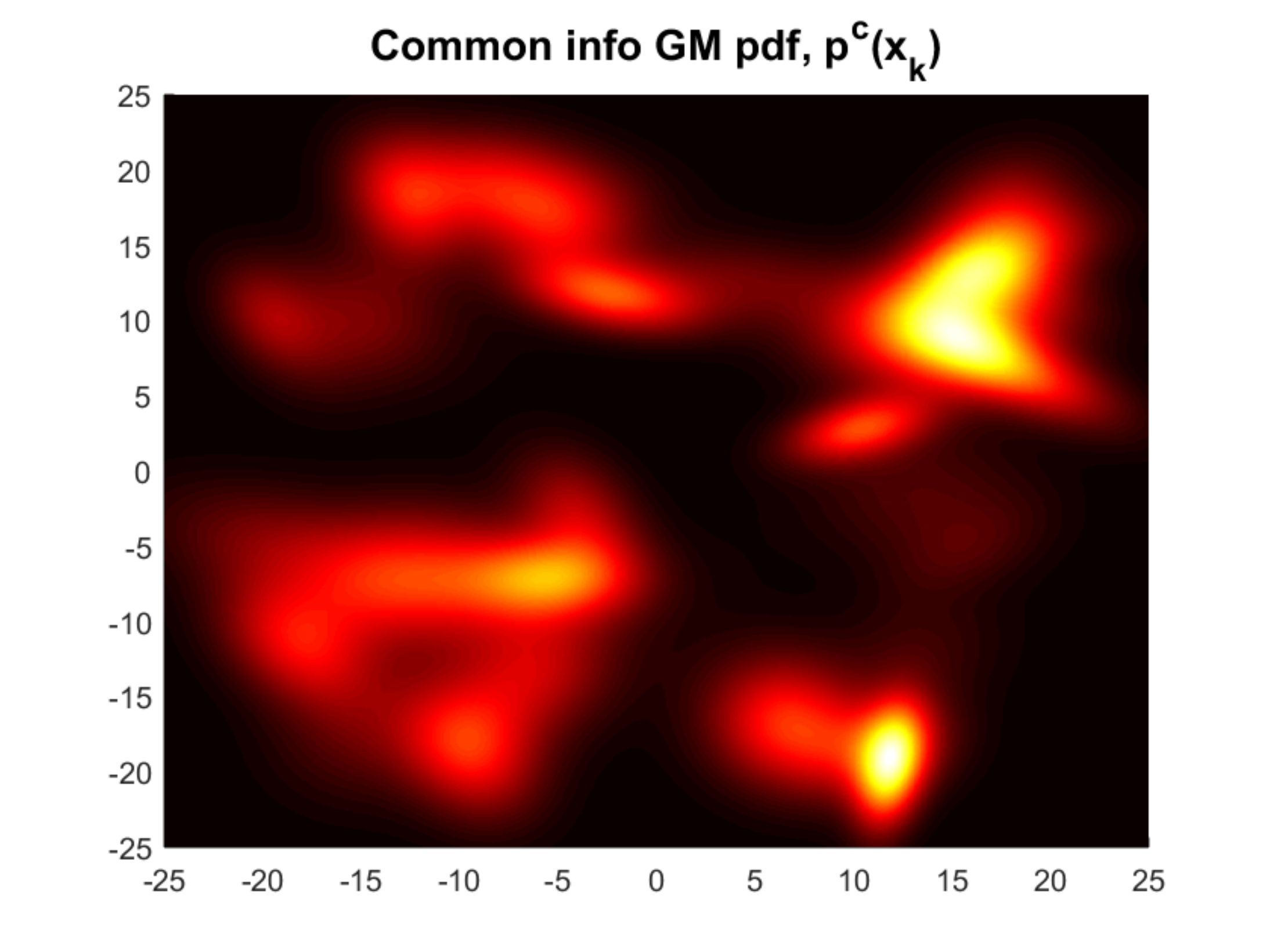} \\
\scriptsize (a) & \scriptsize (b) & \scriptsize (c) 
\end{tabular} 
\caption{GMs for fusion example: (a)-(b) $p^i(x_k)$ and $p^j(x_k)$ with $M^i=M^j=14$; (c) common information pdf $p^{c,ij}(x_k)$ with 40 components.}
\label{fig:gm2Ddemosetup}
\end{figure}

\begin{figure}[t]
\centering
\newcommand{\figsize}{4.5cm}
\begin{tabular}{@{}c@{}c@{}c@{}}
\includegraphics[width=\figsize]{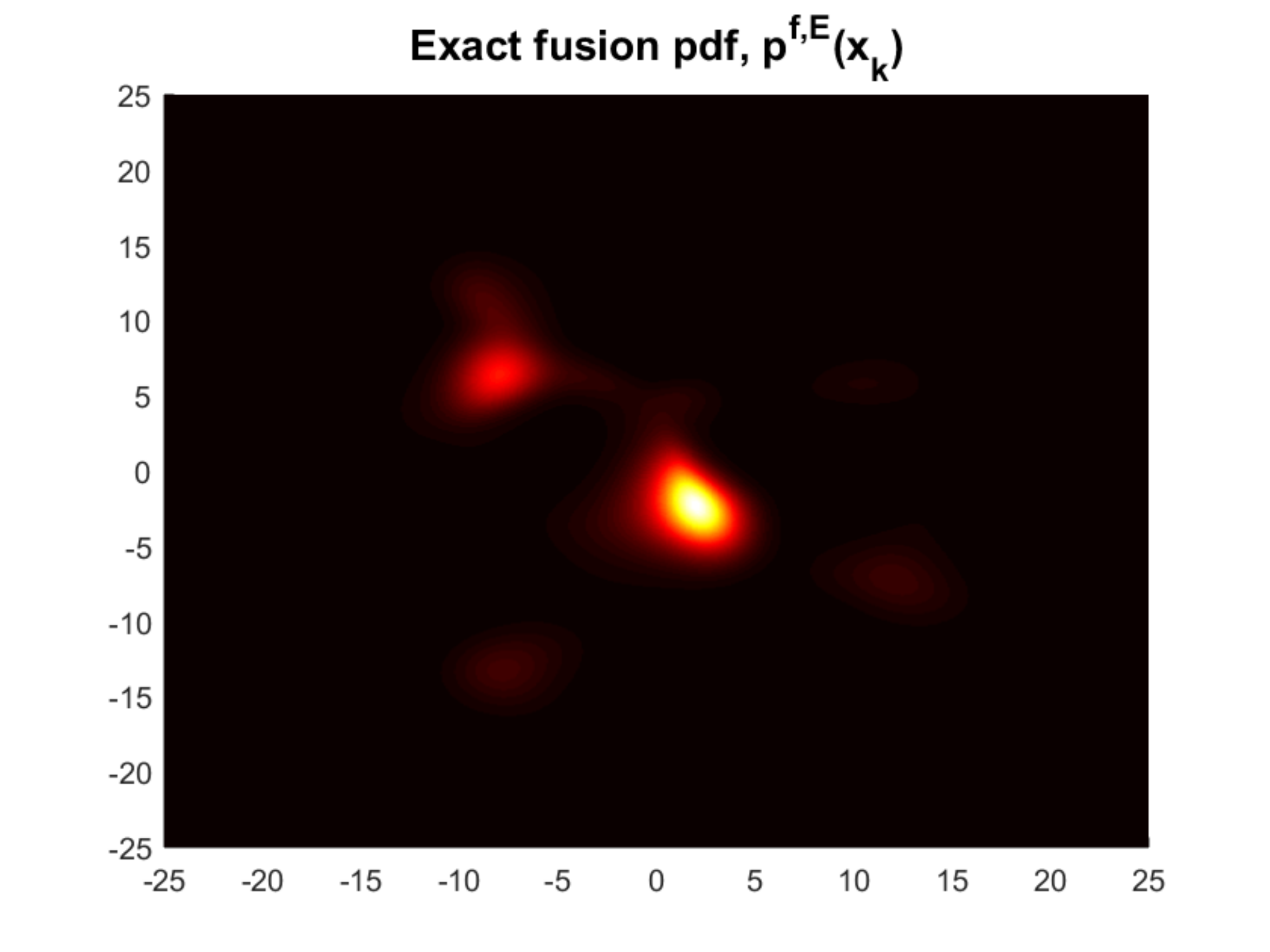} &
\includegraphics[width=\figsize]{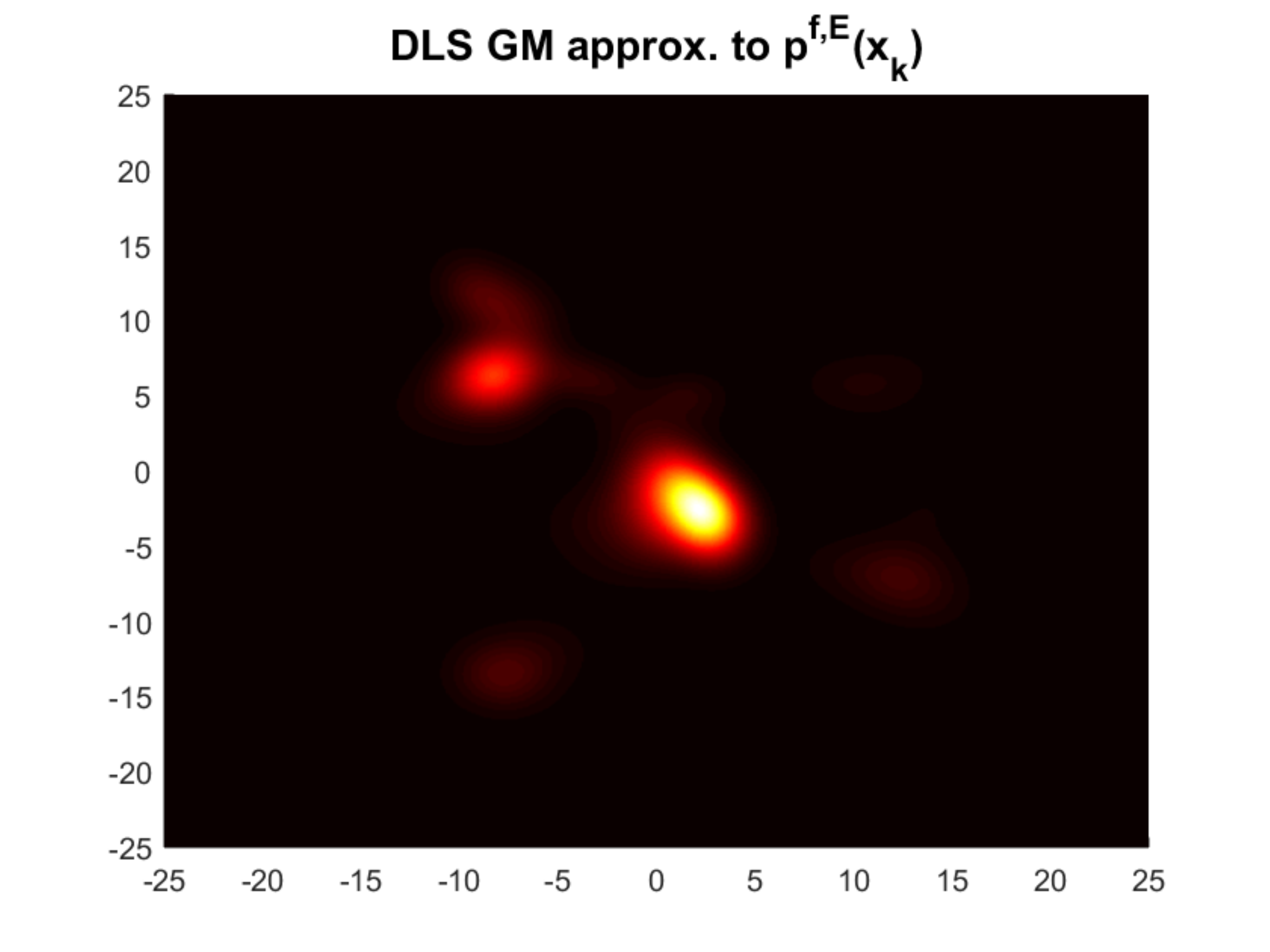} &
\includegraphics[width=\figsize]{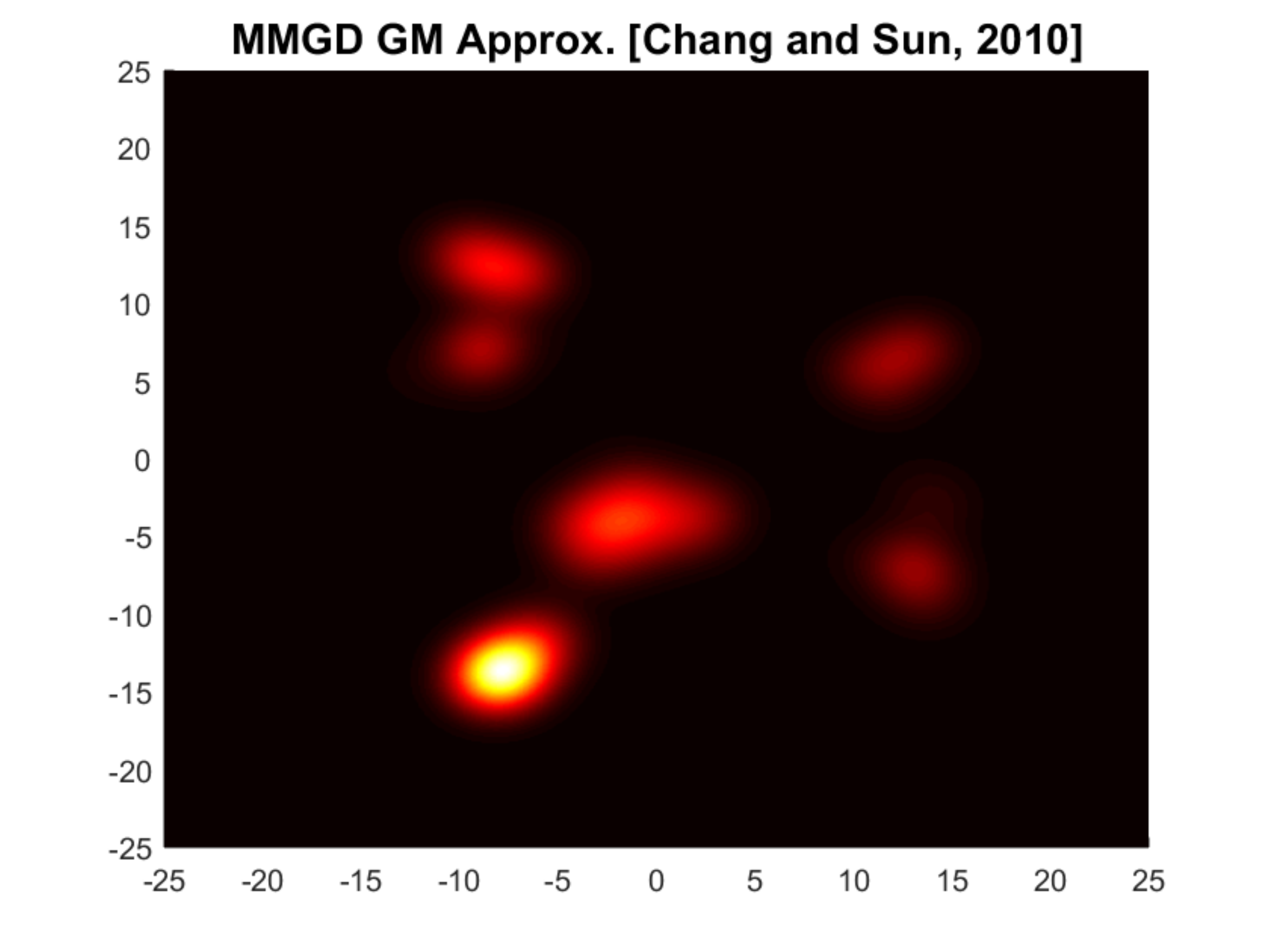} \\
\scriptsize (a) & \scriptsize (b) & \scriptsize (c) \\
\includegraphics[width=\figsize]{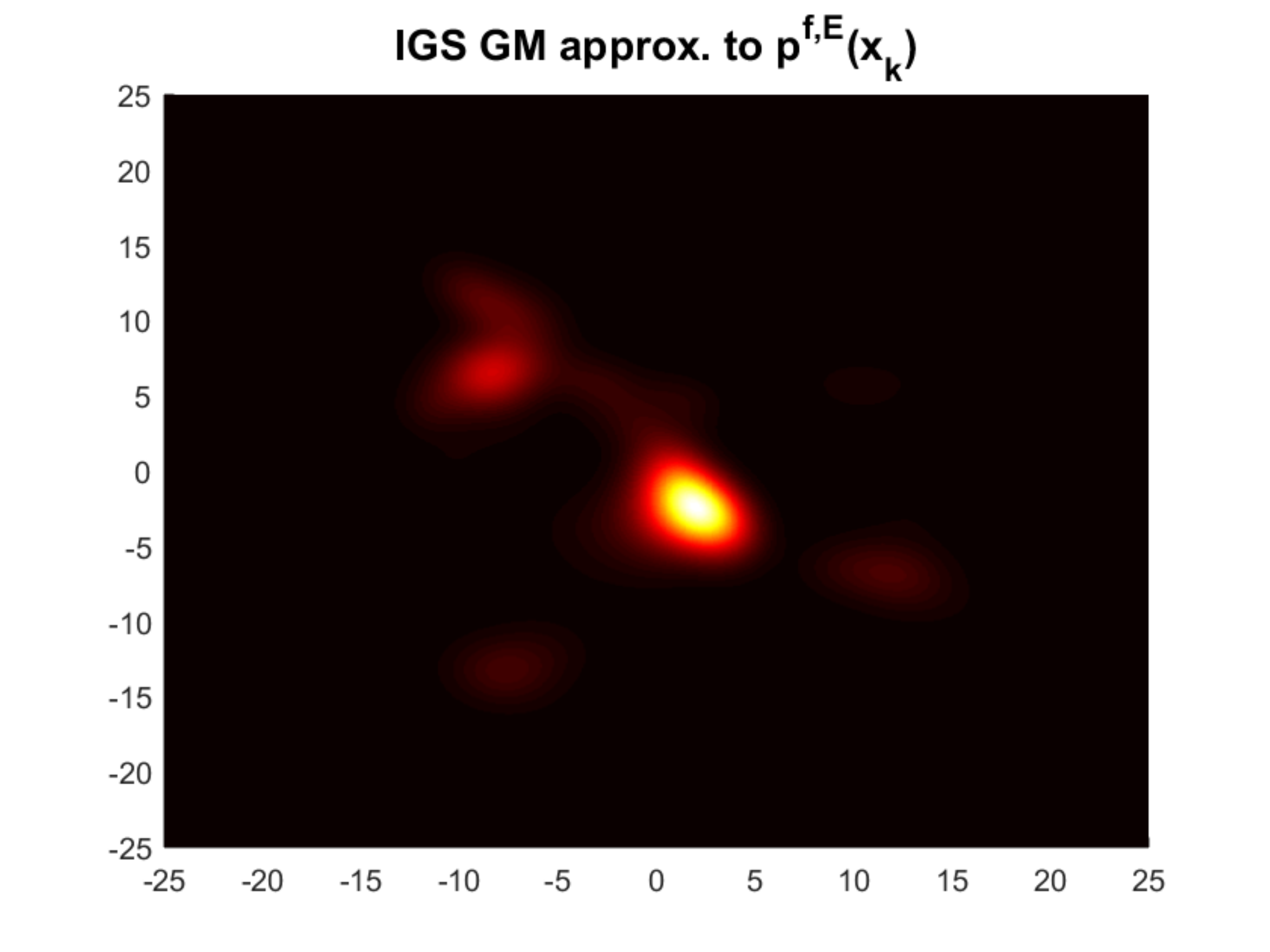} &
\includegraphics[width=\figsize]{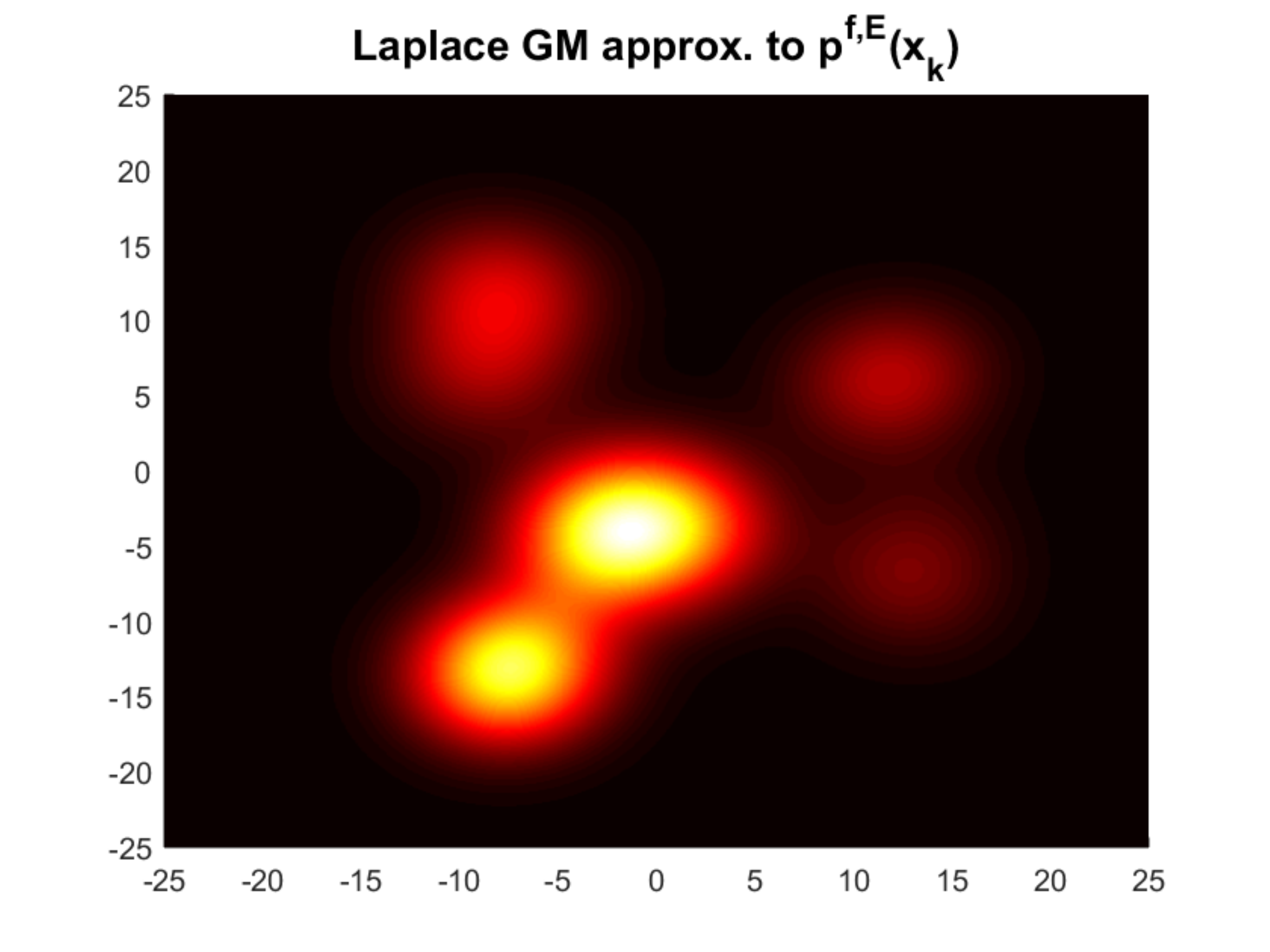} &
\includegraphics[width=\figsize]{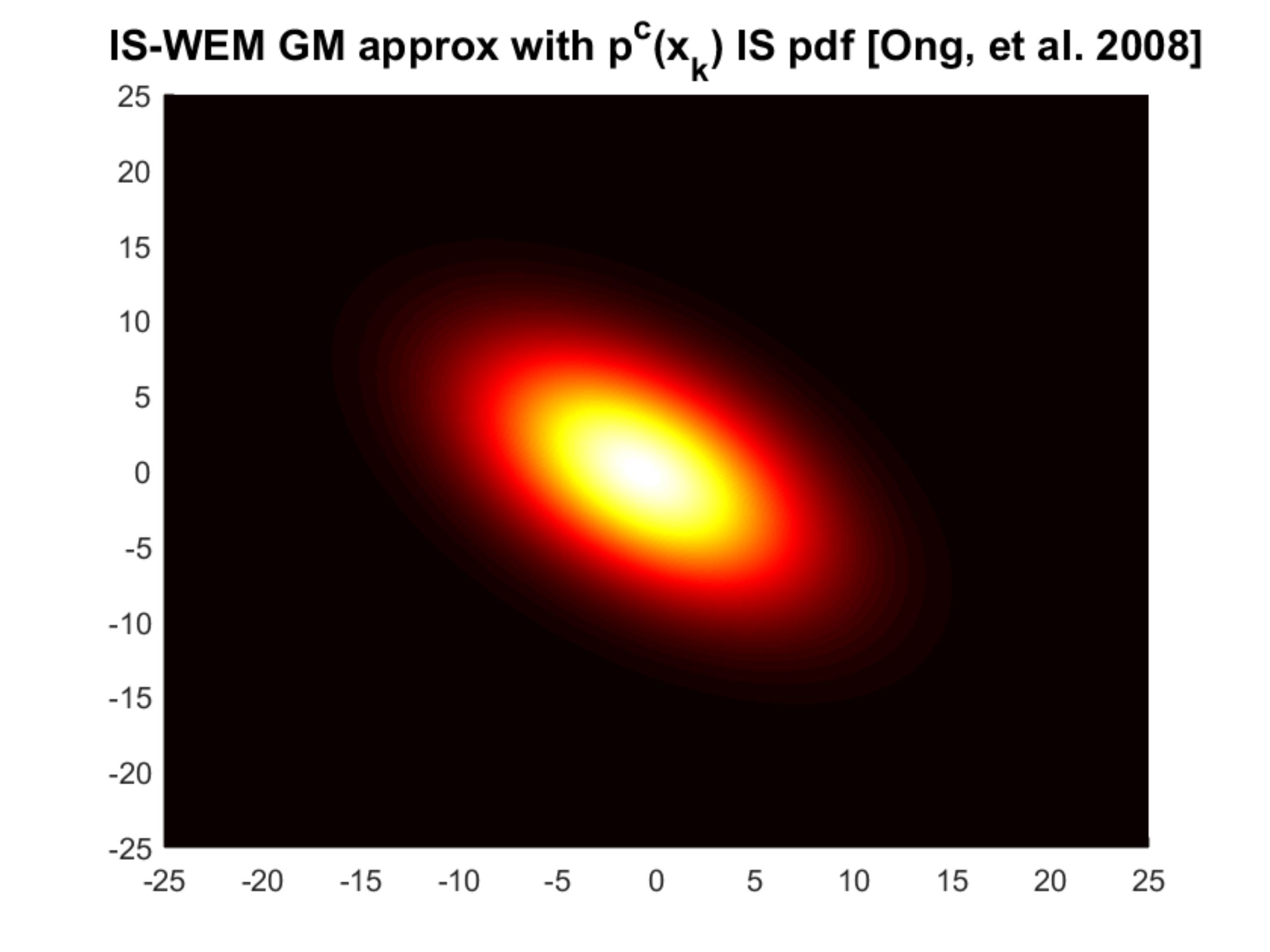} \\
\scriptsize (d) & \scriptsize (e) & \scriptsize (f) \\
\end{tabular}
\caption{\scriptsize Results for approximating $\pexactddf{x_k}$ for GMs in Fig. \ref{fig:gm2Ddemosetup}: 
(a) exact fusion result (computed on grid); (b) DLS approximation (INGIS); (c) MMGD approximation; 
(d) IGS approximation; (e) Laplace approximation GM pdf; (f) EM condensation of IS samples. 
}
\label{fig:exact2Ddemo}
\end{figure}

Figure \ref{fig:gm2Ddemosetup} shows synthetically generated GMs for $p^i(x_k)$, $p^j(x_k)$, and $p^{c,ij}(x_k)$. Figure \ref{fig:exact2Ddemo} (a) shows a grid approximation to the exact DDF result. Also shown are the fused GM obtained by: DLS (b, using INGIS with $\alpha = 5$, $N_s=500$ per mixand); the moment-matched Gaussian denominator (MMGD) approximation of \cite{Chang-ICARV-2010} (c, eq.\ref{eq:ChangApprox}); DLS (d, using $N_s=1000$ total samples); the mixture Laplace approximation (e); and the IS technique of \cite{Ong-FUSION-2008} (f, using $p^{c,ij}(x_k)$ as the proposal pdf, followed by EM compression of 5000 weighted samples to a GM with 14 components).  
Both DLS (b, KLD from truth = 0.0104 nats) and IGS (d, KLD = 0.0281 nats) do an excellent job of accurately capturing both the dominant and weaker modes of the true fusion pdf, unlike the approximations in (c), (e), and (f) (KLDs of 1.1264 nats, 1.0023 nats, and 1.0505 nats, respectively). In (c), the approximation of $p^{c,ij}(x_k)$ by a Gaussian leads to significant information loss. Likewise, the mixture Laplace approximation in (e) does not accurately capture the covariances or relative weightings of the fused mixture modes, although the result still provides a reasonable estimate for use in the IS approximations of DLS and IGS. In contrast, the use of $p^{c,ij}(x_k)$ as an IS proposal pdf does not lead to good results in (f), since $p^{c,ij}(x_k)$ and $\pexactddf{x_k}$ are quite distinct. 


\begin{figure}[t]
\centering
\newcommand{\figsize}{4.5cm}
\begin{tabular}{@{}c@{}c@{}c@{}}
\includegraphics[width=\figsize]{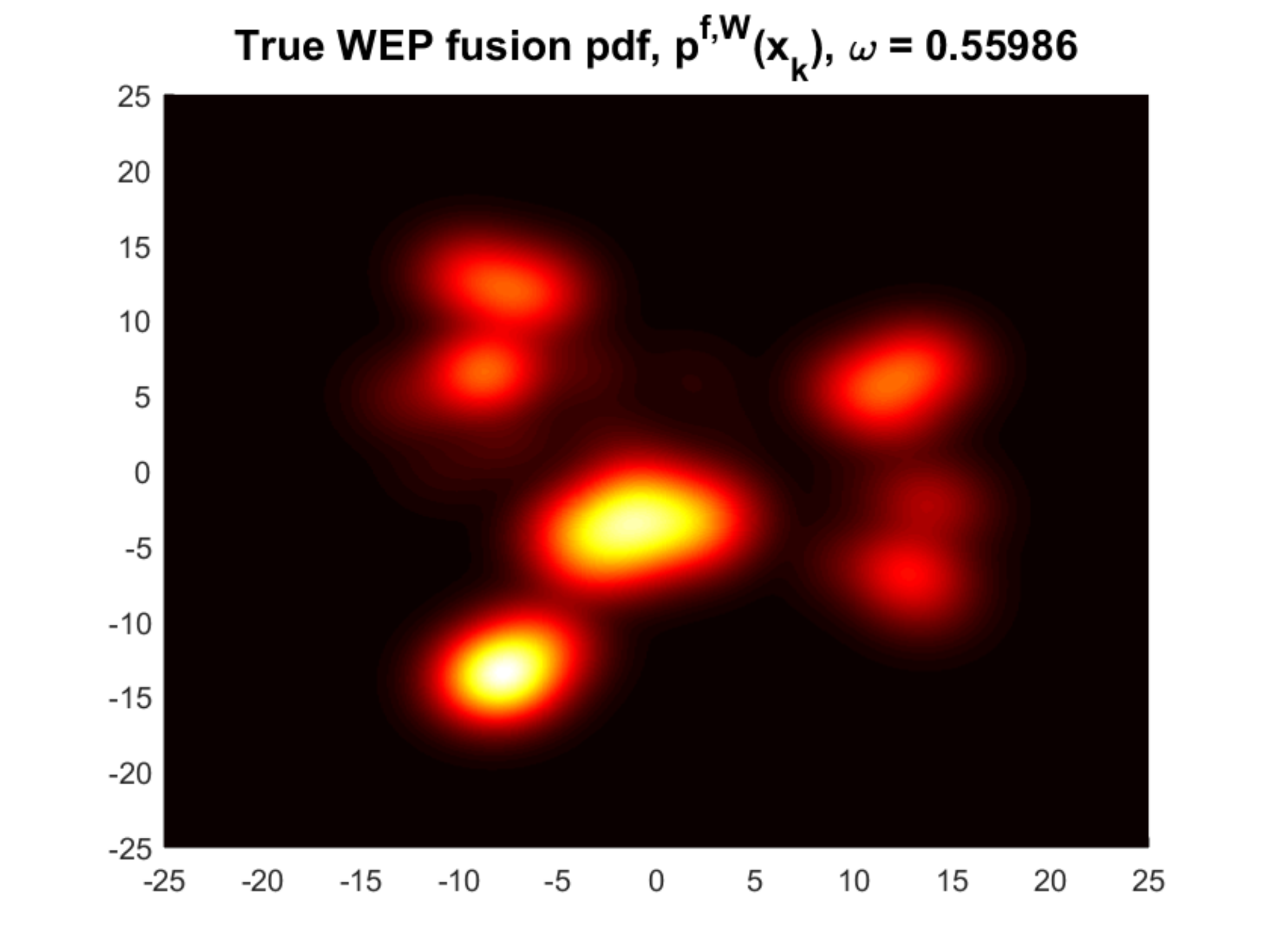} &
\includegraphics[width=\figsize]{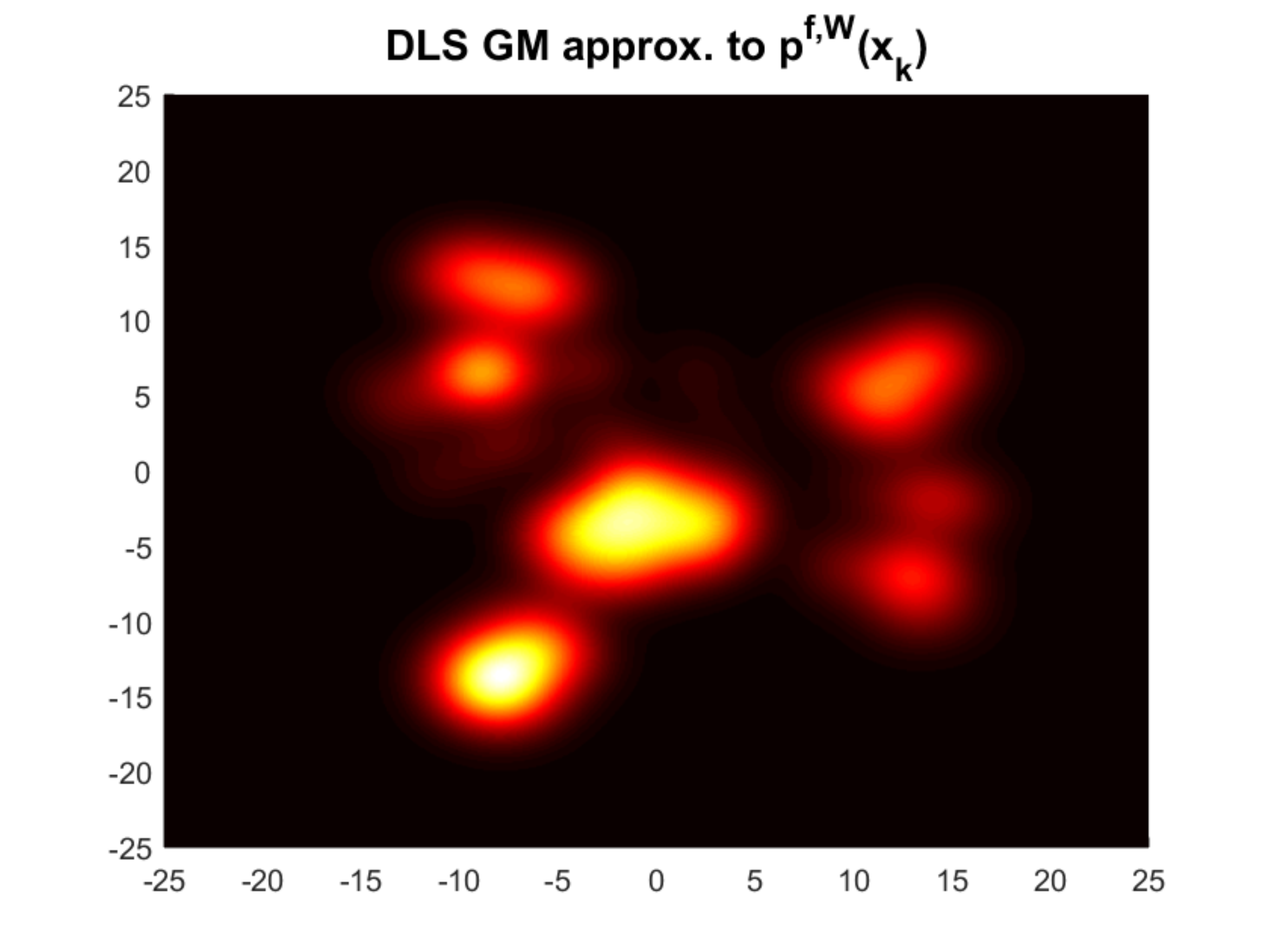} &
\includegraphics[width=\figsize]{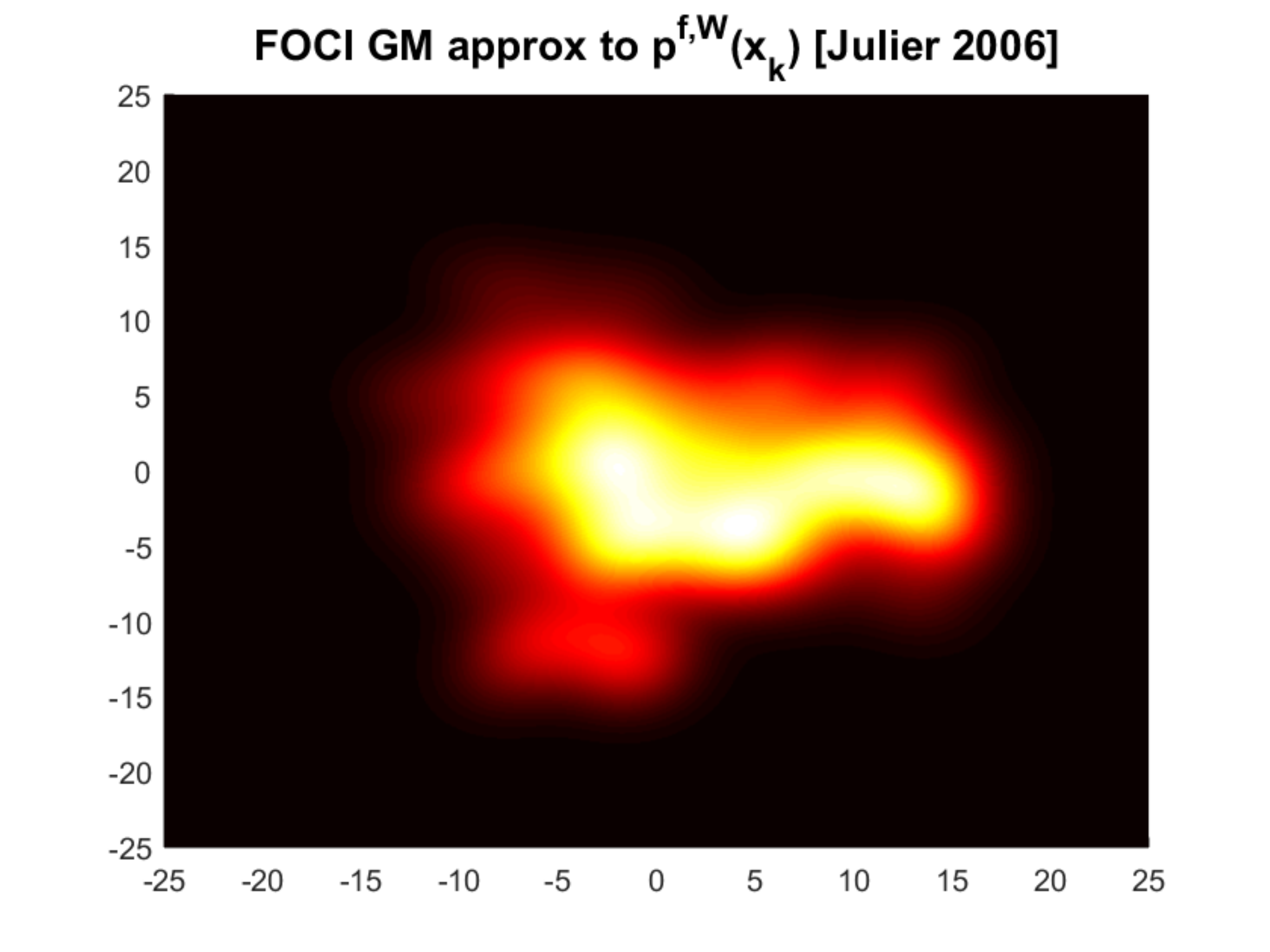} \\
\scriptsize (a) & \scriptsize (b) & \scriptsize (c) \\
\includegraphics[width=\figsize]{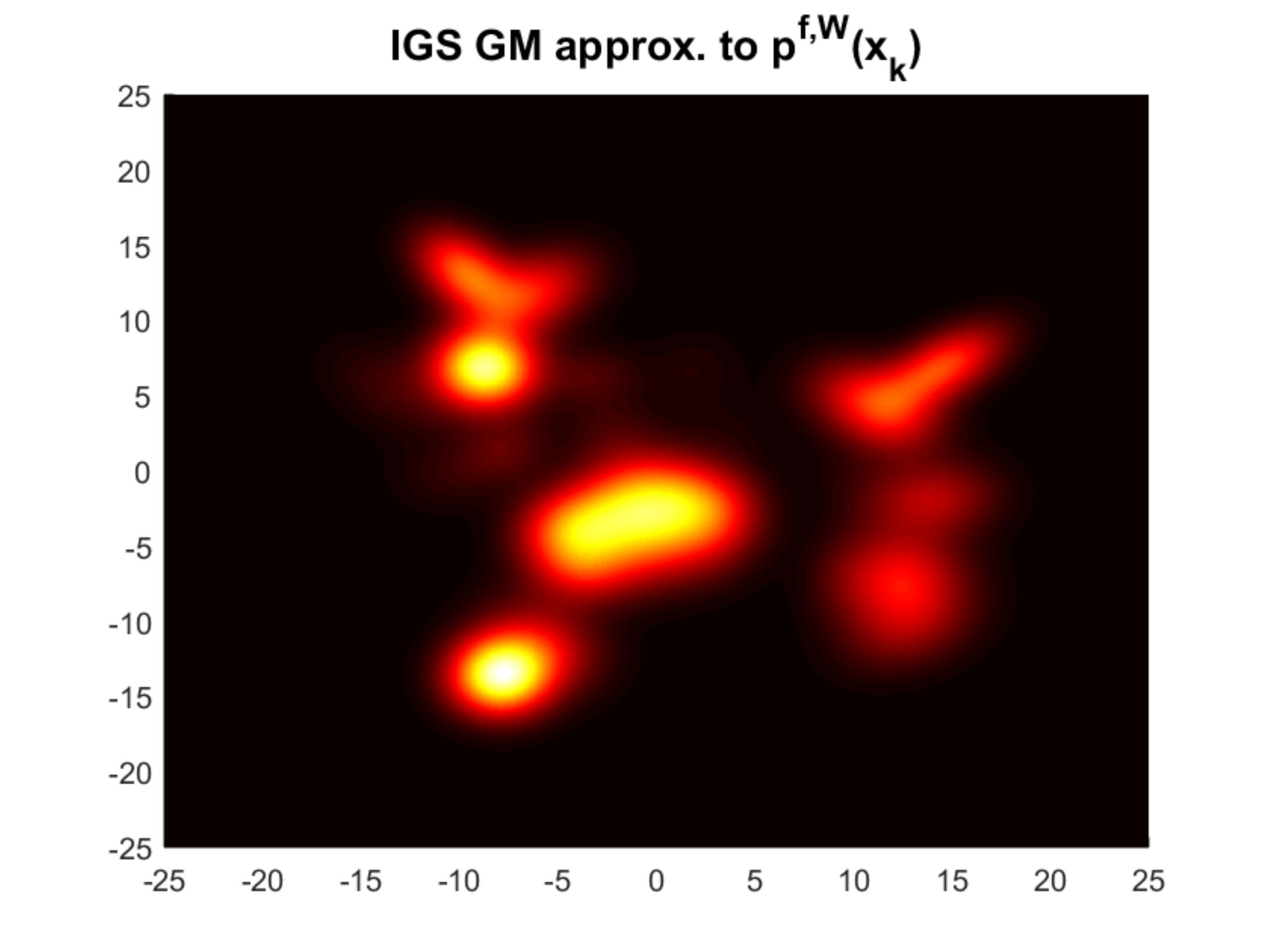} &
\includegraphics[width=\figsize]{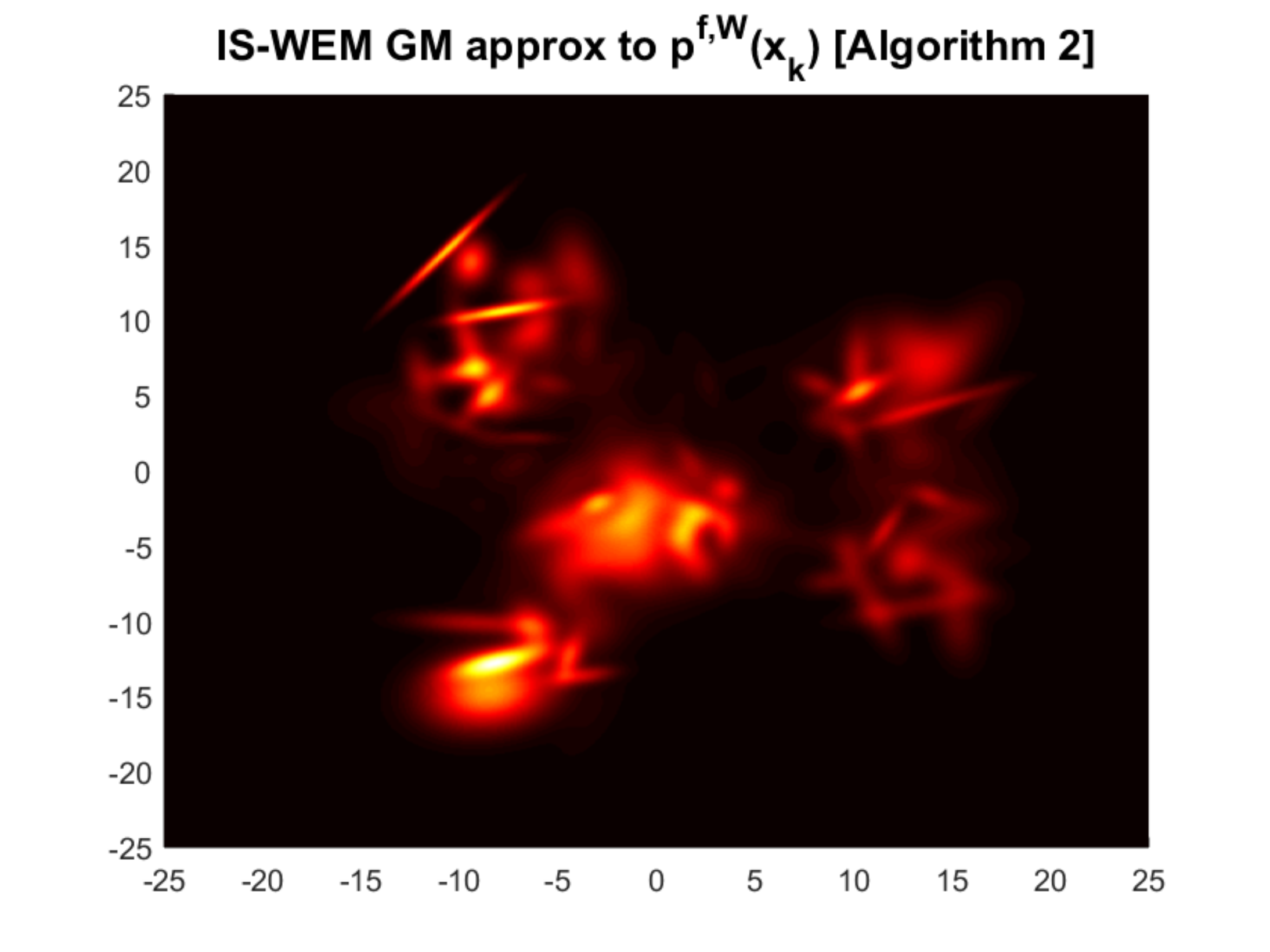} &
\includegraphics[width=\figsize]{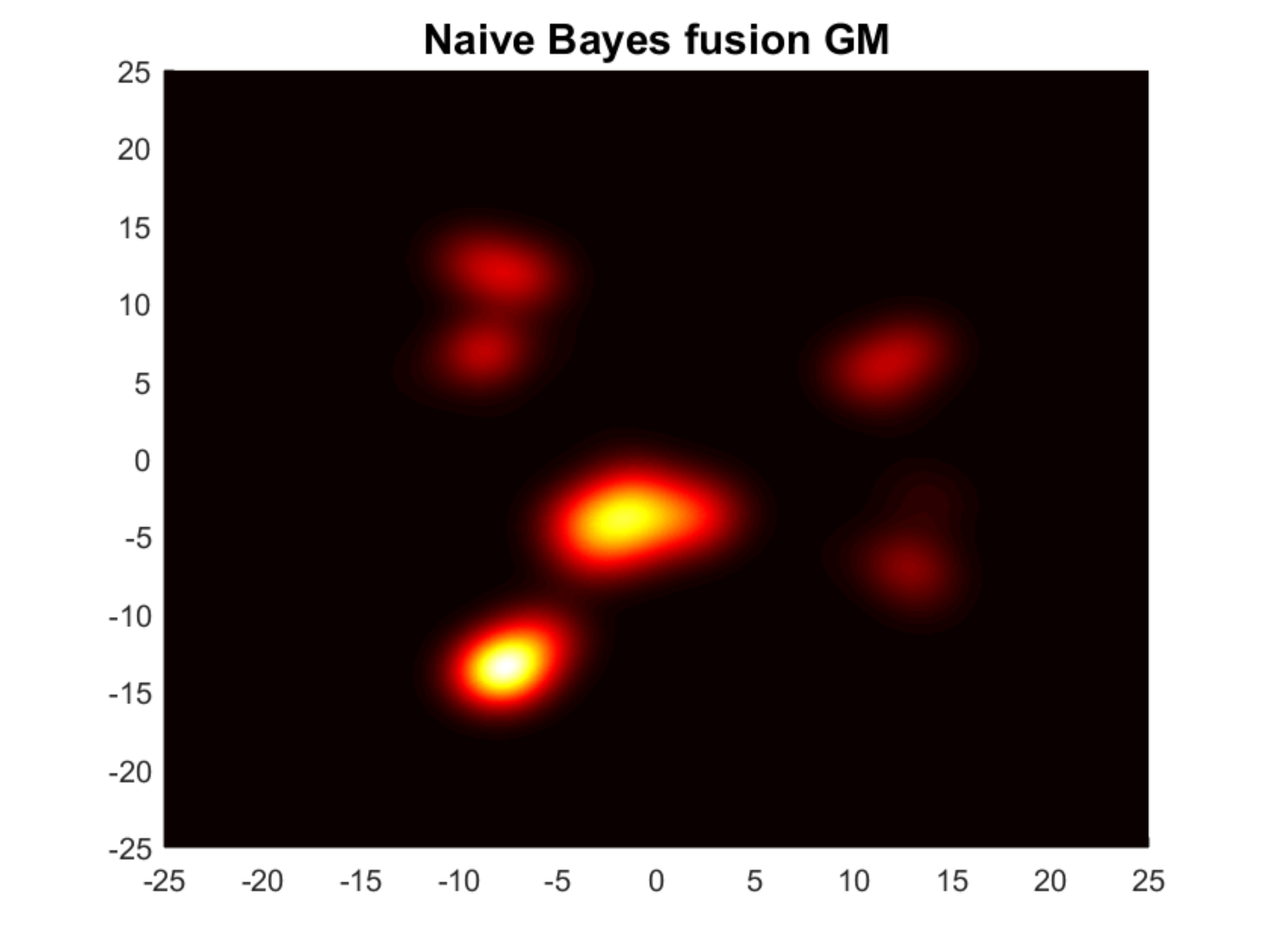} \\
\scriptsize (d) & \scriptsize (e) & \scriptsize (f) \\ 
\end{tabular}
\caption{\scriptsize Results for approximating $\pwepddf{x_k}$ for GMs in Fig. \ref{fig:gm2Ddemosetup} for $\omega = 0.4436$: (a) grid-based fusion result; (b) DLS approximation (INGIS); (c) FOCI approximation; (d) IGS approximation; (e) IS-WEM approximation; (f) Naive Bayes fusion result.
}
\label{fig:wep2Ddemo}
\end{figure}

Figure \ref{fig:wep2Ddemo} (a) shows the WEP fusion pdf for the same example using $\omega = 0.4436$, which was found via Algorithm \ref{alg:ISOptAlg} using 5000 samples with the minimax information loss rule. Also shown are the GMs obtained by: DLS (b, using INGIS with same settings as for exact fusion); the `first order covariance intersection' (FOCI) approximation of \cite{Julier-FUSION-2006} (c, eq. \ref{eq:FOCIApprox}); IGS (d, using the same settings as for exact fusion); the `IS + weighted EM' (IS-WEM) technique of \cite{Ahmed-TSP-2012, ASC-RSS-2012} (e, applying Algorithms \ref{alg:ISOptAlg} and \ref{alg:WEMLearning} in succession with $N_s=5000$); and Naive Bayes fusion (f), which takes the product of the GMs $p^i(x_k)$, $p^j(x_k)$ assuming the absence of $p^{c,ij}(x_k)$. DLS again accurately captures the dominant and weak modes of the true fusion posterior pdf (KLD from truth = 0.0029 nats). IGS does slightly worse but generally maintains the correct overall shape and relative mixand weightings for the overall fused pdf (KLD= 0.0848 nats). The small shape errors apparent for the lower weighted mixands are attributable to `effective sample loss' following Algorithm \ref{alg:ISOptAlg}, which obtains an effective sample size of 284 from $N_s=1000$. FOCI (KLD = 0.6975 nats) fails to resolve the modes of the fusion posterior, as it ignores higher order information from $\hat{p}^{c,ij}(x_k)$. IS-WEM (KLD = 0.1035 nats) preserves this missing information through importance sampling, but converges to a poor local minimum of the negative log-likelihood function for condensing the samples into a GM. Despite using considerably fewer samples than IS-WEM, the resulting IGS result is noticeably better and ultimately far more stable. The Naive Bayes GM (KLD = 0.30 nats) provides optimistic estimates for the mixand covariances and severely underestimates several of the smaller mixand weights. 

Table \ref{table:time2DResults} shows the typical times required to run non-parallelized implementations of each fusion method for the exact and WEP DDF cases (Matlab 9.1, Windows 10, Intel i7-8550U 1.80 GHz CPU with 16 GB RAM). The DLS, IS-WEM and FOCI times for WEP fusion include a `worst case' 0.64 sec time required to run Algorithm \ref{alg:ISOptAlg} with $N_s=5000$ samples (based on the IS-WEM implementation); the time reported for IGS represents the combined $N_s=1000$ run time for Algorithm \ref{alg:ISOptAlg} (0.30 secs) and Algorithm \ref{alg:SSWEMLearning}. Since it does not require estimation of $\omega$, the Naive Bayes result represents the time required to construct the product pdf only. These results show that DLS and IGS generally require only a modest increase in computation time relative to the other less accurate existing fusion techniques. A significant portion of the increased time cost is incurred by: construction of the Naive Bayes pdf for both DLS and IGS in all cases; construction of the mixture importance sampling pdf for DLS in both the exact and WEP cases; and by Algorithm \ref{alg:ISOptAlg} to optimize $\omega$ for both DLS and IGS in the WEP case. The results shown here also represent unoptimized software implementations of DLS and IGS, whereas significant performance gains could, for instance, be obtained via parallelization of Algorithms \ref{alg:ISOptAlg}, \ref{alg:SSWEMLearning} and \ref{alg:DLSGMAlg}, as well as mixture condensation before or during fusion operations. 

%
%
\begin{table}
\centering
\caption{Typical GM fusion execution times for 2D example.}
\begin{tabular}{@{}ccc@{}} 
\hline
Method & Exact DDF Time (secs) & WEP DDF Time (secs) \\
\hline
DLS & 0.46 & 1.09  \\ 
IGS & 0.60 & 0.74 \\ 
MMGD & 0.07 & -  \\ 
Laplace  & 0.19 & - \\
IS-WEM  & 1.72 & 1.64\\ 
Naive Bayes & - & 0.09 
\end{tabular}
\label{table:time2DResults}
\vspace{-0.25in}
\end{table}

The accuracies and execution times of these methods were also studied in a larger set of 100 simulated `one shot' exact and WEP GM fusion problems with randomly constructed 2-dimensional pdfs. In each simulation, the platform GMs $p^i(x_k)$ and $p^j(x_k)$ were constructed with 10-11 components each, with component means drawn uniformly in each dimension between -14 and 14, component weights drawn from a uniform distribution and then renormalized, and component covariances drawn from a Wishart pdf with 10 degrees of freedom and a base covariance scale factor of 0.75. For exact fusion, the common information GM $\hat{p}^{c,ij}(x_k)$ was constructed similarly, except with between 40-41 components and component means drawn uniformly in each dimension between -20 and 20. The MMGD, DLS (INGIS), and IGS methods were applied for exact fusion, while the FOCI, DLS (INGIS), and IGS methods were applied for WEP fusion. The true $p^{f,E}$ and $p^{f,W}$ pdfs in each case were also constructed via grid approximation, with Algorithm \ref{alg:ISOptAlg} again applied to approximate $\omega^*$ as before. To better assess the accuracy-computation tradeoff for DLS and IGS, both methods were implemented with multiple $N_s$ values, with $N_s \in \set{10,50,100,200}$ per fused mixand component for DLS and $N_s \in \set{100,500,1000,2000}$ total mixture samples for IGS. 

\begin{figure}[t]
\centering
\newcommand{\figsize}{7.6cm}
\begin{tabular}{@{}c@{}c@{}}
\hspace*{-0.2 in}
\includegraphics[width=\figsize]{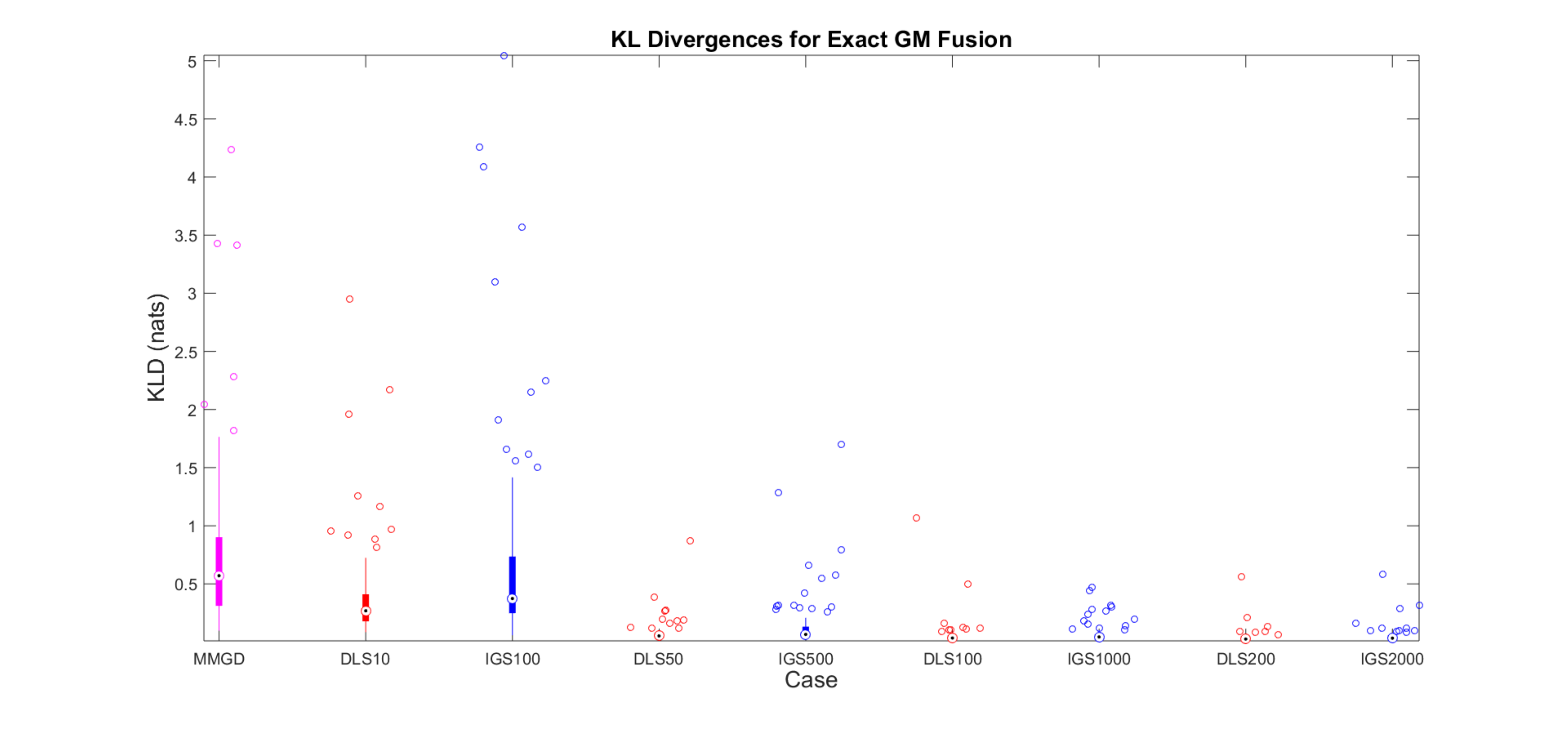} &
\hspace*{-0.2 in}
\includegraphics[width=\figsize]{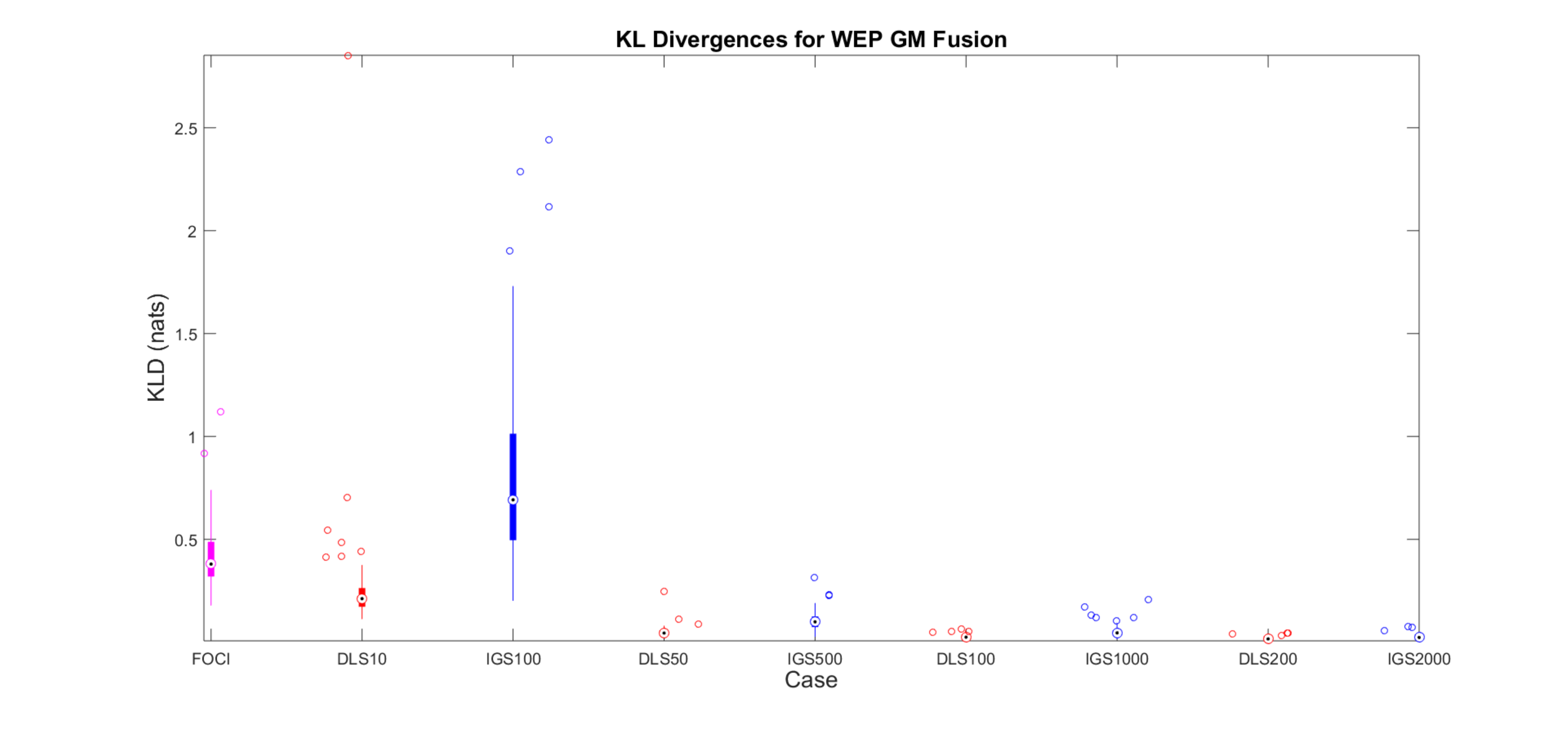} \\
\scriptsize (a) & \scriptsize (b) \\
\hspace*{-0.2 in}
\includegraphics[width=\figsize]{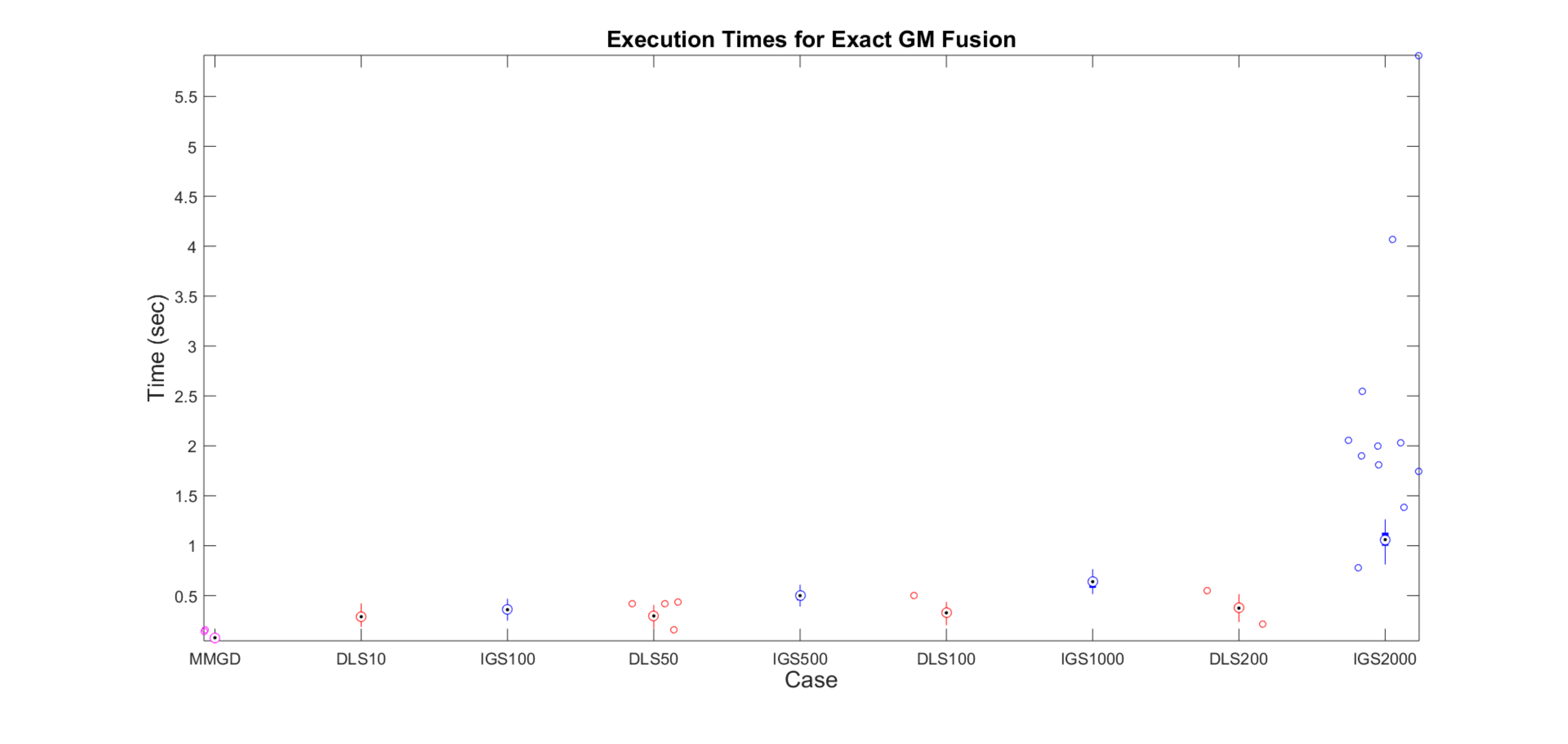} &
\hspace*{-0.2 in}
\includegraphics[width=\figsize]{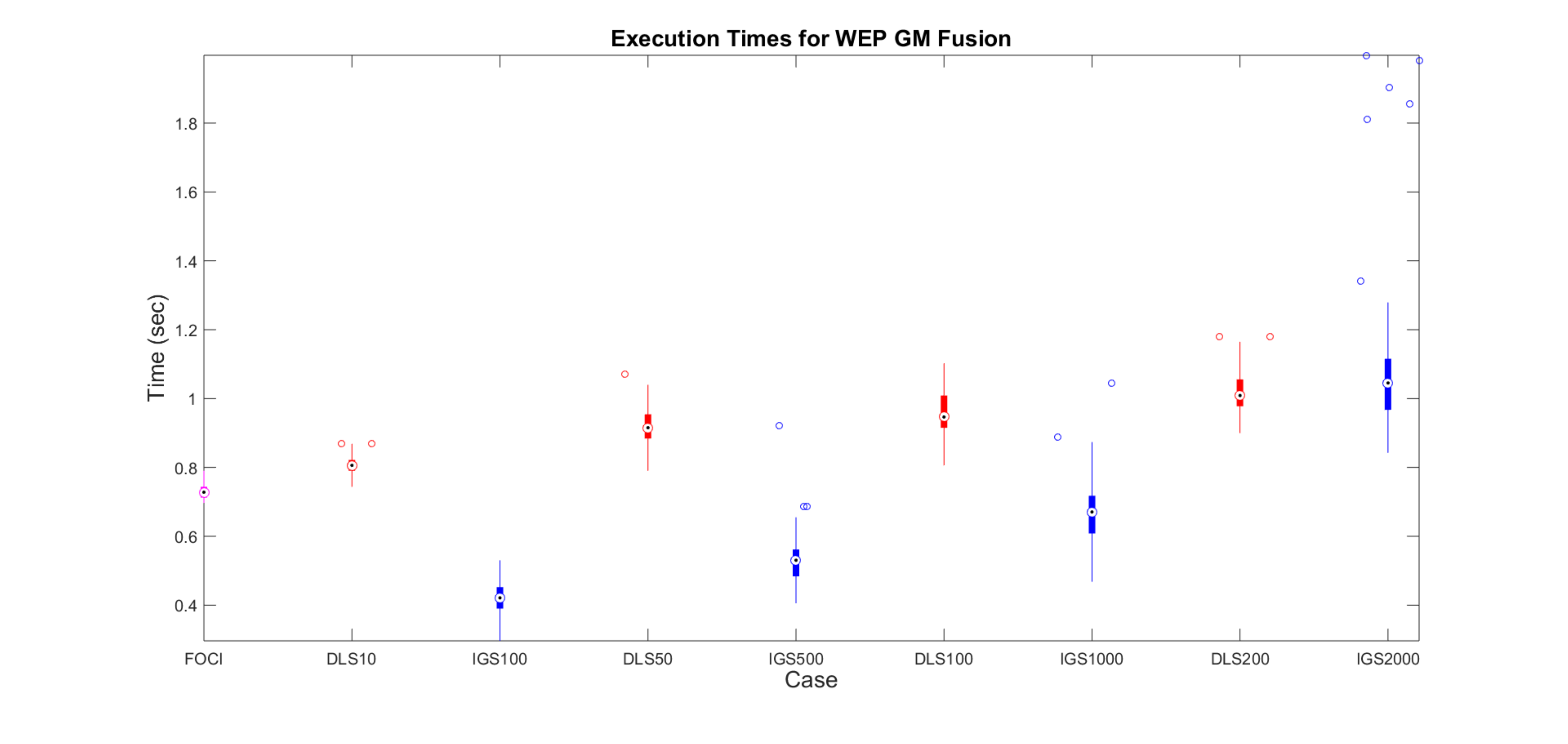} \\
\scriptsize (c) & \scriptsize (d) 
\end{tabular}
\caption{\scriptsize Kullback-Leibler divergences (a)-(b) and execution times (c)-(d) for 100 randomly generated 2D GM Exact and WEP simulated fusion problems.}
\label{fig:montecarlo_2Dresults}
\end{figure}

Figure \ref{fig:montecarlo_2Dresults} shows the resulting Kullback-Leibler divergences for each approximate fusion method relative to the ground truth grid approximations, along with the resulting execution times. As expected, the accuracy and variance of both DLS and IGS improve significantly as more importance samples are used, with corresponding modest increases in required computation time. The improved accuracy of DLS relative to IGS in nearly all cases can be attributed to the fact that the statistics for each posterior fusion pdf mixand are estimated via a set of $N_s$ exclusive samples in DLS, whereas all mixands must `share' the same $N_s$ samples in IGS. Given that each mixture considered in these simulations could have anywhere between 100-121 total posterior fusion mixands, DLS could use anywhere between 1000-24,200 total importance samples to approximate the GM fusion pdf, compared to only 100-2000 total samples for IGS. The larger total sample sizes lead to noticeably higher DLS execution times for WEP fusion (which includes time to execute Algorithm \ref{alg:ISOptAlg} with 5000 samples), but only smaller corresponding time penalties for exact fusion compared to IGS. The time increases for IGS in the exact case can be attributed to the fact that IGS must sample from whole mixture, compute IS weights, and then compute SS-WEM responsibilities via Algorithm \ref{alg:SSWEMLearning} for 100-2000 samples 100-121 times before estimating component statistics, resulting in 10,000-242,000 operations (i.e. an order of magnitude more than needed for only computing IS weights for DLS before estimating approximate fusion mixand statistics). It is also worth noting that the 5000 samples used for WEP represents a typical upper bound on sampling size needed for 2D optimization via Algorithm \ref{alg:ISOptAlg}; the times for DLS in WEP fusion are about same as IGS if the same number of samples are used for $\omega$ optimization. As will be shown later for the 4D maneuvering target tracking problem, the gap between the execution times for FOCI and IGS/DLS approximations also drops significantly when fewer mixands need to be fused at each platform. 

\subsection{Example 2: INGIS, LAGIS and Mixture IS pdfs} 

\begin{figure}[t]
\centering
\newcommand{\figsize}{3.8cm}
\begin{tabular}{@{}c@{}c@{}c@{}c@{}}
\hspace*{-0.15 in}
\includegraphics[width=\figsize]{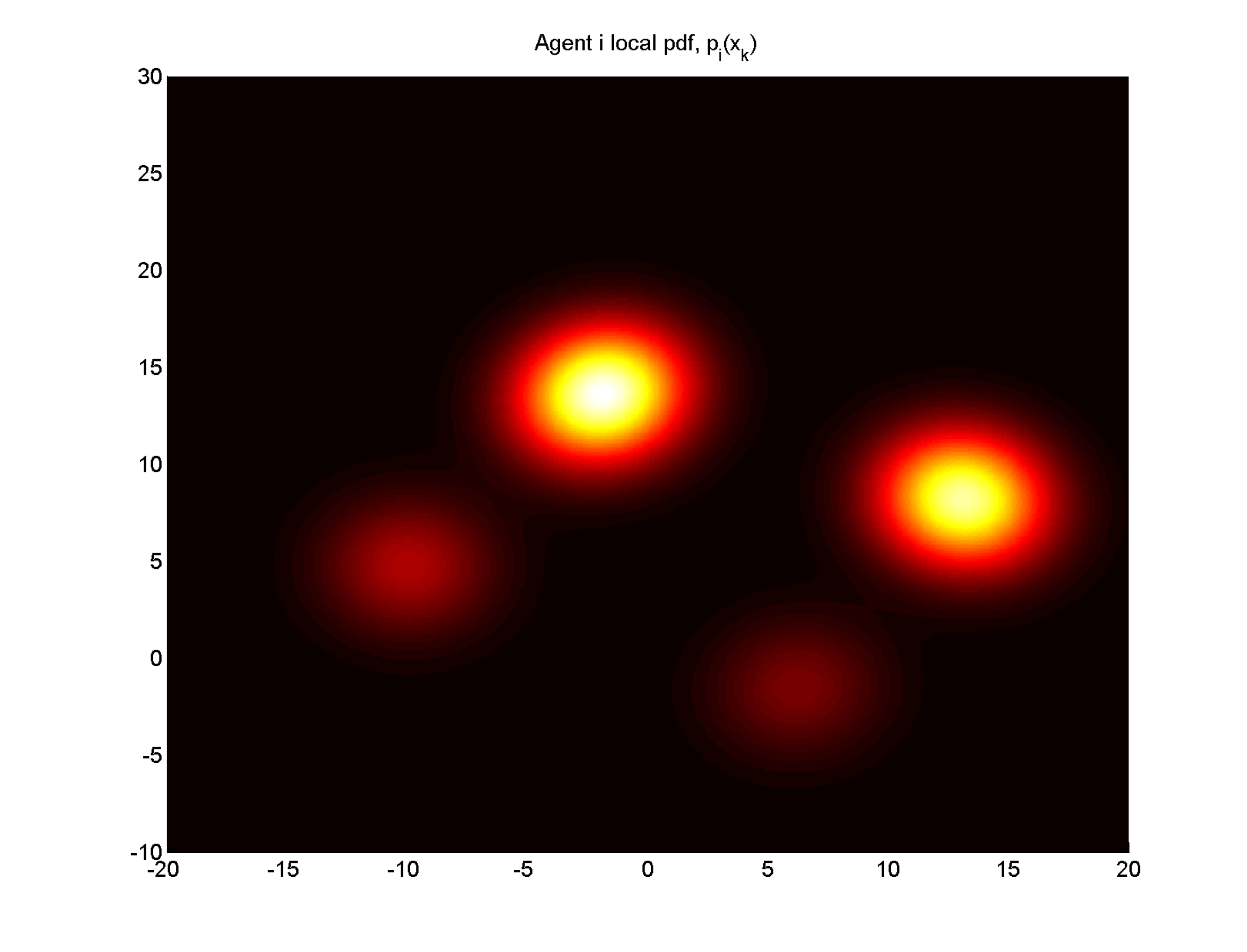} &
\hspace*{-0.15 in}
\includegraphics[width=\figsize]{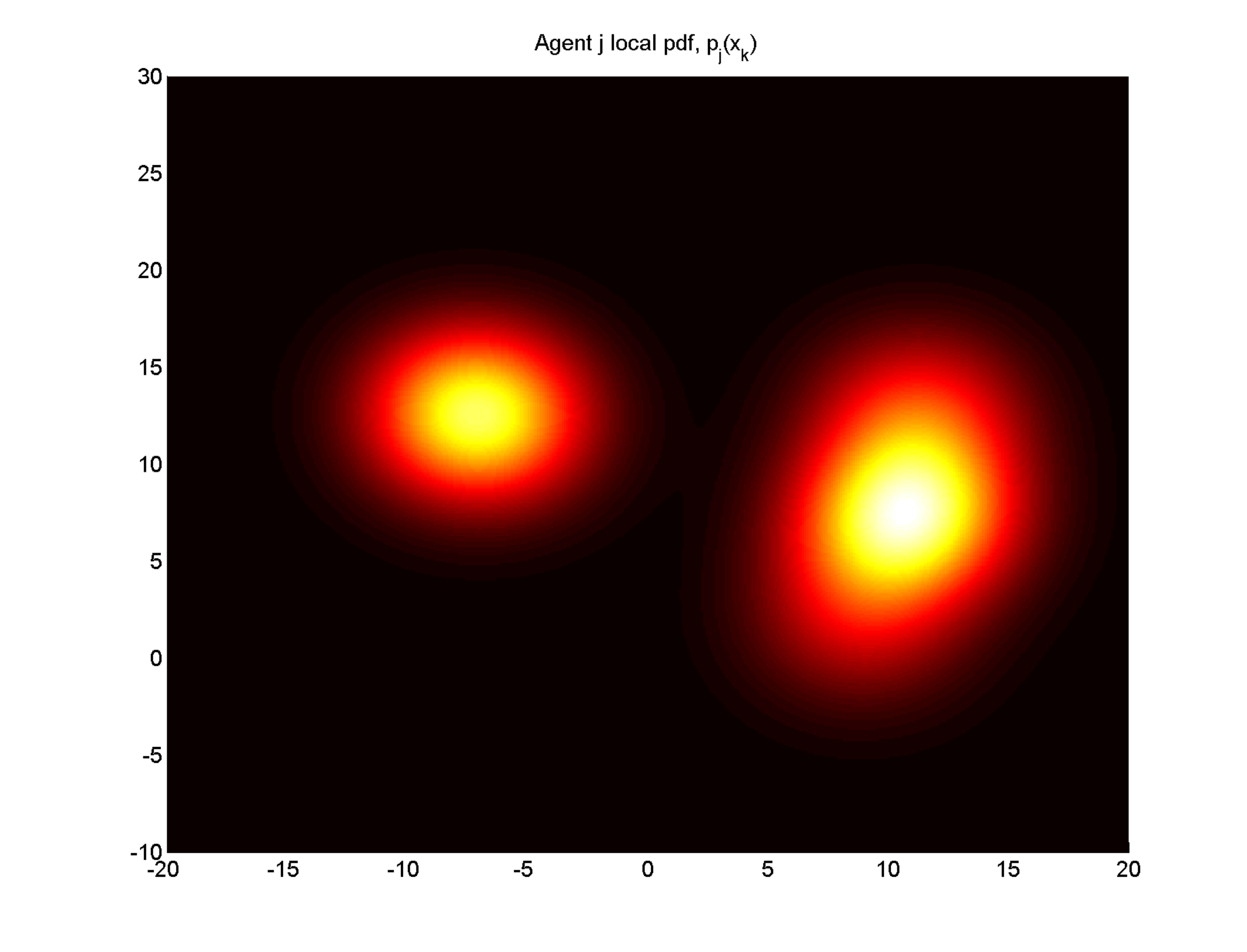} &
\hspace*{-0.15 in}
\includegraphics[width=\figsize]{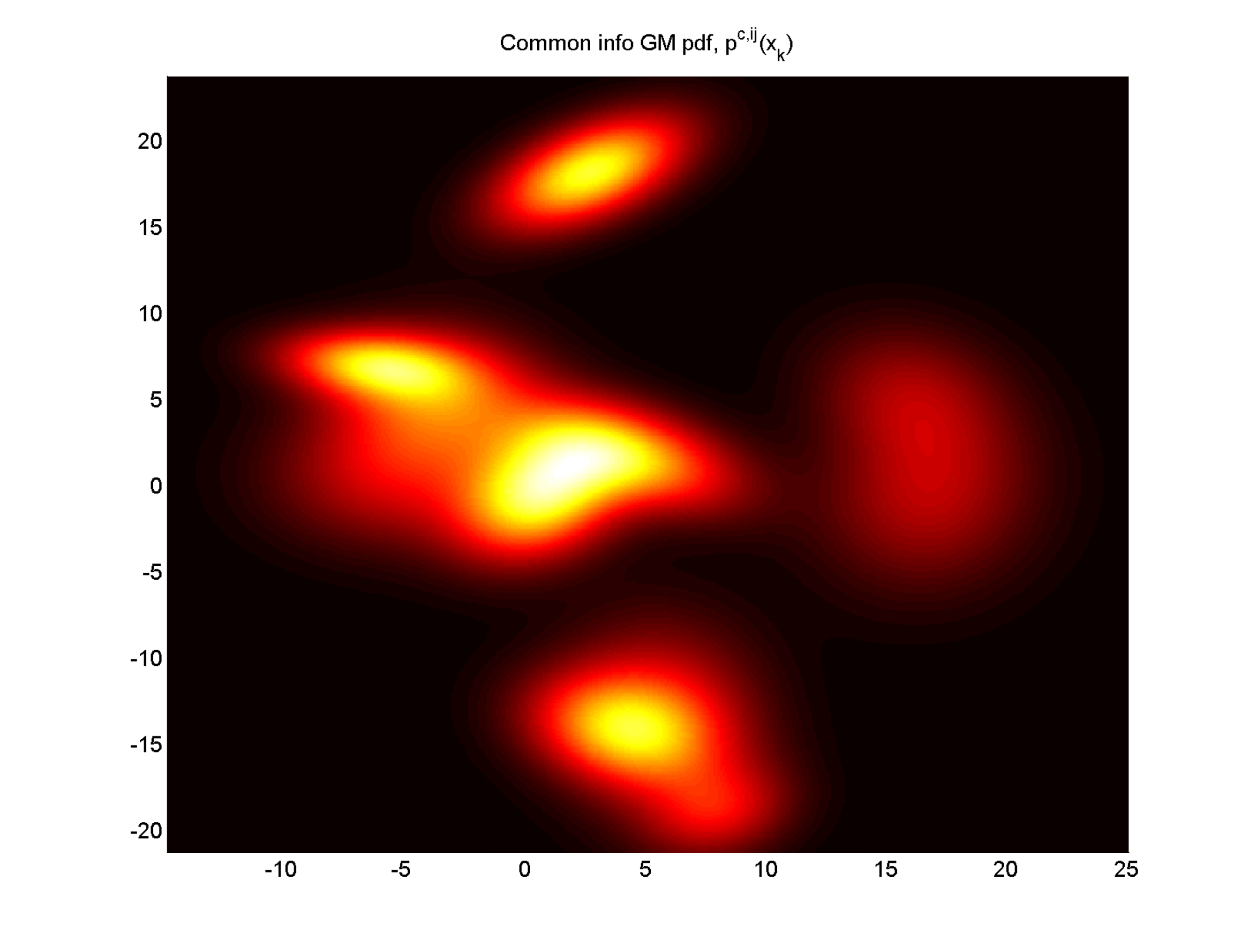} &
\hspace*{-0.15 in}
\includegraphics[width=\figsize]{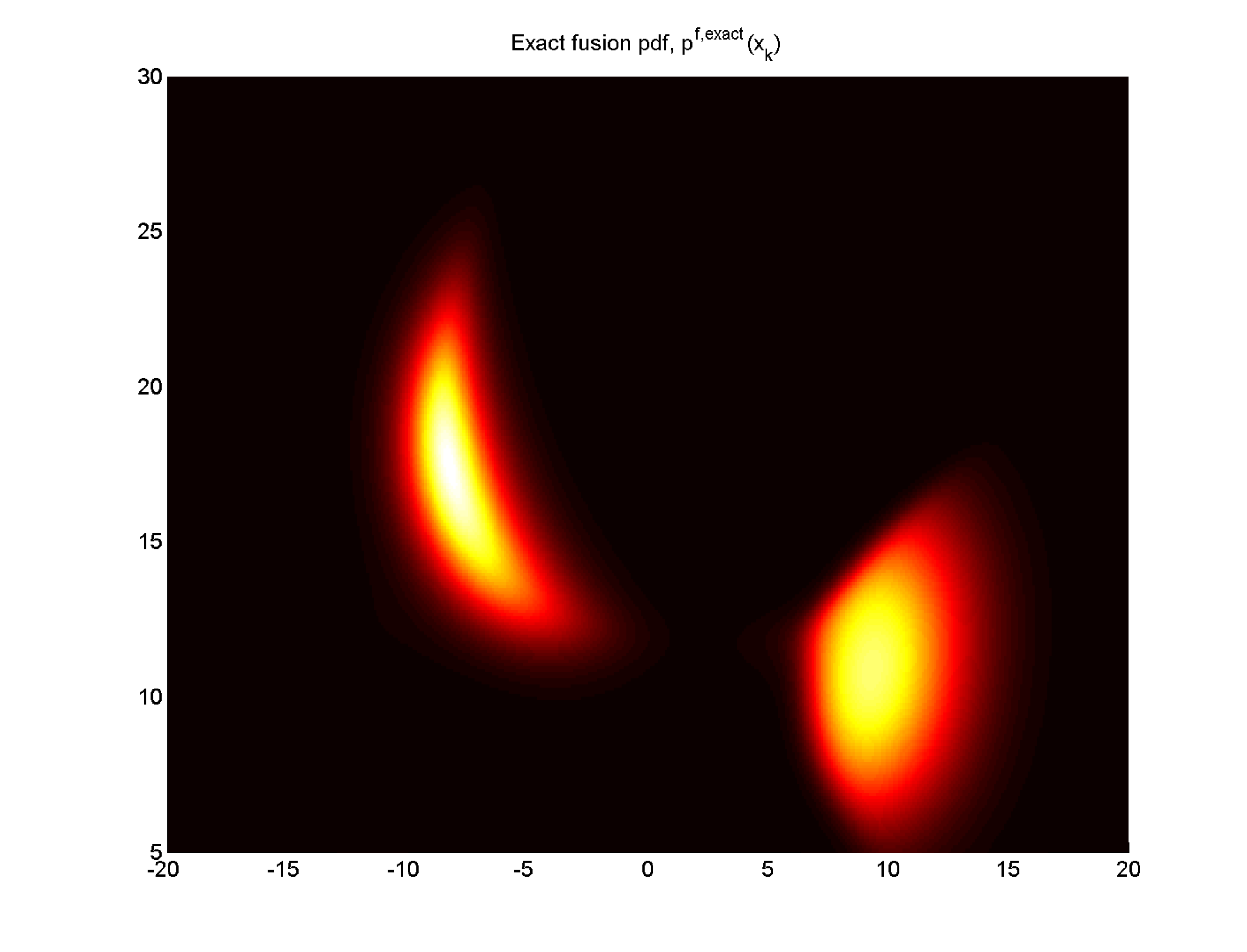} \\
\scriptsize (a) & \scriptsize (b) & \scriptsize (c) & \scriptsize (d) \\
\hspace*{-0.15 in}
\includegraphics[width=\figsize]{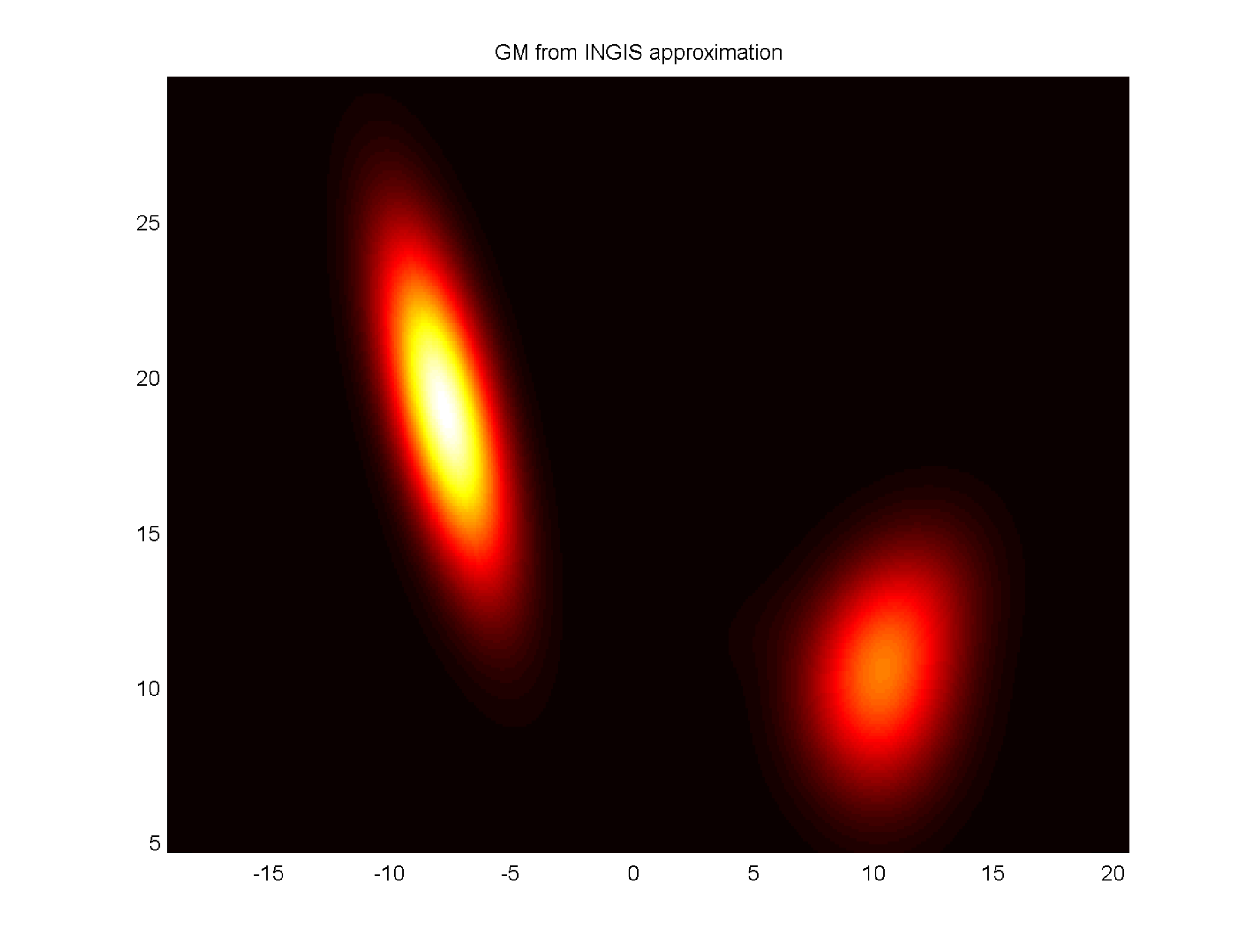} &
\hspace*{-0.15 in}
\includegraphics[width=\figsize]{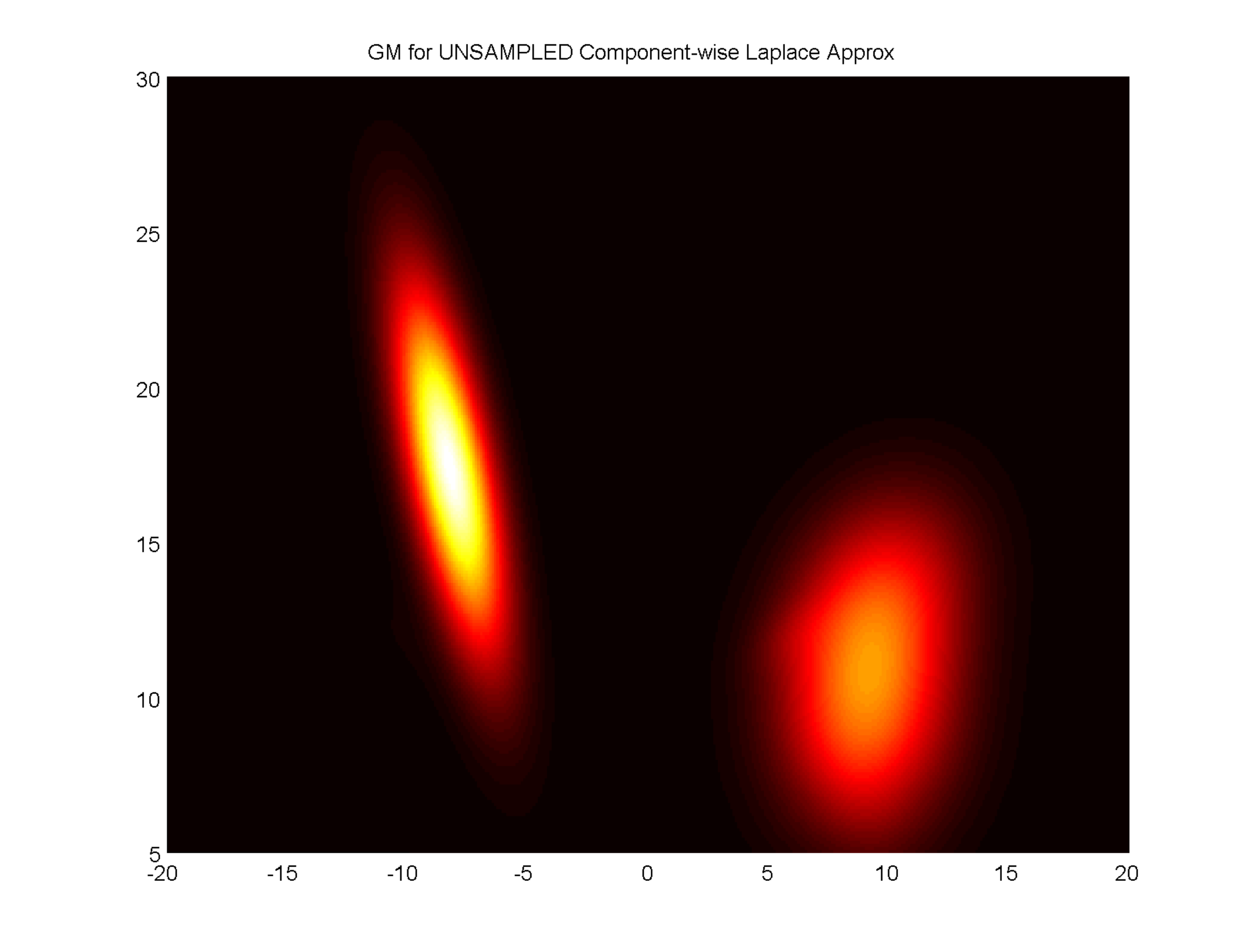} &
\hspace*{-0.15 in}
\includegraphics[width=\figsize]{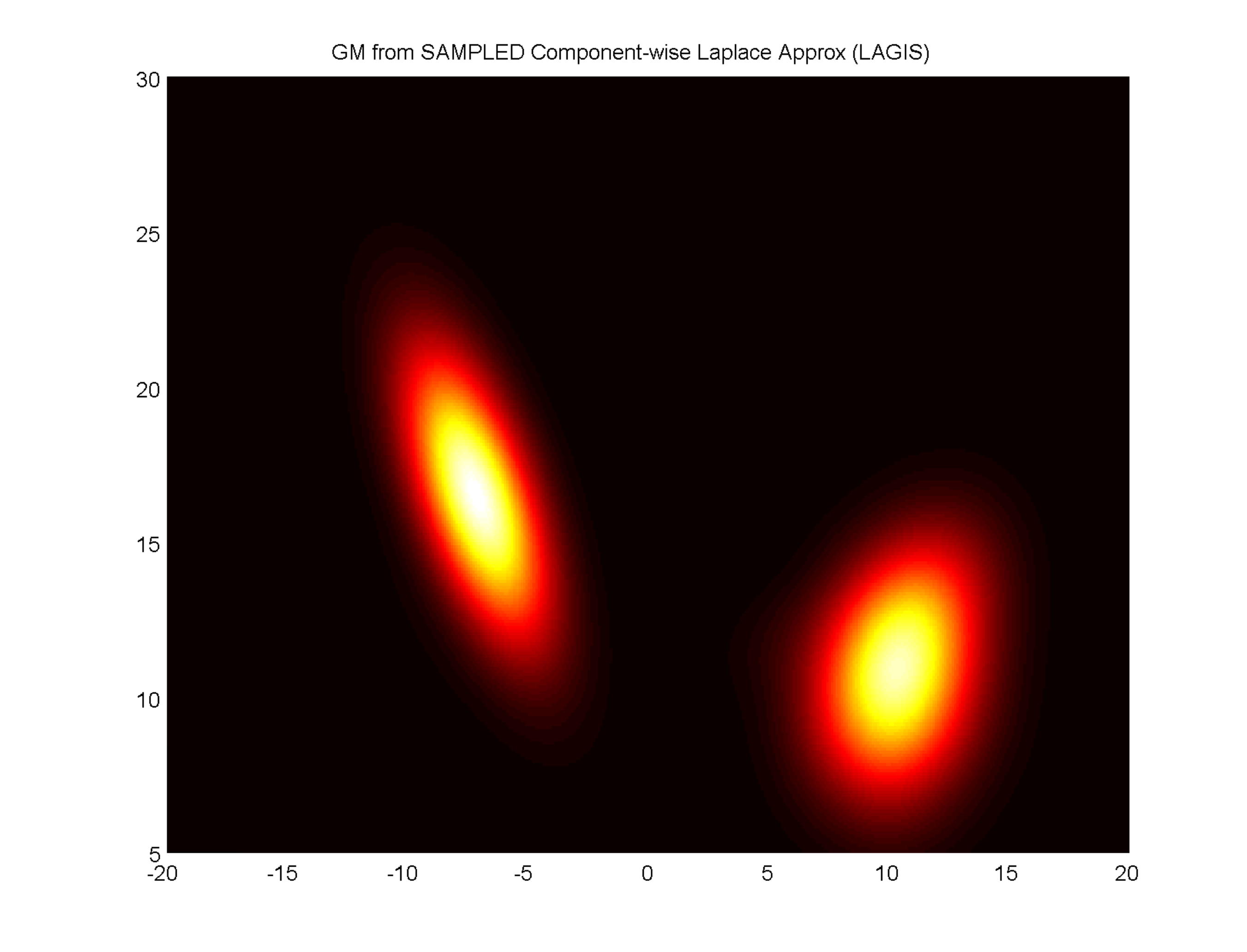} &
\hspace*{-0.15 in}
\includegraphics[width=\figsize]{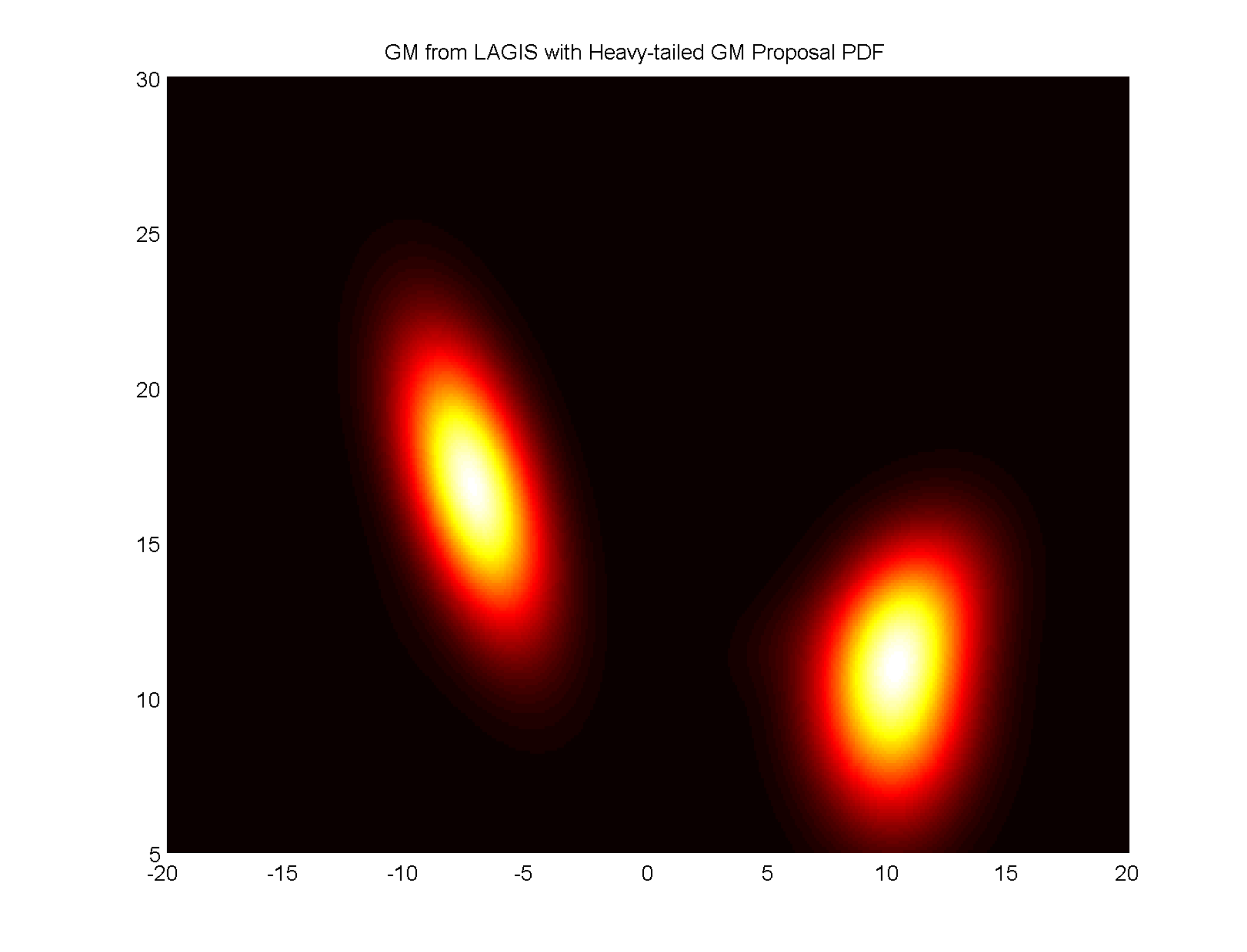} \\
\scriptsize (e) & \scriptsize (f) & \scriptsize (g) & \scriptsize (h)
\end{tabular}
\caption{\scriptsize Comparison of component-wise GM fusion approximation techniques for exact DDF (with KLD from true fusion result): (a) agent $i$ GM pdf, (b) agent $j$ GM pdf, (c) common information GM pdf, (d) exact DDF fusion result (approximated on high density grid); (e) INGIS approximation (KLD = $0.2555$ nats); (f) direct component-wise Laplace approximation without sampling correction (KLD = $0.3763$ nats); (g) LAGIS approximation (KLD = $0.1054$ nats); (h) heavy-tailed LAGIS approximation (KLD = $0.0910$ nats).}
\label{fig:ex2setupResults}
\end{figure}

This example examines the tradeoff between accuracy and computation cost for the INGIS, LAGIS, and heavy tail mixture importance sampling variants. Only the DLS approximation is considered here, as the main insights are similar for the IGS approximation. The effective sample size (ESS) defined in eq. \pareqref{ESS} is useful here in assessing the quality of the different IS approximations with respect to each of the fusion pdf mixands in the case of DLS. 

Fig. \ref{fig:ex2setupResults} (a)-(c) show randomly generated pdfs for $p^i(x_k)$, $p^j(x_k)$ and $p^{c,ij}(x_k)$; Fig. \ref{fig:ex2setupResults} (d) shows the exact DDF result with 16 non-Gaussian mixands. Figs. \ref{fig:ex2setupResults} (e)-(h) respectively show the approximate GMs obtained via different IS methods with $N_s=500$ per mixand and no mixture compression: INGIS (e, $\alpha = 5$); non-sampling Laplace mixture approximation (f, where the Laplace proposal pdf directly approximates each $\tilde{m}_{vr}(x_k)$ \emph{without sampling}); LAGIS (g); and heavy-tail mixture IS (h, which uses maximum scale determined by $\Sigma^{t}_{vr}$ from \pareqref{upperboundDef} for $M^h=5$ mixdands). The KLDs with respect to the true fusion pdf in (d) are also provided. The approximations in (e) and (f) capture the broad features of (d), but give only rough estimates of the covariance for the `banana' shaped mode in the upper left side and the mixand weight for lower right mode. The LAGIS and heavy-tail methods correct these issues, where the latter better accounts for the `slanted top' of the lower right mode in (d).


\begin{figure}[t]
\centering
\newcommand{\figsize}{6.5cm}
\begin{tabular}{@{}c@{}c@{}}
\hspace*{-0.25 in}
\includegraphics[width=\figsize]{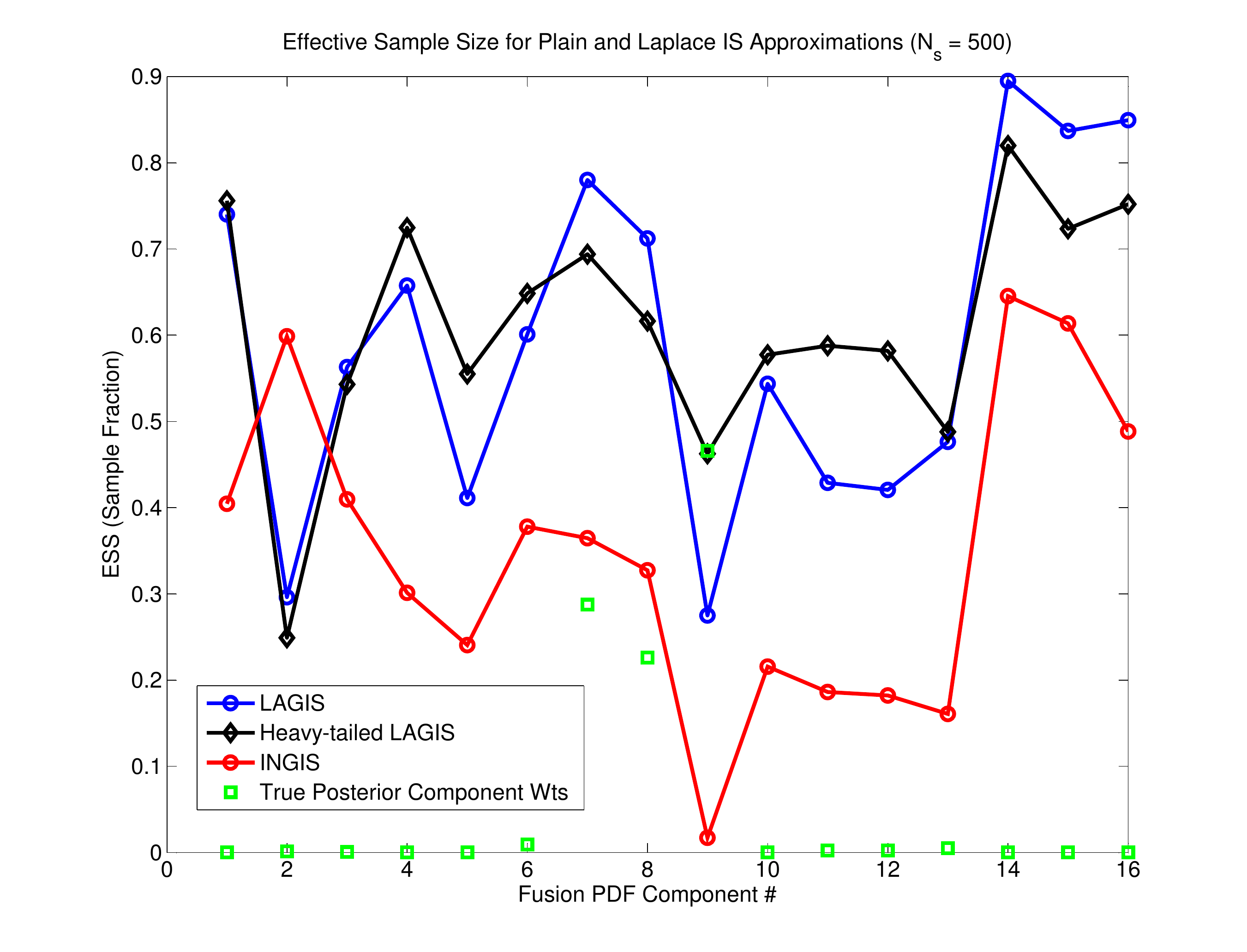} &
\hspace*{-0.25 in}
\includegraphics[width=\figsize]{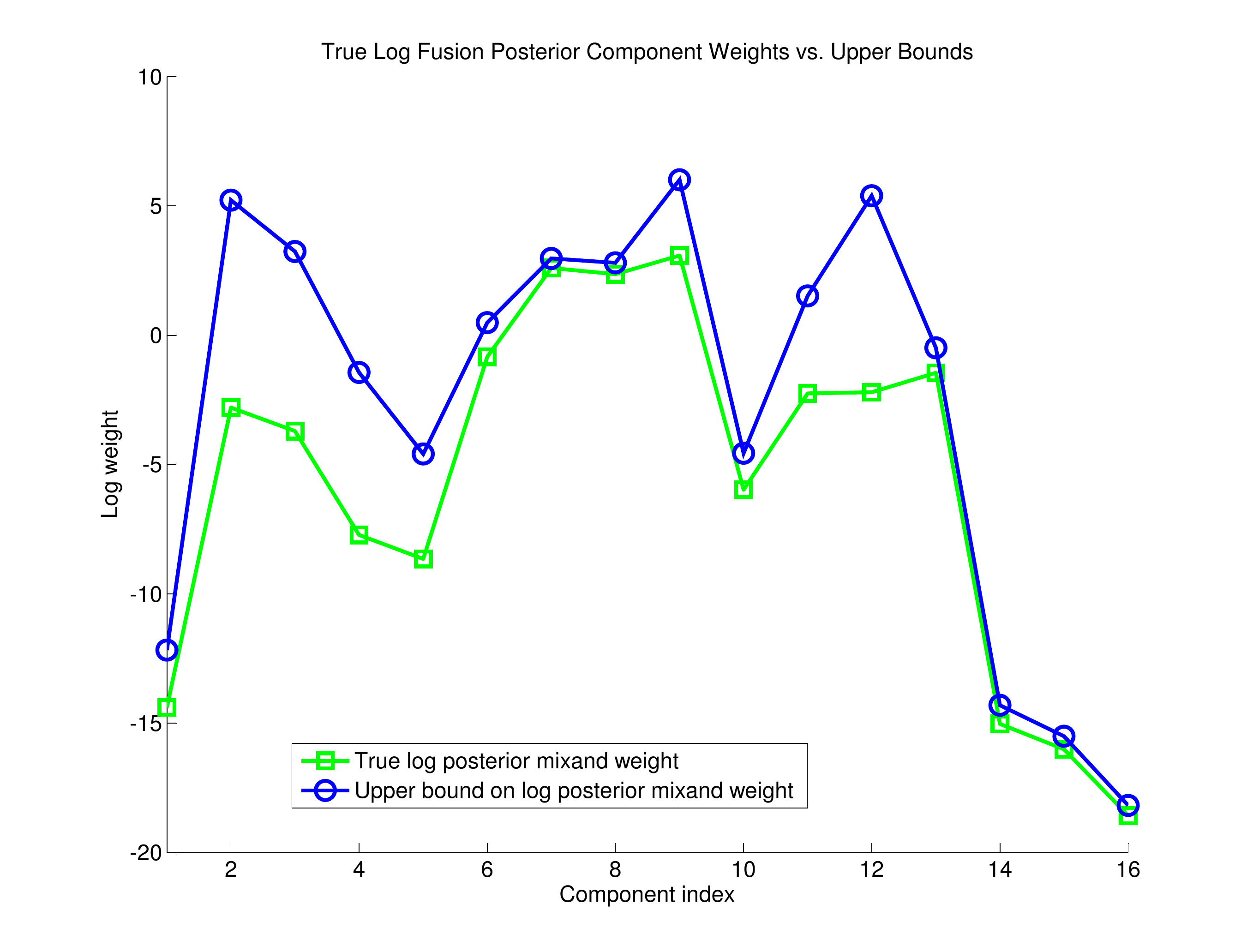} \\
\scriptsize (a) & \scriptsize (b) \\
\hspace*{-0.25 in}
\includegraphics[width=\figsize]{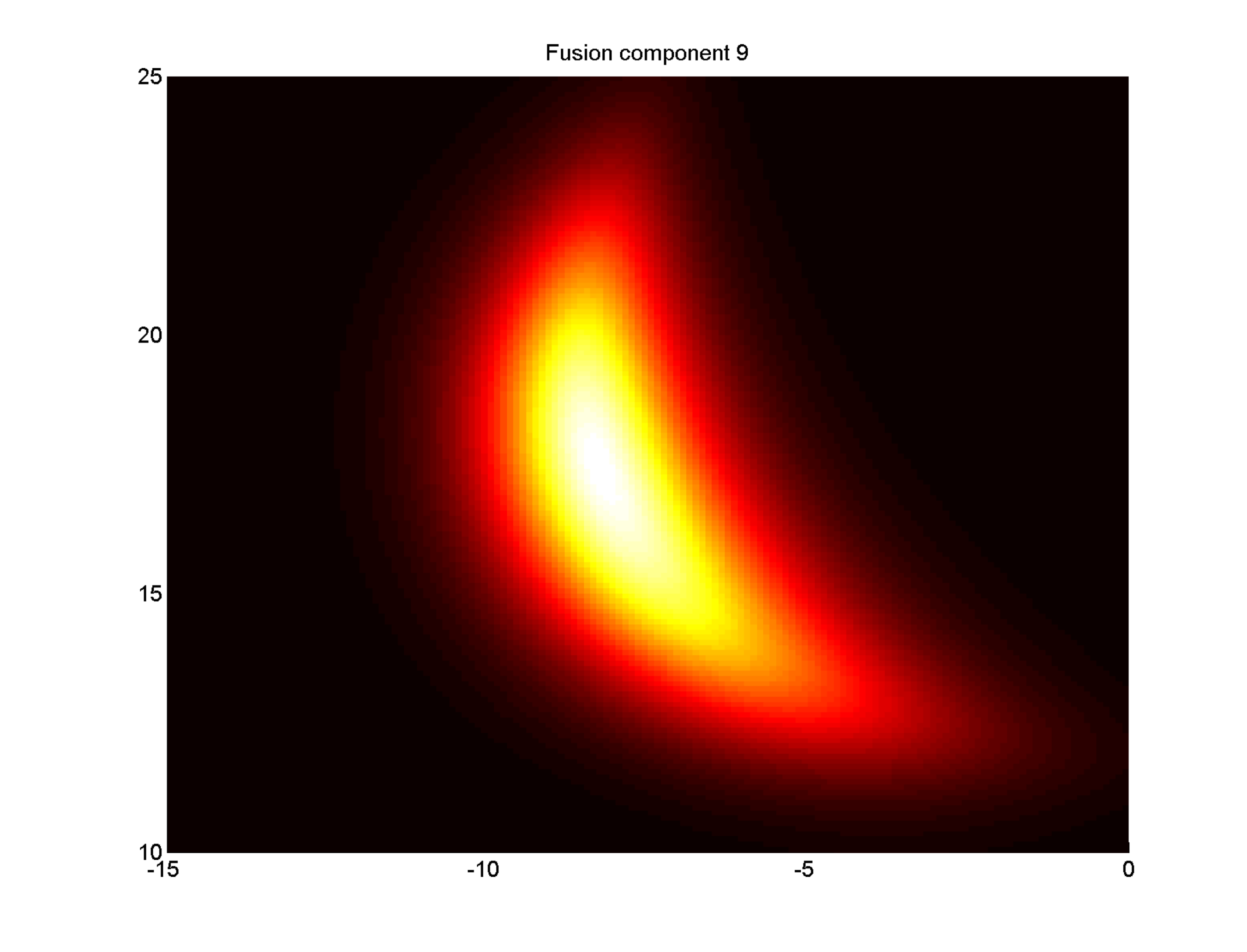} &
\hspace*{-0.25 in}
\includegraphics[width=\figsize]{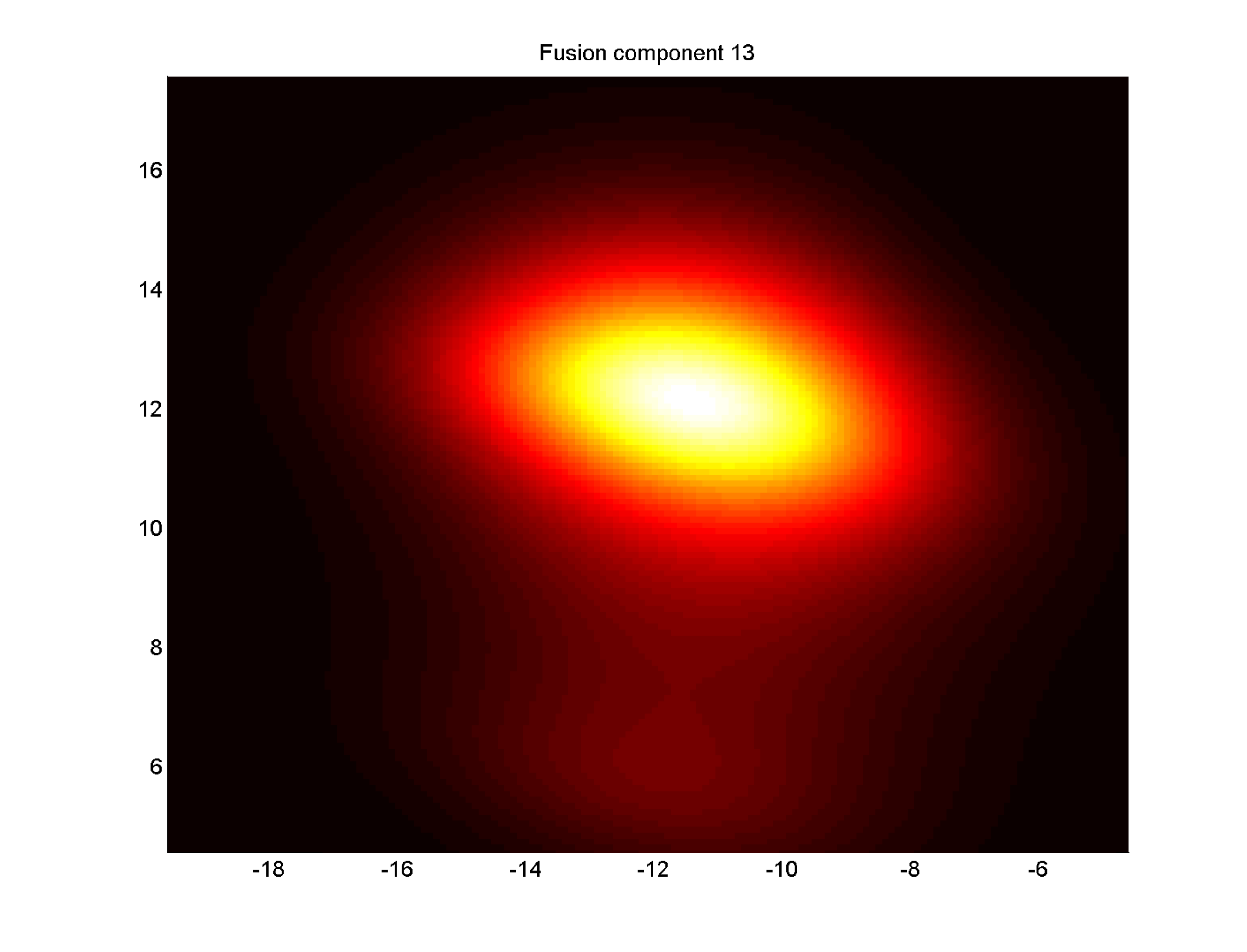} \\
\scriptsize (c) & \scriptsize (d)
\end{tabular}
\caption{\scriptsize (a) Single run component-wise ESS results for exact GM fusion problem in Fig. \ref{fig:ex2setupResults} using INGIS (red), LAGIS (blue) and heavy-tailed LAGIS (black), shown along with true posterior fusion mixture component weights (green); (b) logarithm of true component weights (green) and upper bounds given by $\kappa_{qr}$; (c)-(d) components 9 and 13 true exact posterior fusion pdf, showing highly non-Gaussian features that are not easily captured by naive Gaussian approximations.}
\label{fig:ex2ESSweights}
\vspace{-0.5cm}
\end{figure}

Fig. \ref{fig:ex2ESSweights} (a) shows the ESS obtained by each importance sampling method for each mixand of the fusion pdf (plotted here as sample fraction); the true posterior mixand weights are also shown to give a sense of how well each method does on the `important parts' of fusion pdf. The LAGIS and heavy-tail methods generally provide the best ESS across all posterior mixands, even though many of these are highly non-Gaussian, as shown in Fig. \ref{fig:ex2ESSweights} (c) and (d). INGIS struggles the with mixand $\#9$, the `banana-shaped' mixand in Fig. \ref{fig:ex2ESSweights} (c), which has nearly $50\%$ of the posterior probability mass. LAGIS and heavy-tailed IS significantly improve on INGIS here by accounting for the fact that the mean and mode are not co-located. The heavy-tail method also accounts for the non-symmetric mass distribution around the mode. Fig. \ref{fig:ex2ESSweights} (a) also shows that heavy-tailed IS sometimes performs worse than LAGIS, i.e. whenever the scale factors in \pareqref{scaleLAGIS} lead to inefficient sampling too far from the posterior mode. Fig. \ref{fig:ex2ESSweights} (b) compares the log value of the weight upper bound \pareqref{upperboundWt} to the true log posterior weight for each mixand. This bound is loose in many cases, but extremely small values give strong indication of negligible mixands and are thus useful for mixand pruning ahead of importance sampling.


\subsection{Example 3: Multi-platform Target Search} 
\begin{figure*}[t]
\centering
\newcommand{\figsize}{3.8cm}
\begin{tabular}{@{}c@{}c@{}c@{}c@{}}
\hspace*{-0.25 in}
\includegraphics[width=\figsize]{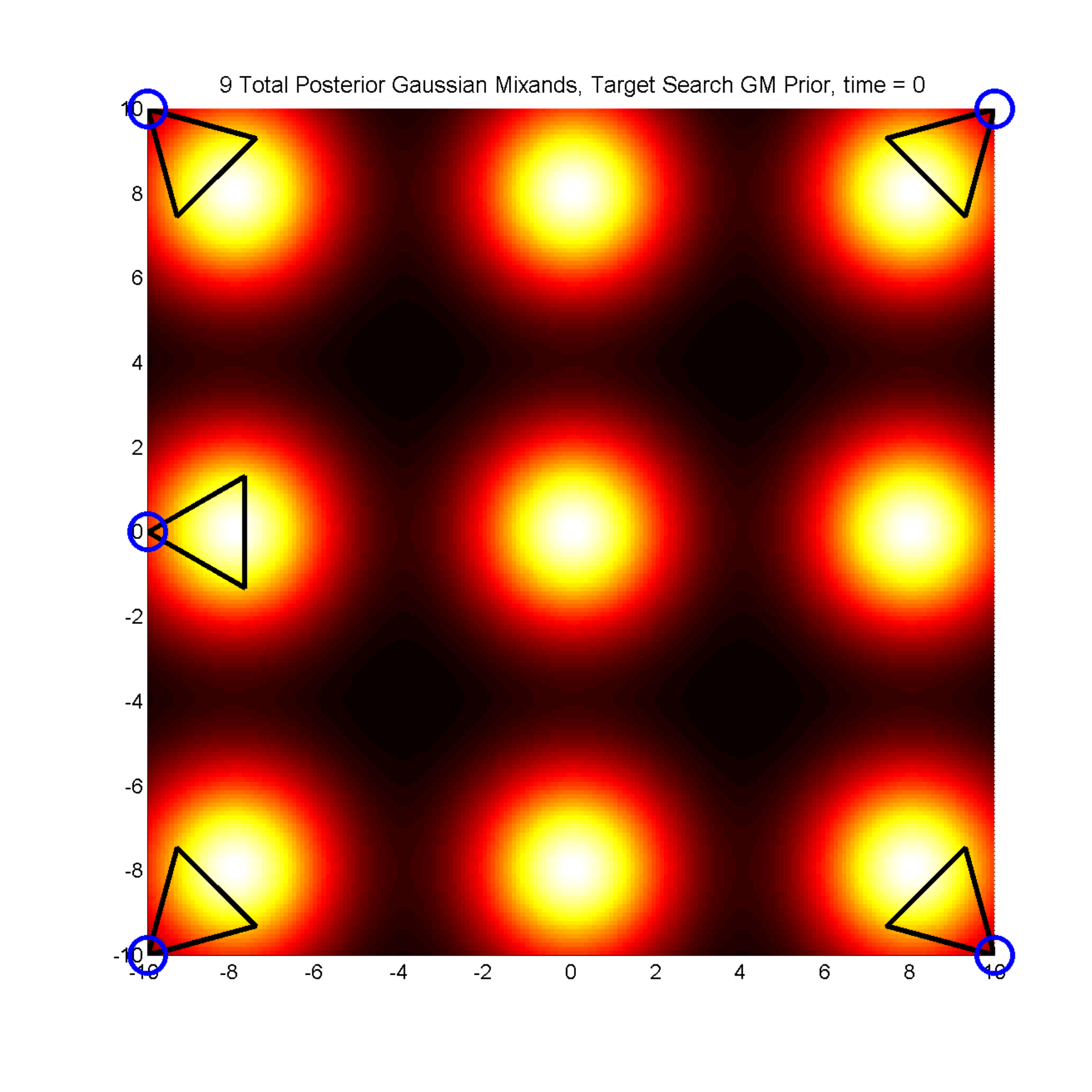} &
\hspace*{-0.25 in}
\includegraphics[width=\figsize]{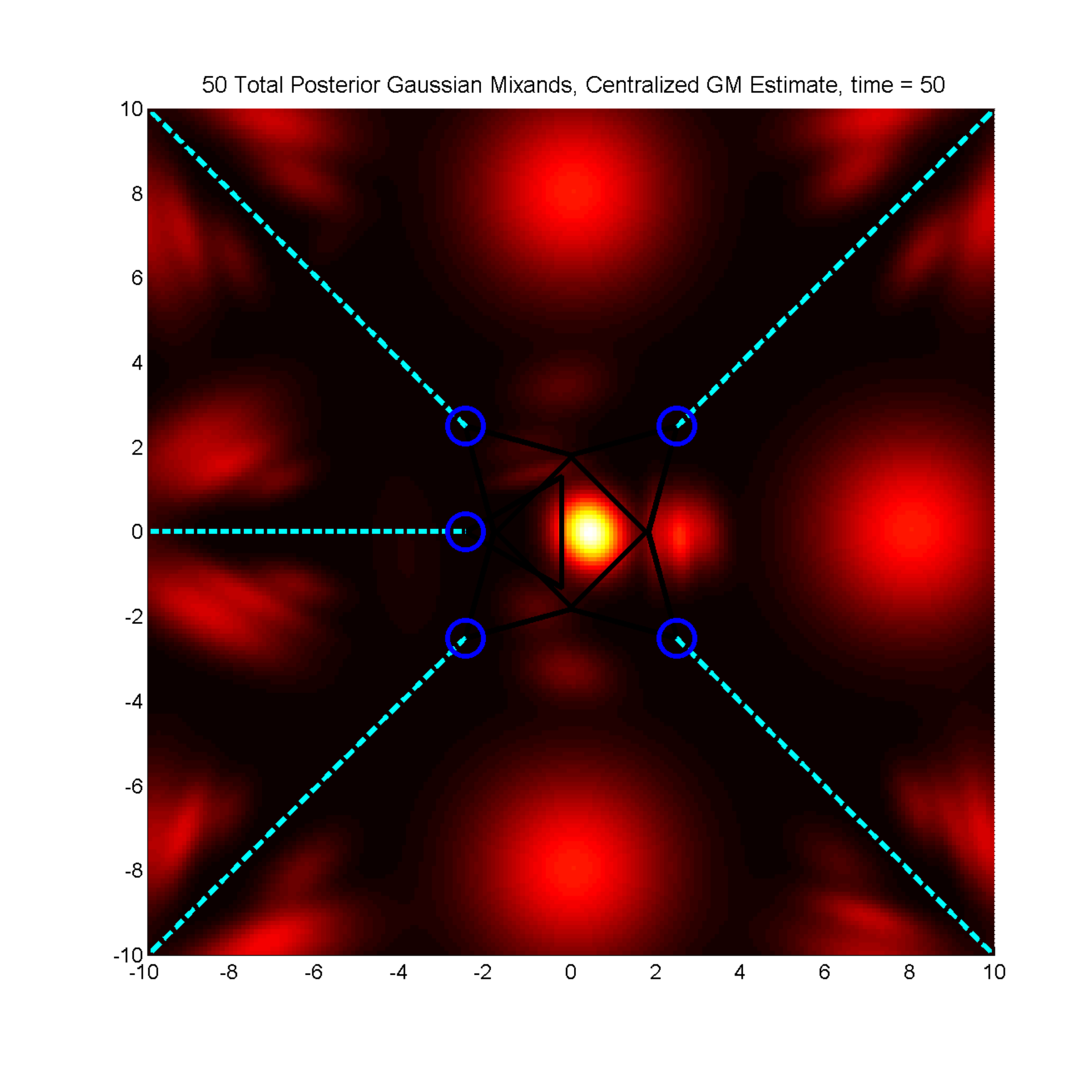} &
\includegraphics[width=\figsize]{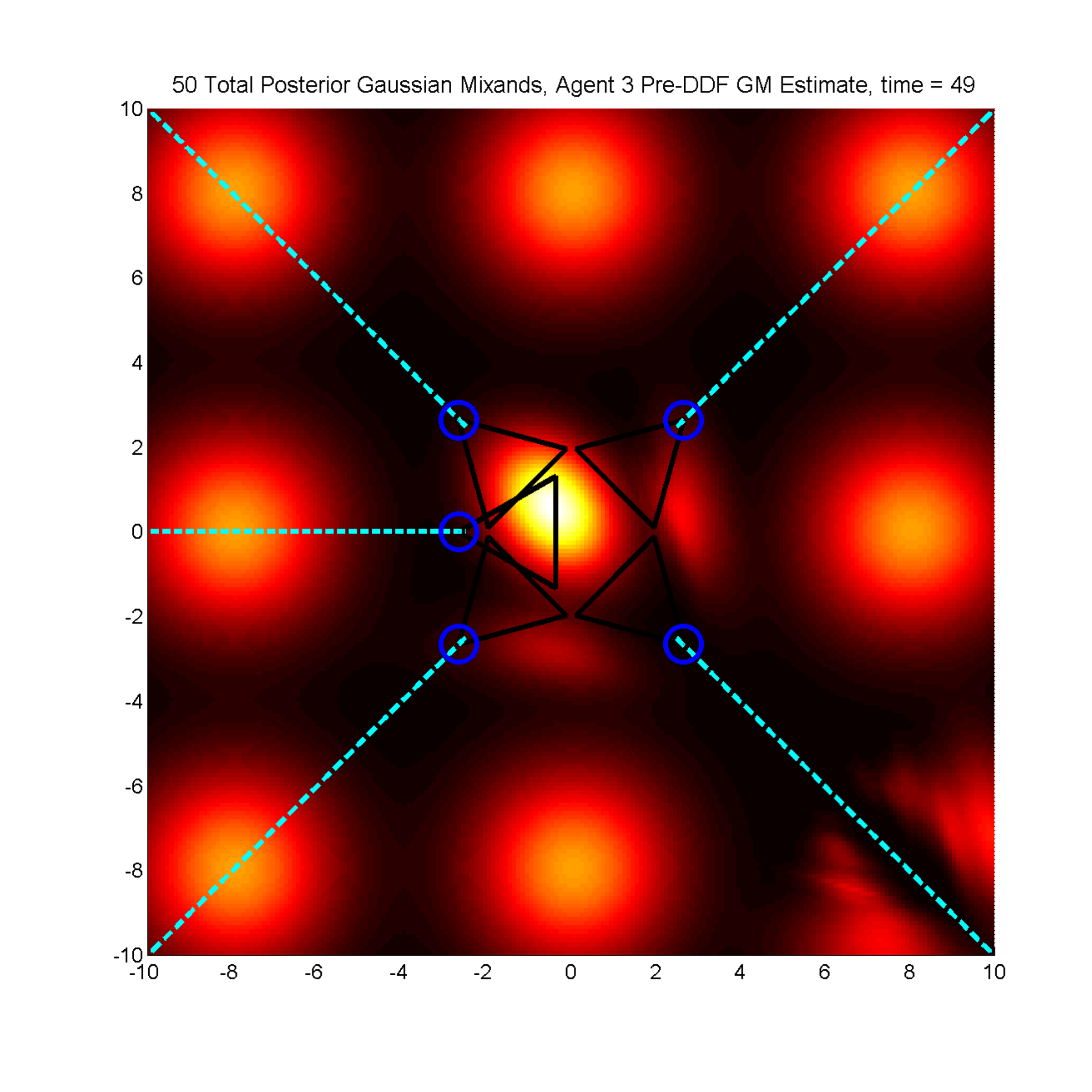} &
\hspace*{-0.25 in}
\includegraphics[width=4.45cm]{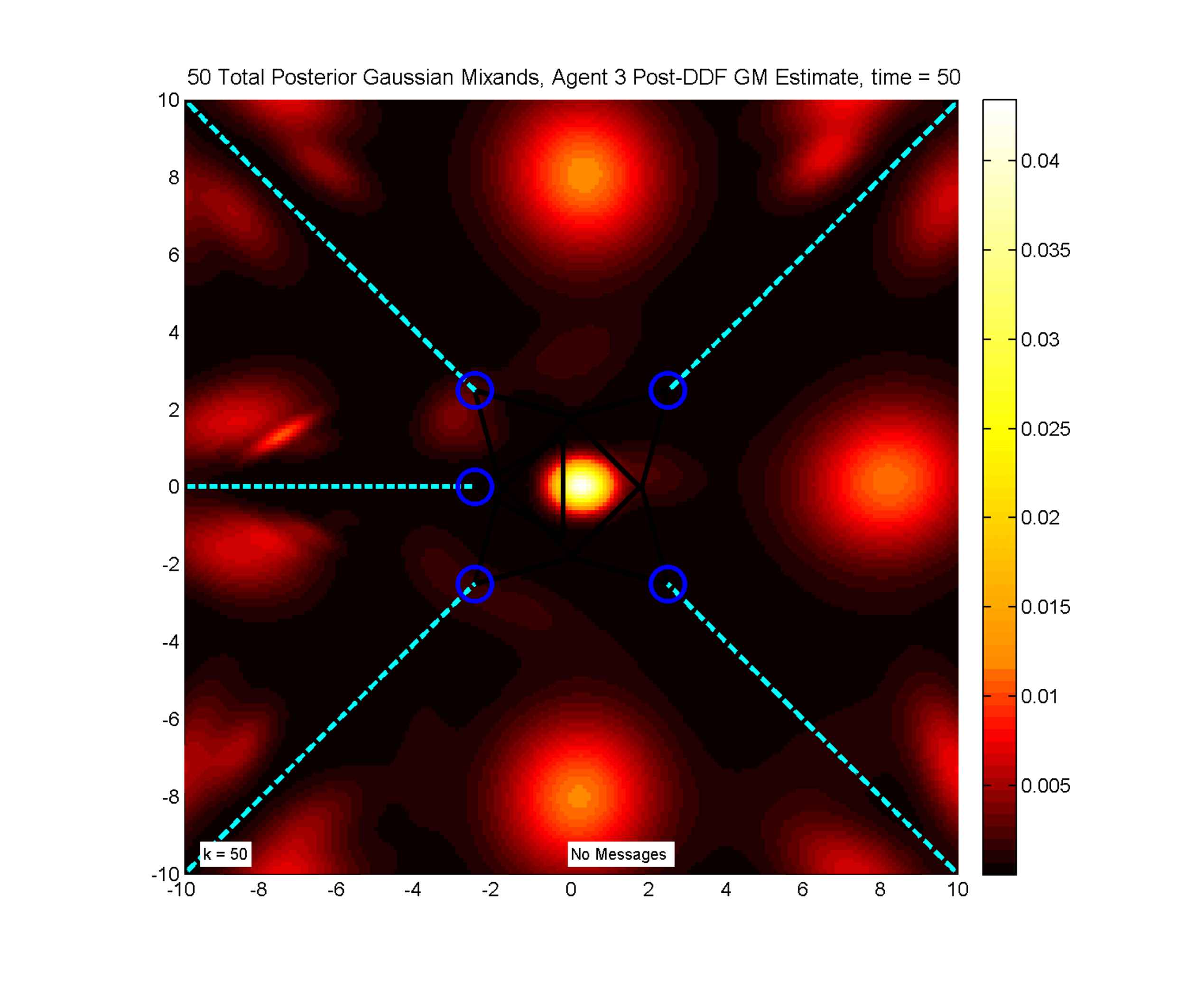} \\
\scriptsize (a) & \scriptsize (b) & \scriptsize (c) & \scriptsize (d)
\end{tabular}
\caption{\scriptsize Comparison of centralized and DDF GM fusion results for 5 robot target search scenario: (a) GM prior and binary visual target detection viewcones for search robots; (b) centralized GM fusion posterior pdf for robot trajectories (shown in cyan) after $k=50$ time steps; (c) local GM fusion result for robot 3 (which started from lower right corner) after $k=49$ steps, prior to DDF update; (d) local GM fusion result for robot 3 at step $k=50$ following DDF update with robots 1,2, 4 and 5 (KLD w.r.t. (b) = $0.1173$ nats).}
\label{fig:ex3searchdemo}
\vspace{-0.35cm}
\end{figure*}

This example shows how the DLS approximation performs over multiple sequential exact fusion instances for a simulated multi-robot target search mission. Fig. \ref{fig:ex3searchdemo} (a) shows the 9 component GM prior used to specify the location of a single static target (located at $(0,0)$, not shown). Also shown are the starting positions and headings of 5 search robots, which are equipped with forward-looking binary visual target detectors (viewcones depicted by the black triangles). Fig. \ref{fig:ex3searchdemo} (b) shows the GM pdf resulting from centralized fusion of all 5 robots' `no detection' measurements collected over 50 consecutive time steps along the indicated trajectories (dashed lines). This GM is compressed to 50 components, but still gives a very close approximation to the true Bayes posterior pdf, which is shows the non-Gaussian `scattering effect' characteristic of negative information fusion in search problems \cite{FredThesis}. 

Fig. \ref{fig:ex3searchdemo} (c) and (d) show the results of applying exact Bayesian DDF to the search problem, where all robots communicate in a star topology with robot 3 at the hub of the network. The robots all start off with the same prior GM shown in Fig. \ref{fig:ex3searchdemo} (a), and do not communicate with each other until time step $k=50$, at which point they share their locally constructed posterior GMs and common information pdfs to perform GM DDF updates. Fig. \ref{fig:ex3searchdemo} (c) shows the GM pdf for robot 3 (located lower right corner) following Bayesian fusion of only its own local `no detection' measurements for the first 49 time steps. Fig. \ref{fig:ex3searchdemo} (d) shows the GM that results from successive fusion of the local GMs from robots 1, 2, 4 and 5, respectively, according to Algorithm 1 (INGIS, with $\alpha = 0.5$ and $N_s = 500$). The resulting KLD between the centralized fusion GM in (b) and the exact DDF GM in (d) is 0.1173 nats. Although the smaller modes around $(0,0)$ are diminished in (d) due to the use of successive GM compression following GM fusion with each other robot (which limits the maximum mixture size to 50 components after each pass), the overall agreement between the centralized and DDF results is still good, especially as the information obtained by the other robots for the modes in the other corners of the search space comes through clearly.


\subsection{Example 4: Maneuvering Target Tracking with Limited Data and Comms}
\subsubsection{Problem setup} 
This example demonstrates GM DDF for a more challenging decentralized dynamic target tracking scenario involving higher dimensional GM pdfs. Consider three independent and static sensing platforms $i \in \set{1,2,3}$ located at East-North positions $(\xi^i,\eta^i)$ that must each track a highly maneuverable aerial target over a large 2D surveillance area. The target's dynamics are given by a 5-mode jump Markov hybrid linear system model with nearly constant velocity kinematics for inertial East-North position and velocity states $x_k = [\xi^t_k,\dot{\xi}^t_k,\eta^t_k,\dot{\eta}^t_k]^T$ discretized at $\Delta T= 1$ sec, 
\begin{align}
x_{k+1} &= F^{m_k} x_k + w_k, \\
w_k &\sim {\cal N}(0,Q^{m_k}), \\ 
F^{m_k} &= \begin{bmatrix}
1 & \Delta T & 0 & 0 \\
0 & 1 & 0 & 0 \\
0 & 0 & 1 & \Delta T \\
0 & 0 & 0 & 1 
\end{bmatrix}, \ \mbox{for \ } m_k = 1 \\
F^{m_k} &= \begin{bmatrix}
1 & \frac{\sin(\Omega^{m_k} \Delta T)}{\Omega^{m_k}} & 0 & \frac{-(1-\cos(\Omega^{m_k} \Delta T))}{\Omega^{m_k}} \\
0 & \cos(\Omega^{m_k} \Delta T) & 0 & -\sin(\Omega^{m_k} \Delta T) \\
0 & \frac{(1-\cos(\Omega^{m_k} \Delta T))}{\Omega^{m_k}} & 1 & \frac{\sin(\Omega^{m_k} \Delta T)}{\Omega^{m_k}} \\
0 & \sin(\Omega^{m_k} \Delta T) & 0 & \cos(\Omega^{m_k} \Delta T) 
\end{bmatrix}, \ \mbox{for \ } m_k = 2,3,4,5 
\end{align}
where mode 1 represents straight level flight with no turn, modes 2 and 4 represent starboard turns with $\Omega^2=-0.05$ rad/s (wide turn) and $\Omega^4=-0.15$ rad/s (tight turn), and modes 3 and 5 represent port-side turns with $\Omega^3=0.05$ rad/s and $\Omega^5=0.15$ rad/s. The discrete time process noise matrix $Q^{m_k}$ for each mode is found by applying Van Loan's method \cite{Brown2012KFintroduction} to continuous time white noise process accelerations with intensity $2$ (m/s$^2$)$^2$. 
The stochastic mode switching dynamics are governed by a 5-state Markov chain for the discrete random variable $m_k \in \set{1,...,5}$, with state transition probabilities encoded in matrix $A \in \mathbb{R}^{5 \times 5}$
\begin{align}
\pi_{k+1}(m_{k+1}) &= A \pi_{k}(m_k), \\
A(i,j) &= \begin{cases}
0.85, \mbox{ \ if \ } i=j, \\
0.0375, \mbox{ \ if \ } i\neq j,
\end{cases}
\end{align}
where $\pi_k = [P(m_k=1),\cdots, P(m_k=5)]^T$ is the modal probability vector at time $k$ such that $\sum_{m}\pi_k(m)=1$ and $\pi_0$ is assumed given. 

\begin{figure}[t]
\centering
\newcommand{\figsize}{8.5cm}
\includegraphics[width=\figsize]{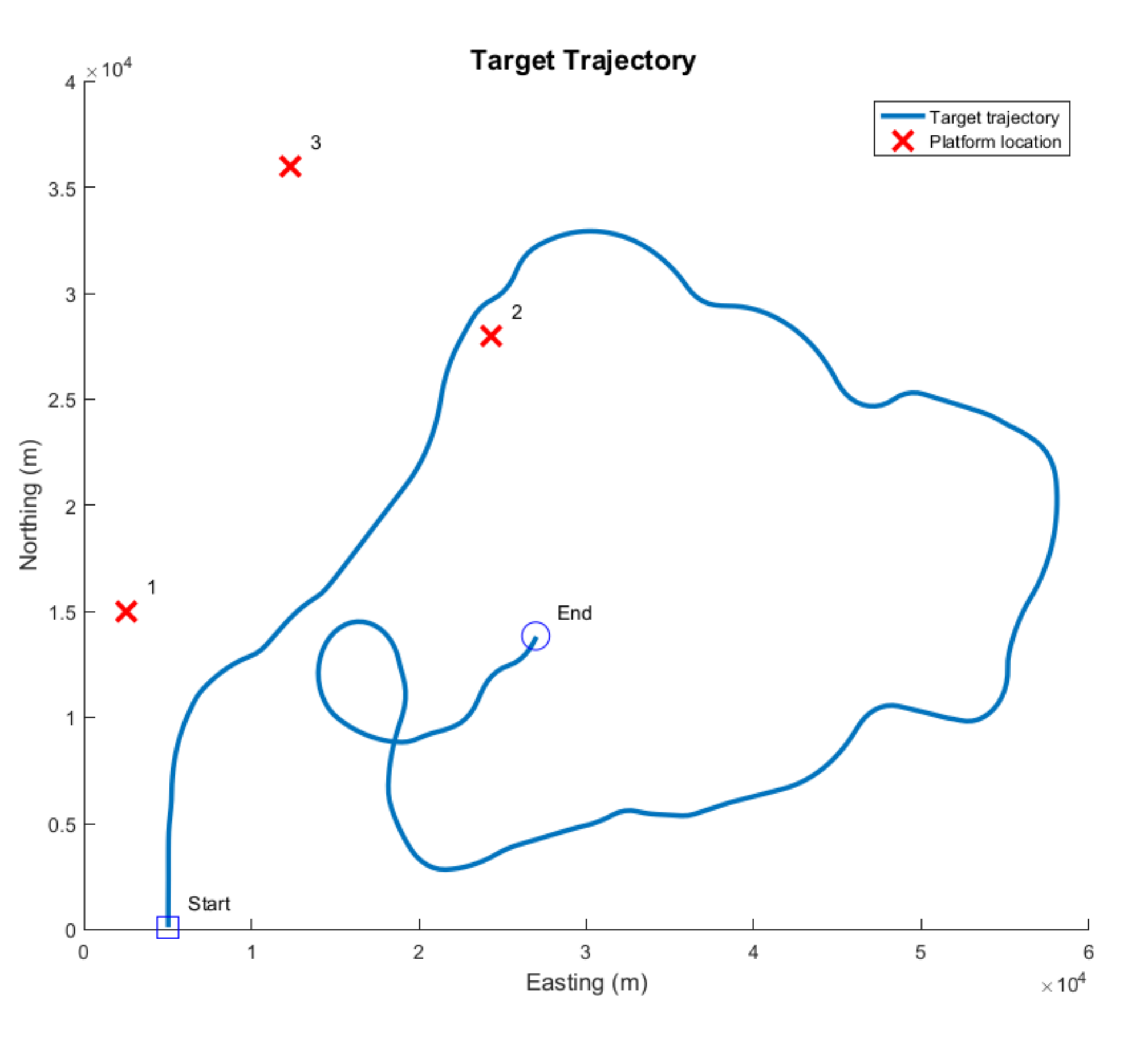}
\caption{\scriptsize Typical true target trajectory and true sensing platform locations for maneuvering range-only tracking scenario.}
\label{fig:ex4trackingsetup}
\end{figure}

Figure \ref{fig:ex4trackingsetup} shows the relative geometry of the sensor platforms and a typical true 2D target trajectory for a 7 min tracking scenario. For simplicity, each $i$ is assumed to have unlimited sensing range and fixed sensor noise characteristics for synchronous measurements that are converted to noisy 2D pseudo-ranges and range rates, 
\begin{align}
y^i_k &= \begin{bmatrix}
\rho^i_k \\
\dot{\rho}^i_k
\end{bmatrix} \\
\rho^i_k &= \sqrt{ (\xi^t_k - \xi^i)^2 - (\eta^t_k - \eta^i)^2 } + v^{i,\rho}_k, \\
\dot{\rho}^i_k &= \frac{(\xi^t_k - \xi^i)(\dot{\xi}^t_k - \dot{\xi}^i)-(\eta^t_k - \eta^i)(\dot{\eta}^t_k - \dot{\eta}^i)}{\rho^i_k} + v^{i,\dot{\rho}}_k, \\
v^{i,\rho}_k &\sim {\cal N}(0,R^{i,\rho}), \ \ v^{i,\dot{\rho}}_k \sim {\cal N}(0,R^{i,\dot{\rho}}),
\end{align}
where $R^{i,\rho}=400$ m$^2$ and $R^{i,\dot{\rho}}=1$ (m/s)$^2$. 

\subsubsection{Bayesian estimators for data fusion} 
Interactive multiple model (IMM) filtering strategies are well-suited to the hybrid stochastic dynamics for this problem. 
In IMM filtering, recursive Bayesian estimates are sought for the joint posterior mode and state distribution $p(x^t_k,m_k|y_{1:k})$ given all available platform observations $y_{1:k}$. This joint posterior can be computed via a two stage update: the first to compute the posterior mode conditional state pdfs $p(x^t_k|m_k,y_{1:k})$ for each possible $m_k$, and the second to compute the posterior mode distribution $P(m_k|y_{1:k})$,
\begin{align}
p(x^t_k|m_k,y_{1:k}) = \frac{p(x^t_k|m_k,y_{1:k})p(y_k|x^t_k,m_k,y_{1:k-1})}{p(y_k|m_k,y_{1:k-1})}\\
P(m_k|y_{1:k}) = \frac{P(m_k|y_{1:k-1})p(y_k|m_k,y_{1:k-1})}{\sum_{m_k} P(m_k|y_{1:k-1})p(y_k|m_k,y_{1:k-1})} \label{eq:IMMBayesUpdate} \\
p(y_k|m_k,y_{1:k-1}) = \int p(x^t_k|m_k,y_{1:k})p(y_k|x^t_k,m_k,y_{1:k-1}) dx^t_k, 
\end{align}
where $p(x^t_k,m_k|y_{1:k}) = p(x^t_k|m_k,y_{1:k})P(m_k|y_{1:k})$. 
\noindent Following Bayes measurement updates, each distribution is predicted forward through the Markov model switching dynamics, where 
\begin{align}
&P(m_{k+1}|y_{1:k}) = \pi_{k+1}(m_{k+1}) = \sum_{m} P(m_{k+1}|m_{k}=m) P(m_{k}=m|y_{1:k}) \\
&= A [P(m_{k+1}=1|y_{1:k}),\cdots,P(m_{k+1}=5|y_{1:k})]^T = A \pi_k(m_k), \label{eq:IMMModePredict}
\end{align}
and where it is easily shown that
\begin{align}
&p(x^t_{k+1}|m_{k+1},y_{1:k}) = \sum_{m_k} \int p(x^t_{k+1},m_{k+1}|x^t_{k},m_{k}) p(x^t_{k},m_{k}|y_{1:k}) dx^t_k \\
&=\sum_{m_k} P(m_{k+1}|m_{k}) P(m_{k}|y_{1:k})  p(x^t_{k+1}|m_{k},y_{1:k}). \label{eq:IMMStatePredict}
\end{align}
This last expression for $p(x^t_{k+1}|m_{k+1},y_{1:k})$ is naturally a mixture of predicted state pdfs. This mixture uses the predicted mode probabilities $P(m_{k+1}|y_{1:k})$ as weights for the predicted state pdfs $p(x^t_{k+1}|m_{k},y_{1:k})$, which follow from applying the Chapman-Kolmogorov equation to each possible mode hypothesis pdf $p(x^t_{k}|m_{k},y_{1:k})$. The number of mixture terms for $p(x^t_{k+1}|m_{k+1},y_{1:k})$ (and subsequently for $p(x^t_{k+1}|m_{k+1},y_{1:k+1})$ following a Bayesian measurement update) therefore grows geometrically at each time step, as the number of possible mode transition histories grows. This `curse of history' is handled in the conventional IMM by approximating $p(x^t_{k+1}|m_{k+1},y_{1:k})$ with a single Gaussian pdf, whose first and second moments match those of the RHS mixture in \pareqref{IMMStatePredict}. If $p(x^t_{k+1}|m_{k+1},y_{1:k+1})$ is well-approximated by a single Gaussian, then this approach leads to a convenient recursive approximation to the optimal Bayes filter \cite{BarShalomBook}. 
However, if $p(x^t_{k+1}|m_{k+1},y_{1:k+1})$ is highly non-Gaussian, more sophisticated pdf approximations must be used and propagated within the RHS mixture of \pareqref{IMMStatePredict} \cite{Boers-IEEProcRSN-2003}. 

Three different IMM estimation schemes are considered that use Extended Kalman filter Gaussian sum filter (EKF-GSF) GM approximations for the mode conditional predicted pdfs $p(x^t_{k+1}|m_{k+1},y_{1:k})$ and mode conditional posteriors $p(x^t_{k+1}|m_{k+1},y_{1:k+1})$. The first approach is a centralized IMM estimator which process all measurements from all three sensing platforms at every time step. Note that the mode conditional pdfs can each be well-approximated by a single Gaussian in this case, since the target's location can be trilaterated from the full set of $y_k$ data at each time step. This set up therefore provides a high baseline for tracking performance.

The second approach uses an independent IMM at each sensor platform, which processes only that platform's local measurements at each time step via the EKF-GSF and does not fuse any information from the other platforms. 
In this case, the target cannot be easily localized and the mode conditional pdfs become highly non-Gaussian. The mode conditional pdfs are thus modeled as GMs with at most 12 components each, so that the overall marginal pdf for $x^t_k$ (marginalizing out $m_k$) at each platform is a GM with at most 60 components. These independent estimators use Runnalls' algorithm to compress the mode conditional GM pdfs after time update and measurement update steps. This set up provides a low baseline for tracking performance. 

The third approach also deploys independent IMMs at each platform using EKF-GSF GM approximations, but additionally uses WEP DDF to fuse each platform's mode conditional target state GMs every 60 secs according to the asymmetric circular communication topology $1\rightarrow 2 \rightarrow 3 \rightarrow 1$. This austere constraint is representative of operating conditions featuring extended communication blackout periods, e.g. in domains such as persistent undersea and aerial surveillance where lack of reliable interplatform communications, need to conserve onboard energy, etc. must be handled. In this case, each platform communicates the full set of mode conditioned GM pdfs $\set{p^{i}(x^t_{k}|m_{k},Z^i_{k})}_{m_{k}=1:5}$ at some designated fusion time $k$ to its designated recipient, and fuses each GM element of this set with the corresponding mode conditioned GM pdf in the pdf set $\set{p^{j}(x^t_{k}|m_{k},Z^j_{k})}_{m_{k}=1:5}$ sent by its designated sender $j$, 
\begin{align}
p^{f,i}(x^t_{k}|m_{k},y_{1:k}) &= \frac{1}{\eta(m_k)} \bigsquare{p^{i}(x^t_{k}|m_{k},Z^i_{k})}^{\omega} \bigsquare{p^{j}(x^t_{k}|m_{k},Z^j_{k})}^{1-\omega} \label{eq:IMMStateDDF} \\ 
\eta(m_k) &= \int \bigsquare{p^{i}(x^t_{k}|m_{k},y_{1:k})}^{\omega} \bigsquare{p^{j}(x^t_{k}|m_{k},y_{1:k})}^{1-\omega} dx^t_k. 
\end{align}
The mode probabilities are then updated locally by each platform\footnote{in principal, the mode probabilities could also be fused between the platforms using the conditional factorization formulation of DDF described in \cite{Ahmed-MFI-2014}; that approach is not used here for simplicity},  
\begin{align}
P^{+,i}(m_{k}|y_{1:k}) &= \frac{P^{i}(m_{k}|y_{1:k}) \cdot \eta(m_k)}{\sum_{m_k} P^{i}(m_{k}|y_{1:k}) \cdot \eta(m_k)}. 
\end{align}
The minimax information loss rule \pareqref{minimaxInfoLossDef} is used to select $\omega$ in \pareqref{IMMStateDDF}. 
Runnalls' compression to 12 mixands is applied to the fused GM for each $m_k$ after DDF, where mixands are discarded if their weights are numerically indistinguishable from zero. Both IGS (with $N_s=$1000) and FOCI are separately implemented to approximate the resulting WEP fusion pdfs at each platform, where interplatform communication and DDF updates only occur at time steps $k=60,120,180,240,300,360,$ and $420$. 

Fifty Monte Carlo runs of $\sim$7 minute (422 time step) tracking simulations were performed for each fusion method, using an initial target distribution modeled by a mixture of equally weighted 60 Gaussian components (12 equally weighted Gaussians per maneuvering mode), whose means were randomly perturbed about the true target state initial state $x^t_0 = [5\times 10^3 \ m, 0 \ \frac{m}{s}, 1 \times 10^2 \ m, 375 \ \frac{m}{s}]^T$ by zero mean Gaussian random vectors with diagonal covariance $P_0=\mbox{diag}([500 \ m^2, 100 \ (\frac{m}{s})^2, 500 \ m^2, 100 \ (\frac{m}{s})^2])$ and assigned diagonal covariance matrices with independent initial E-N position uncertainties of 2000 m$^2$ and independent velocity uncertainties of 1000 $(\frac{m}{s})^2$. The simulations for all fusion approaches were implemented in Matlab 9.1 on a Windows 10 laptop (Intel i7-8550U 1.80 GHz CPU with 16 GB RAM). 

\subsubsection{Results}

\begin{figure*}[t!]
\centering
\newcommand{\figsize}{5.0cm}
\begin{tabular}{@{}c@{}c@{}c@{}}
\hspace*{-0.2 in}
\includegraphics[width=\figsize]{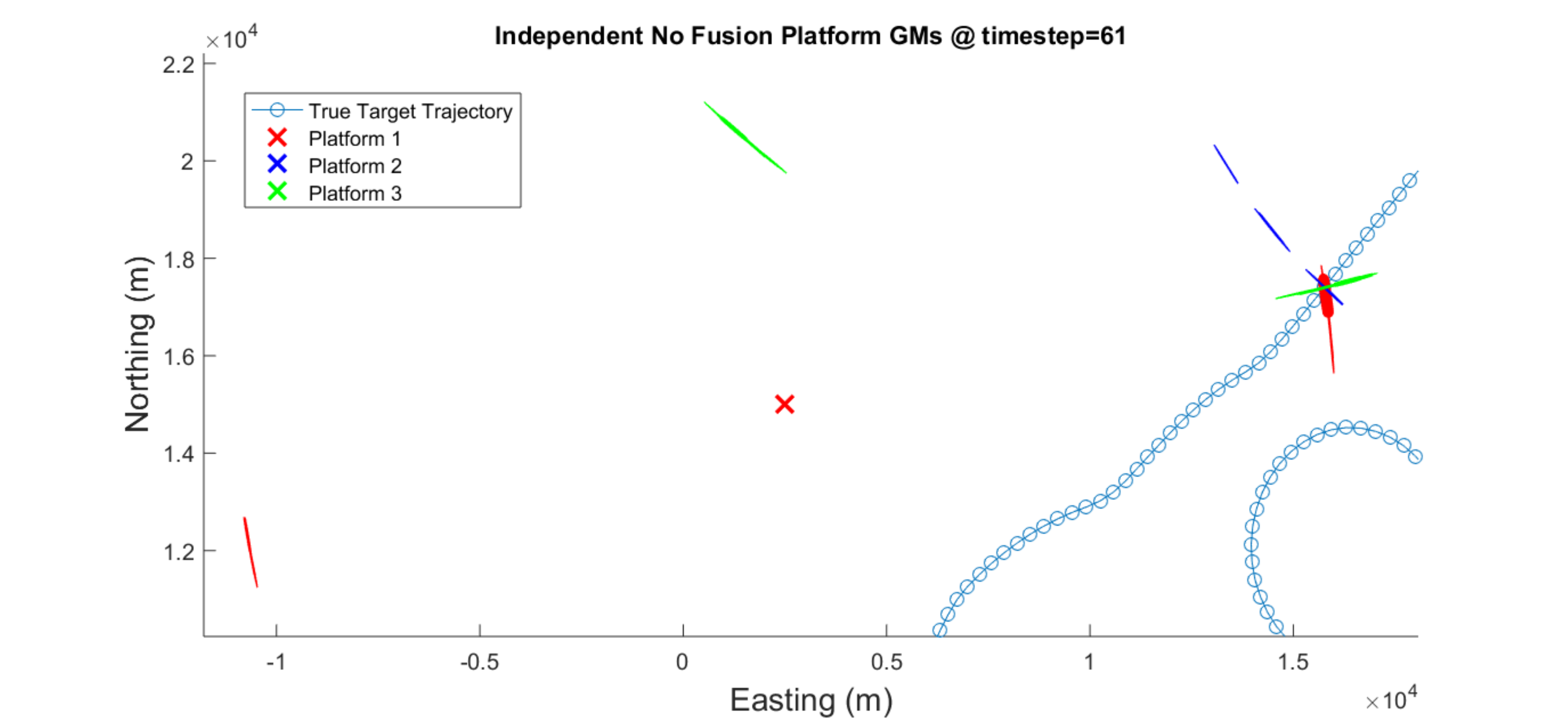} &
\hspace*{-0.2 in}
\includegraphics[width=\figsize]{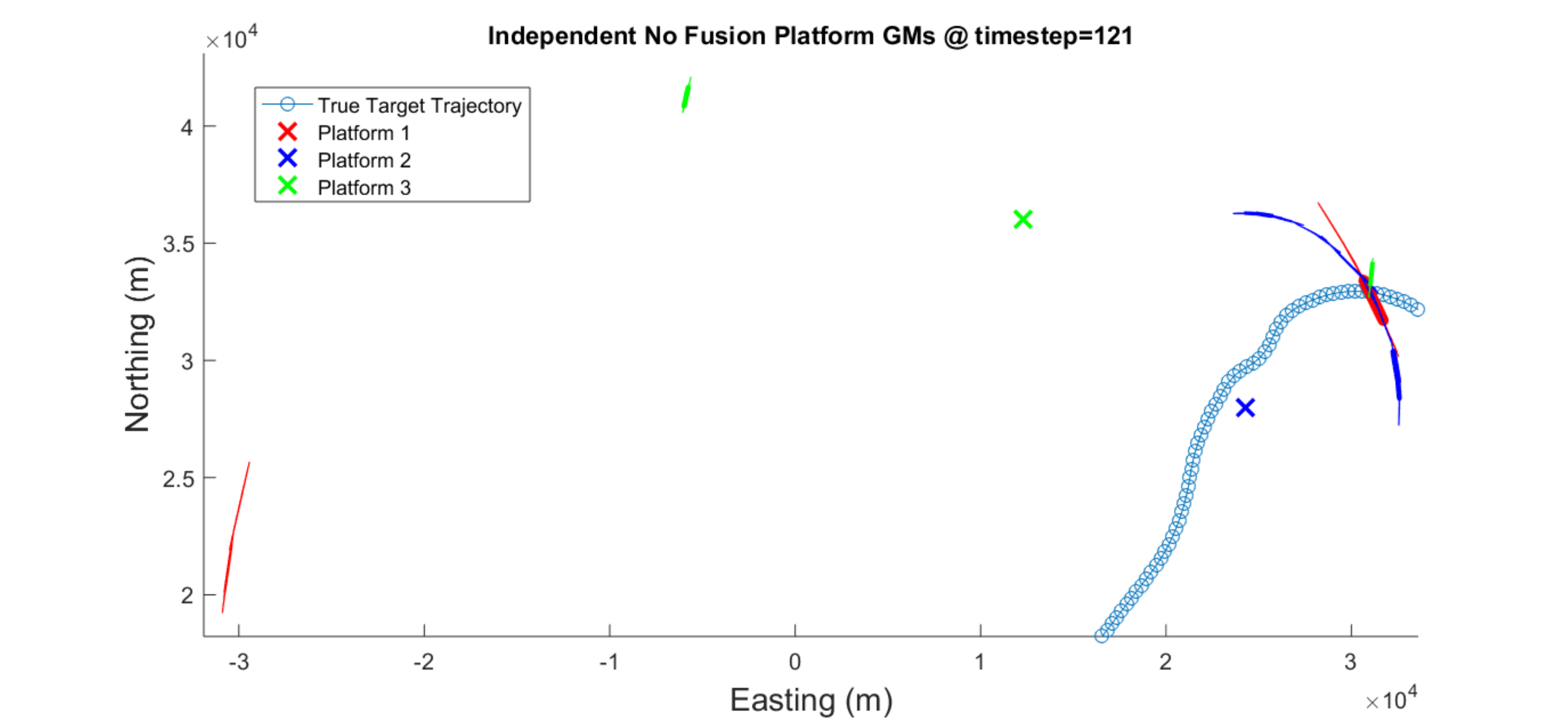} &
\hspace*{-0.2 in}
\includegraphics[width=\figsize]{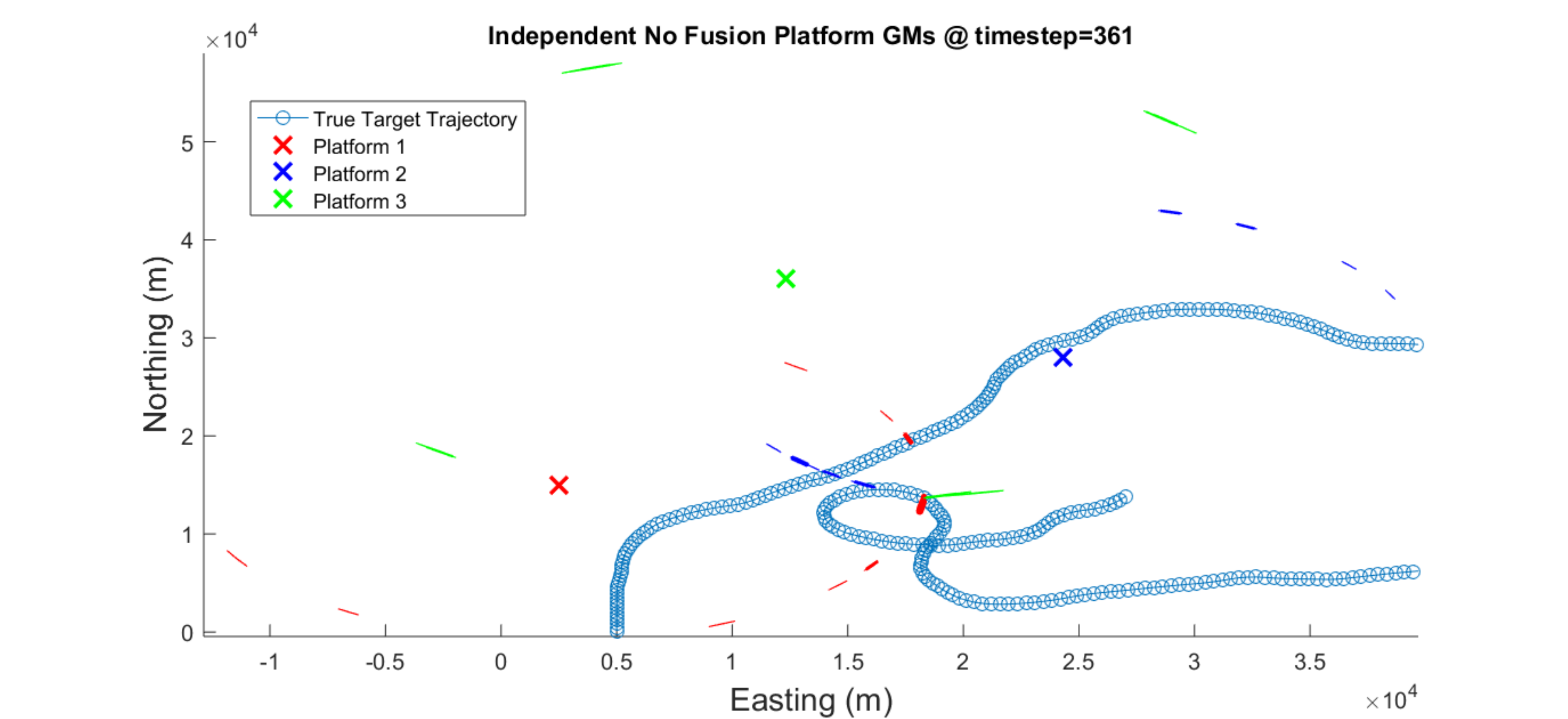} \\
\scriptsize (a) & \scriptsize (b) & \scriptsize (c)
\end{tabular}
\caption{\scriptsize Typical platform marginal GMs for independent non-DDF based tracking across all maneuvering modes for target's estimated E-N position (mixture component 2$\sigma$ ellipses shown, with colors corresponding to platforms).  }
\label{fig:immtracking_IndepGMresults}
\end{figure*}

Figure \ref{fig:immtracking_IndepGMresults} shows the resulting East-North target position pdfs at selected time steps for a typical Monte Carlo tracking run using the non-DDF `independent' estimation scheme at each platform. The ellipses show the $2\sigma$ bounds for local GMs $p^{i}(x^t_{k}|y_{1:k})$, marginalized across discrete maneuvering modes $m_{k}$. These plots clearly show that the position uncertainties in this scenario lead to highly non-Gaussian `ring pdfs' that are characteristic of range-based tracking by a single platform. Since each platform can only carry up to 12 Gaussian mixands per manuevering mode, noticeable gaps appear early on in the local pdfs due to the effect of Runnalls' mixture compression. Each platform generally manages to keep some modal mixands close to the target's true trajectory for a significant portion of the tracking run. However, the combined effects of mixture compression and local non-observability eventually force the GMs for all platforms to deviate significantly from the true trajectory after an extended time. 

\begin{figure}[t!]
\centering
\newcommand{\figsize}{4.95cm}
\begin{tabular}{@{}c@{}c@{}c@{}}
\hspace*{-0.2 in}
\includegraphics[width=\figsize]{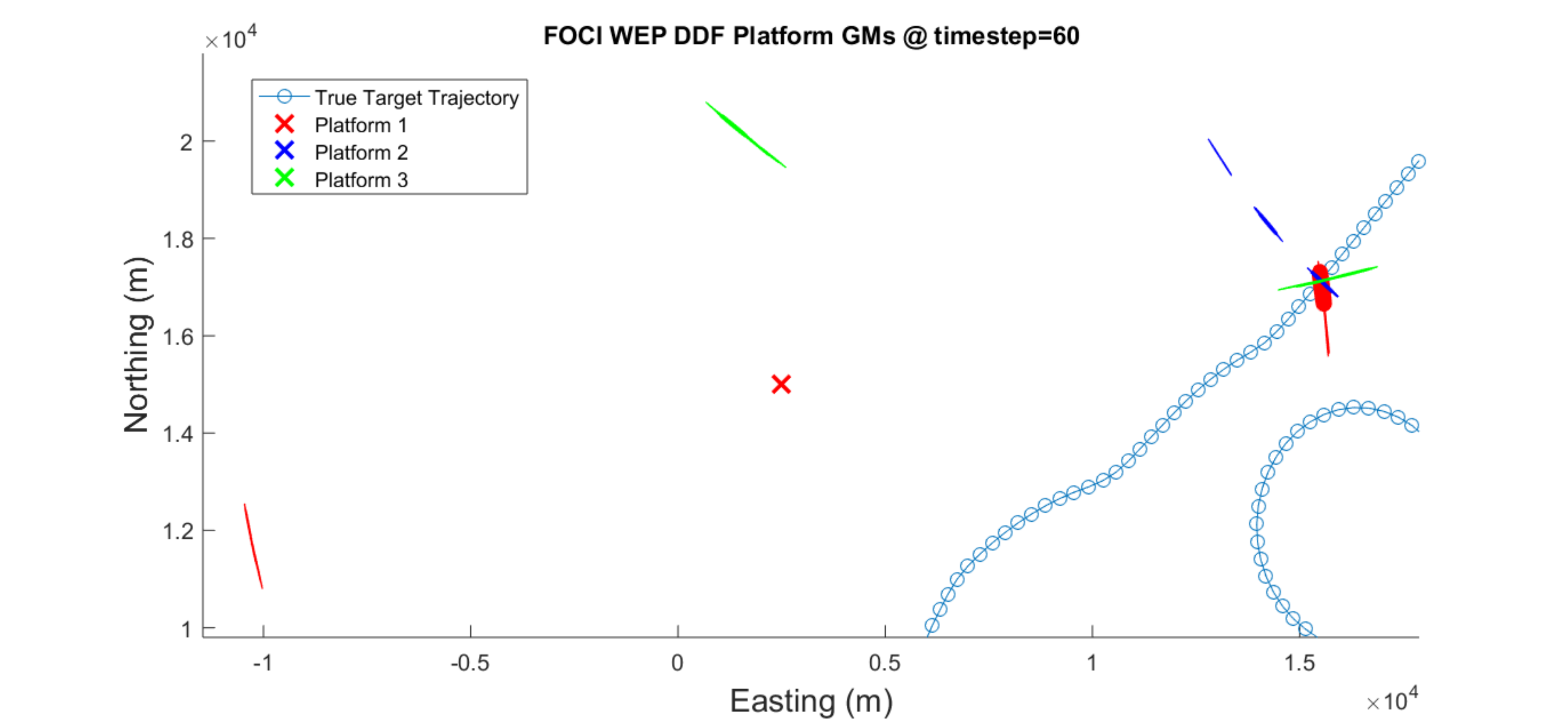} &
\hspace*{-0.2 in}
\includegraphics[width=\figsize]{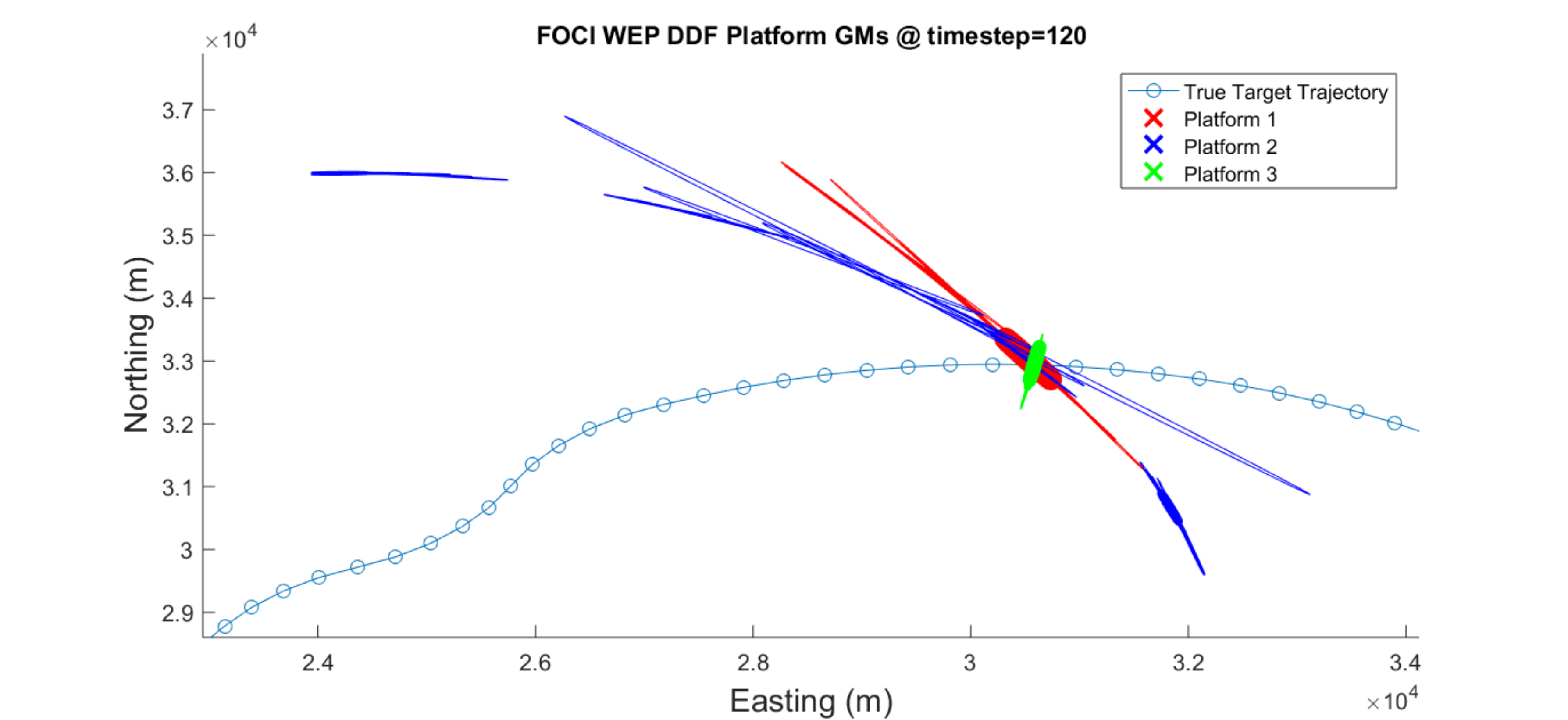} &
\hspace*{-0.2 in}
\includegraphics[width=\figsize]{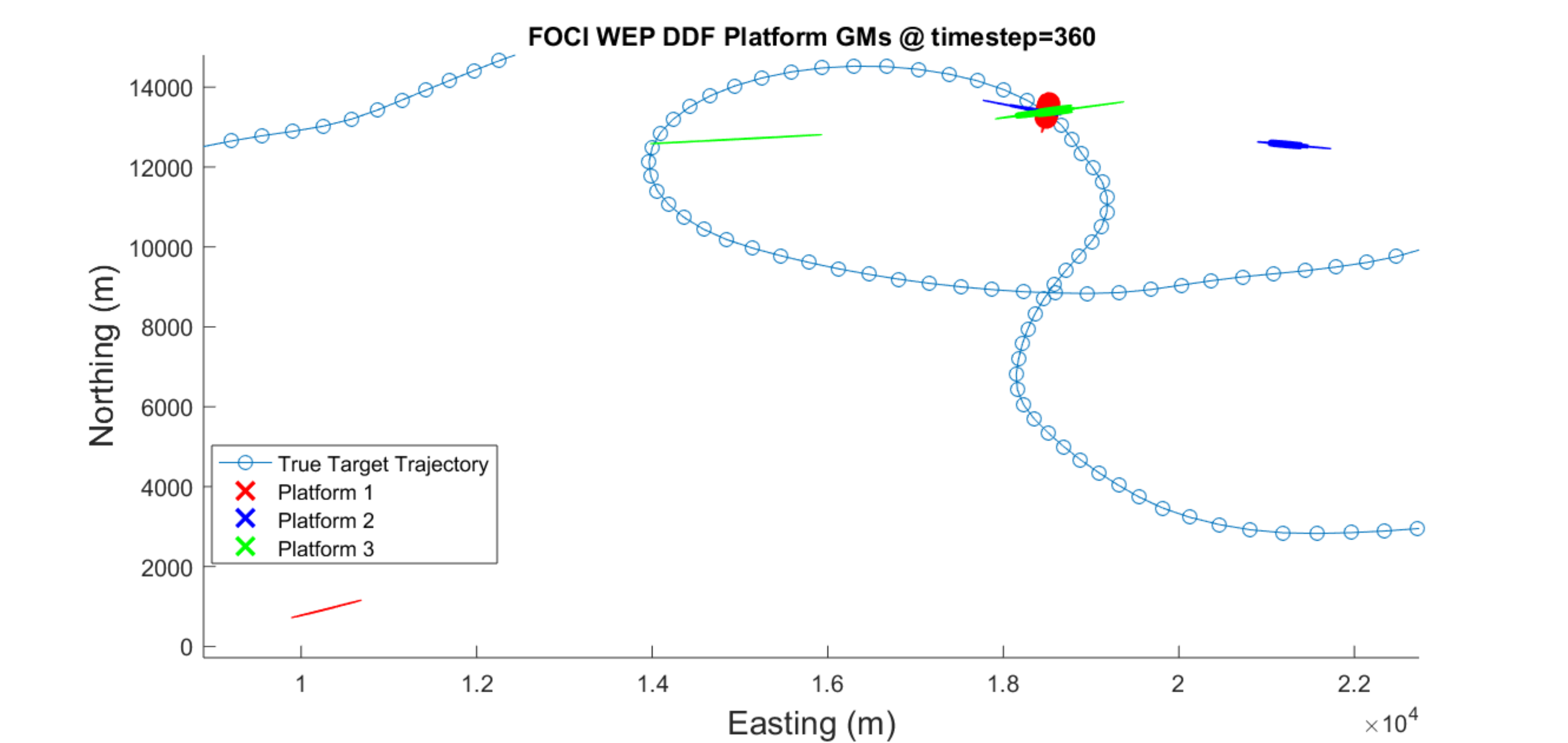} \\
\scriptsize (a) & \scriptsize (b) & \scriptsize (c) \\
\hspace*{-0.2 in}
\includegraphics[width=\figsize]{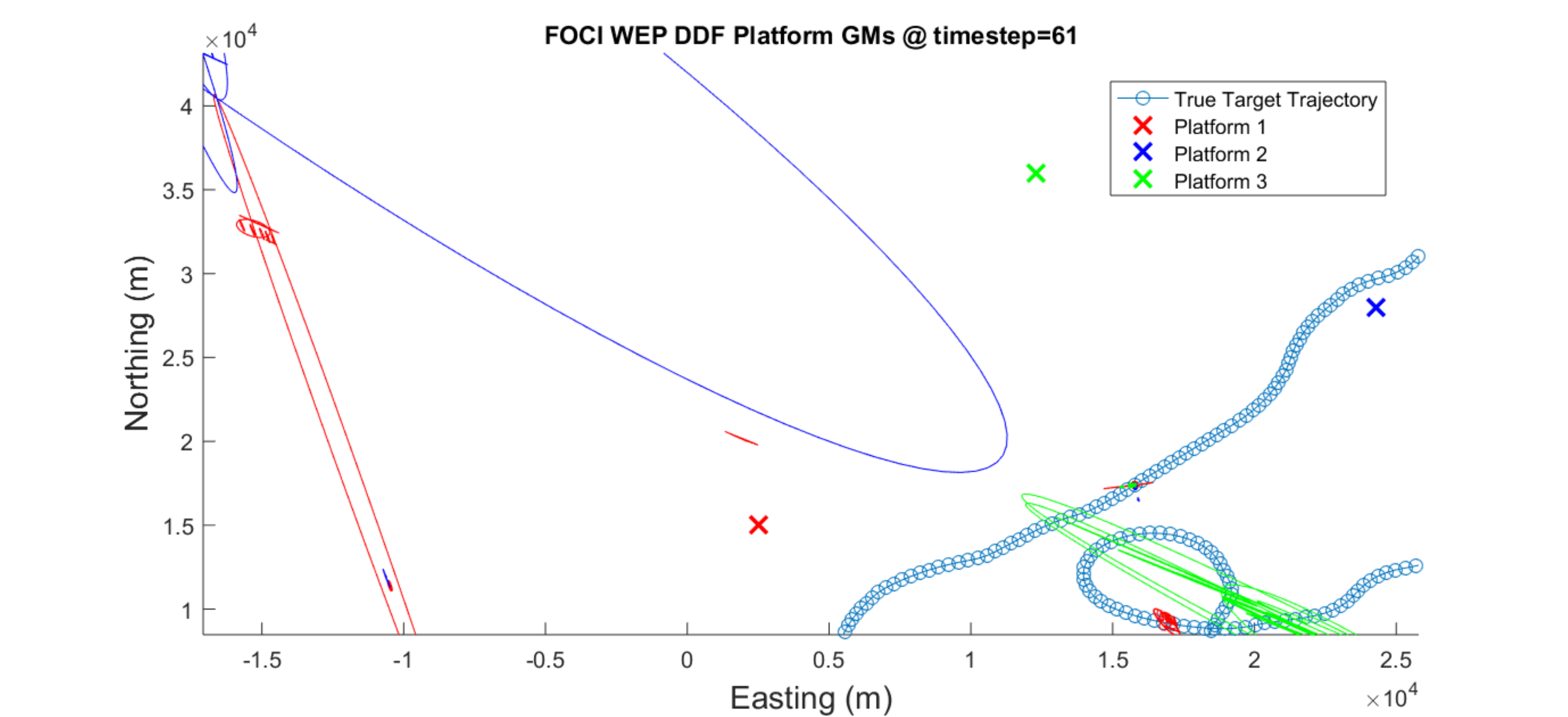} &
\hspace*{-0.2 in}
\includegraphics[width=\figsize]{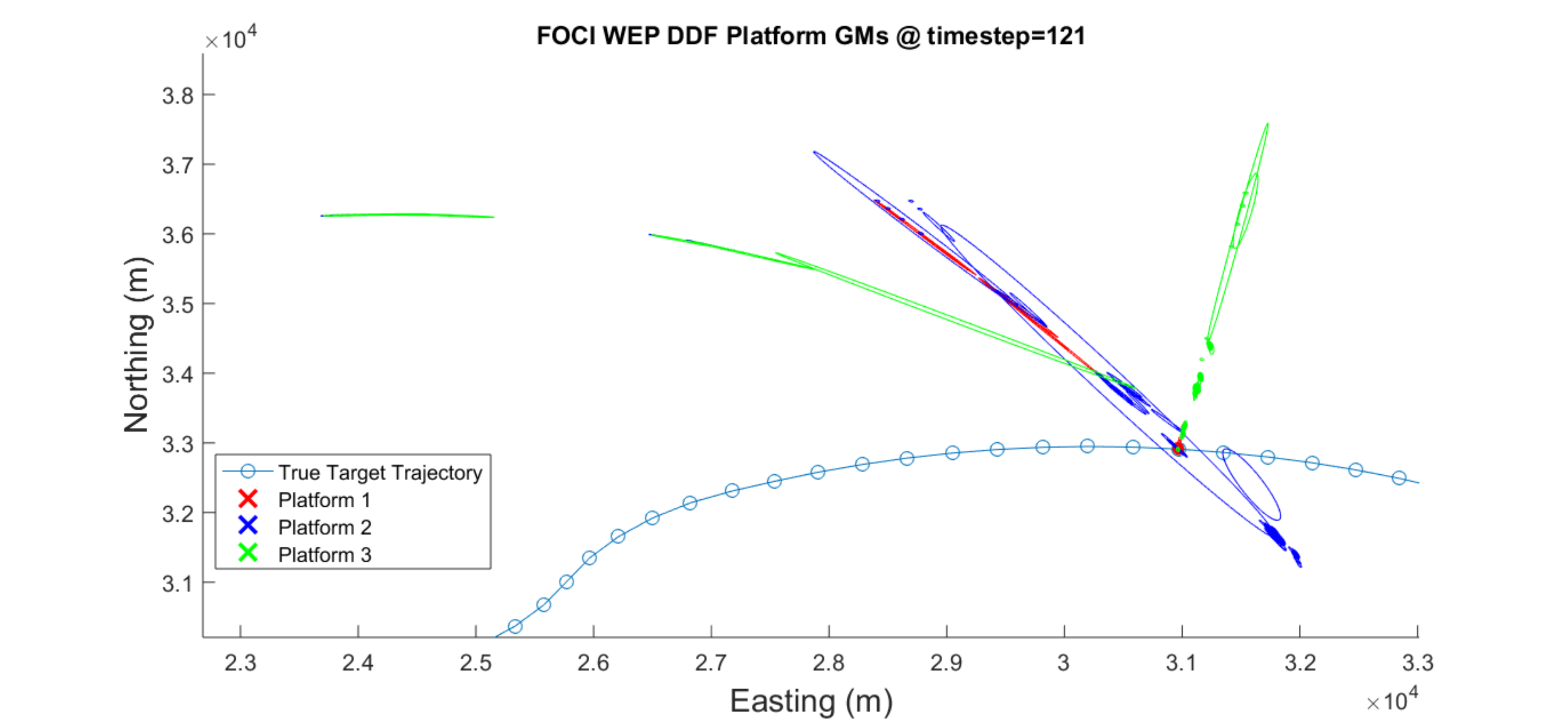} &
\hspace*{-0.2 in}
\includegraphics[width=\figsize]{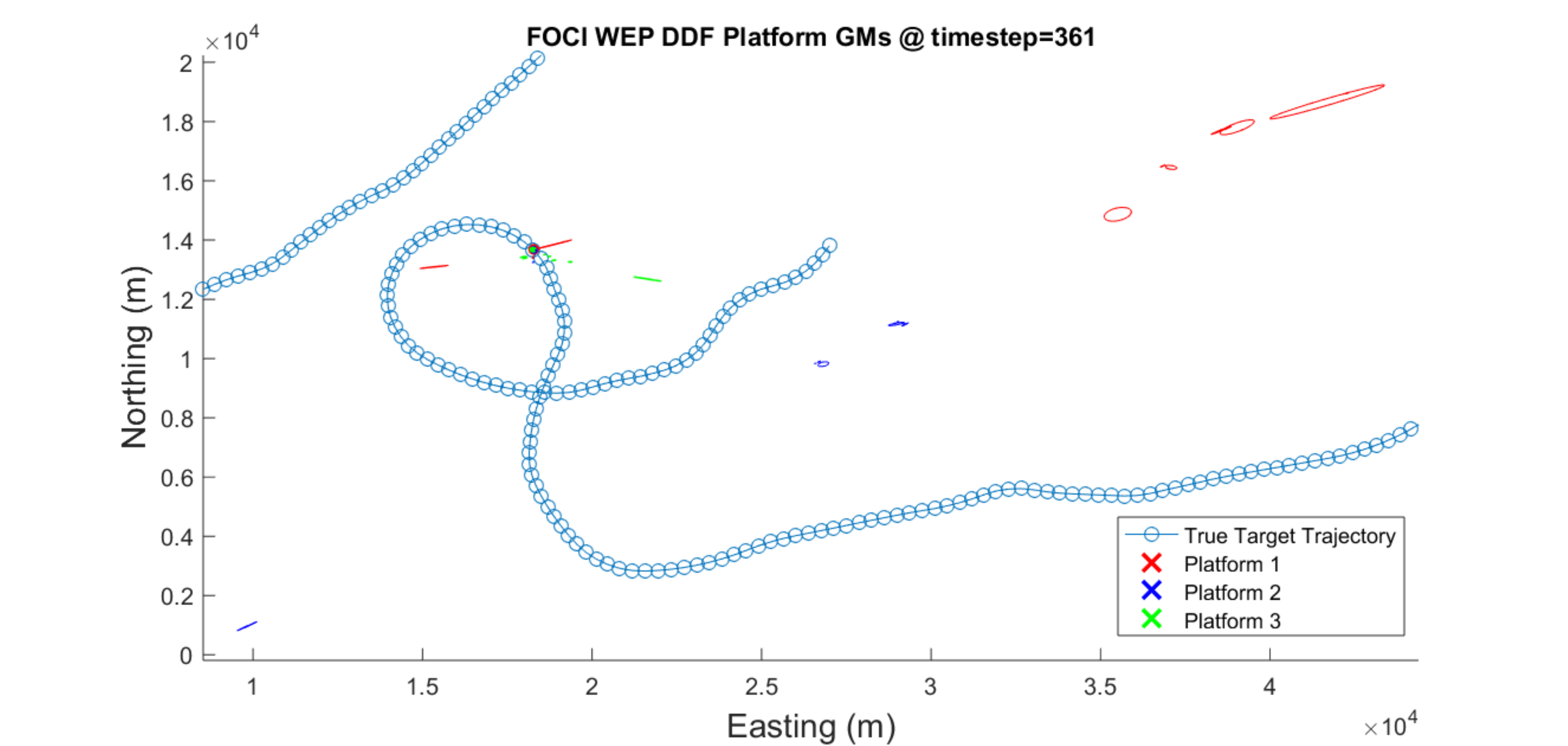} \\
\scriptsize (d) & \scriptsize (e) & \scriptsize (f)
\end{tabular}
\caption{\scriptsize Typical platform marginal GMs for FOCI WEP DDF across all maneuvering modes for target's estimated E-N position: (a)-(c) prior to DDF updates; (d)-(f) following DDF updates (mixture component 2$\sigma$ ellipses shown, with colors corresponding to platforms).  }
\label{fig:immtracking_FOCIGMresults}
\end{figure}
\begin{figure}[t!]
\centering
\newcommand{\figsize}{4.95cm}
\begin{tabular}{@{}c@{}c@{}c@{}}
\hspace*{-0.2 in}
\includegraphics[width=\figsize]{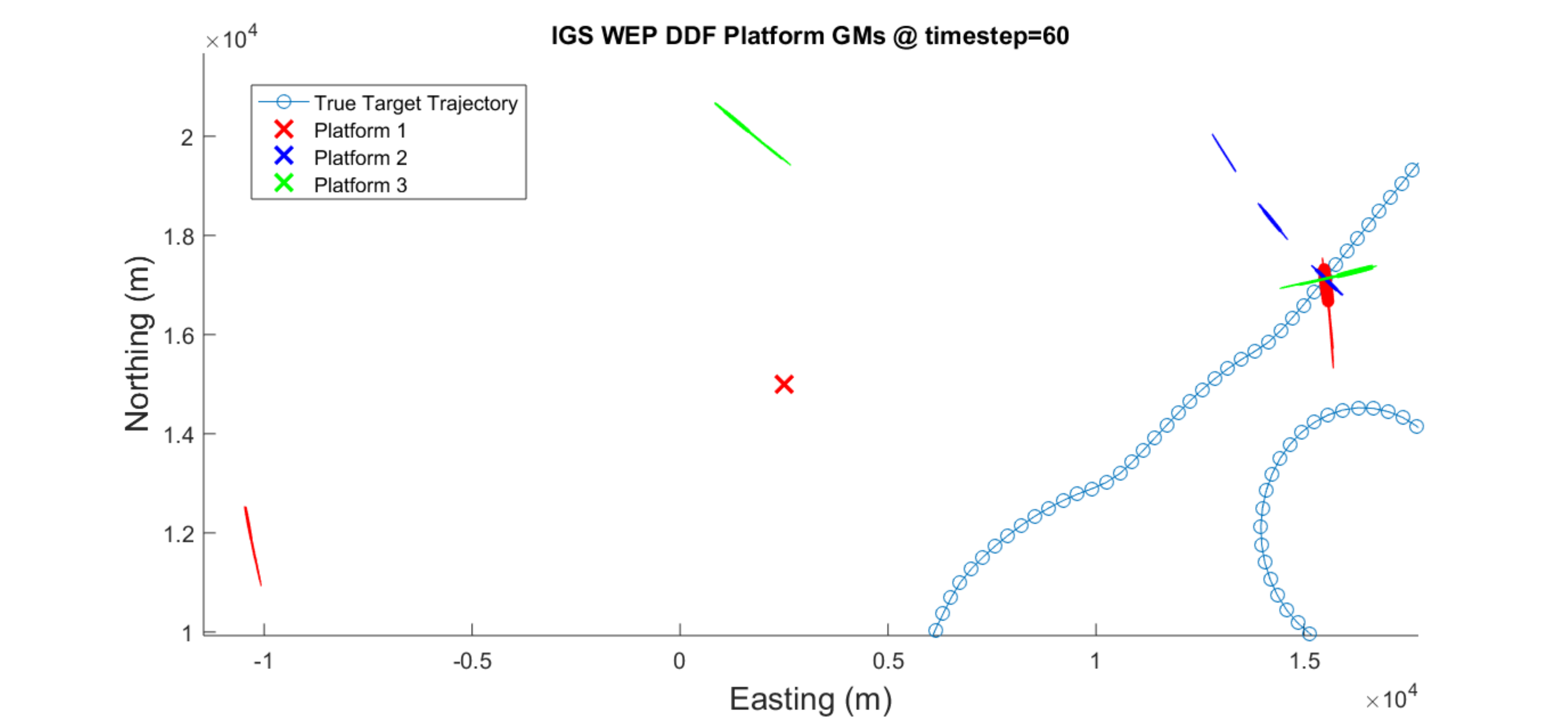} &
\hspace*{-0.2 in}
\includegraphics[width=\figsize]{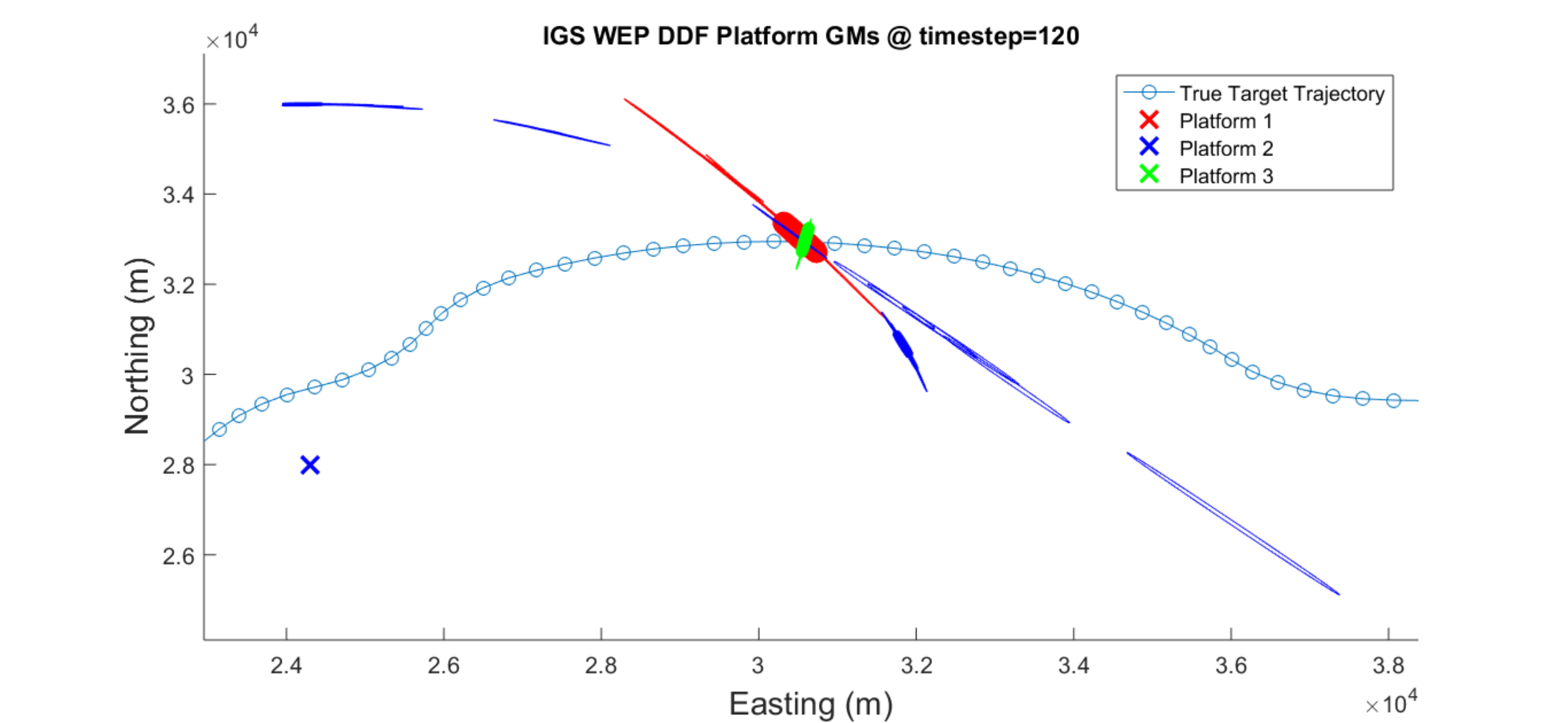} &
\hspace*{-0.2 in}
\includegraphics[width=\figsize]{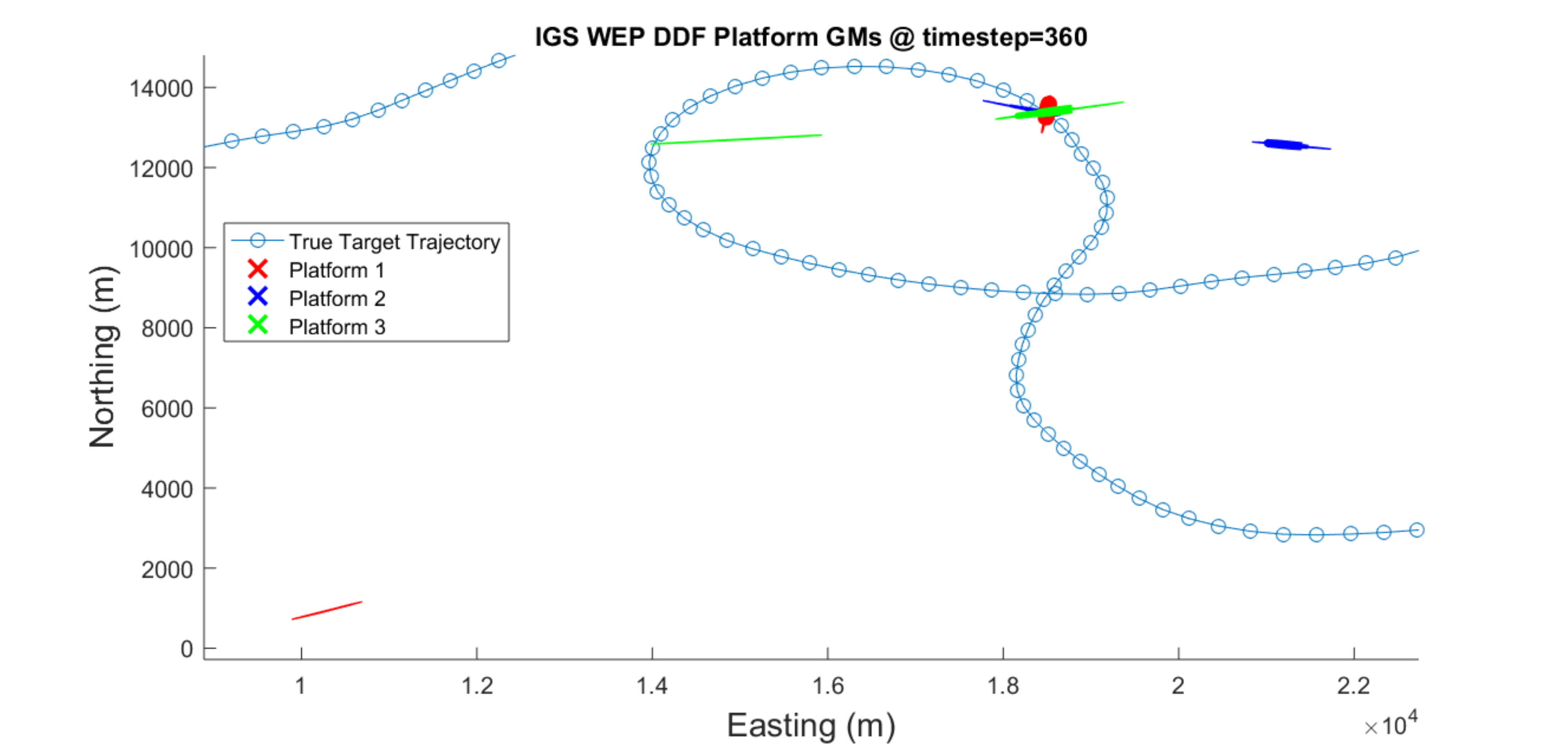} \\
\scriptsize (a) & \scriptsize (b) & \scriptsize (c) \\
\hspace*{-0.2 in}
\includegraphics[width=\figsize]{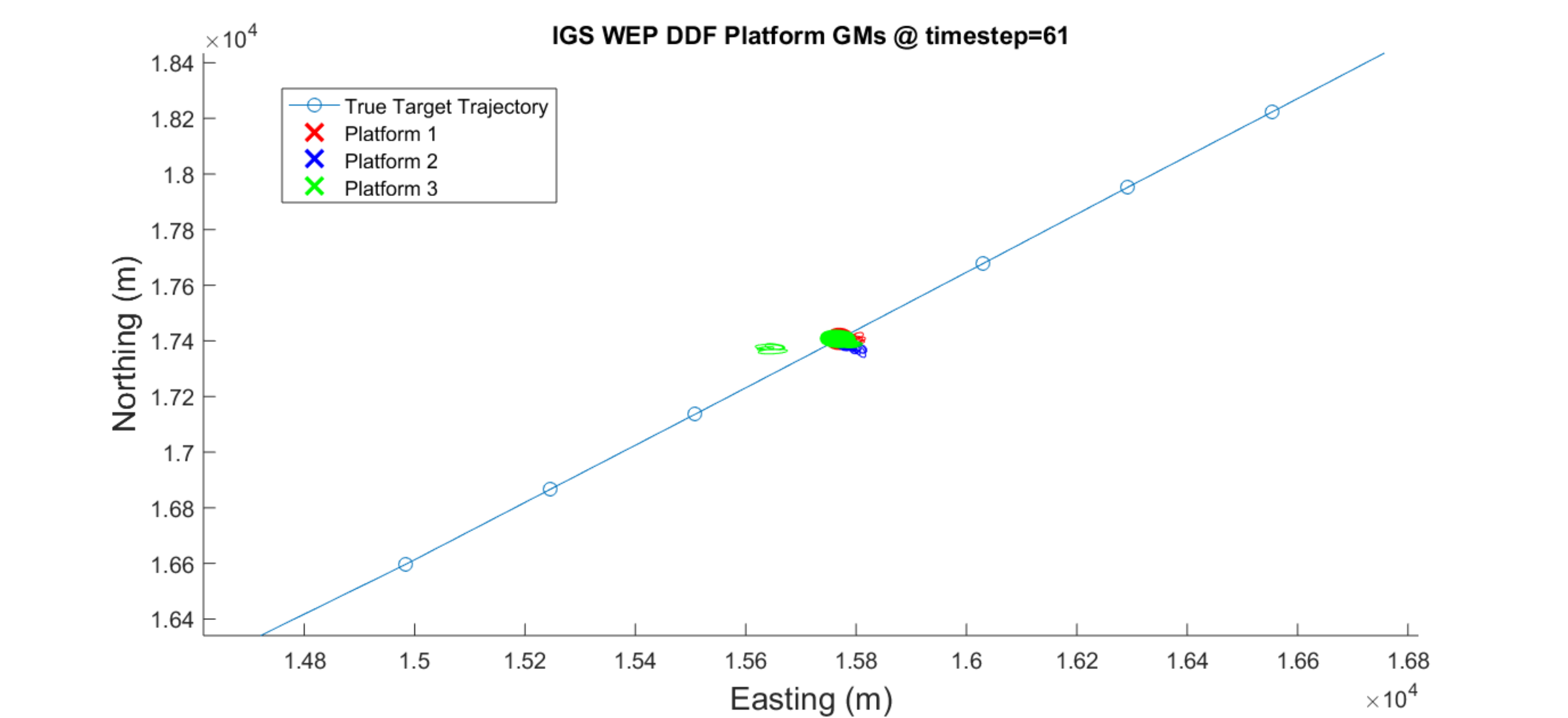} &
\hspace*{-0.2 in}
\includegraphics[width=\figsize]{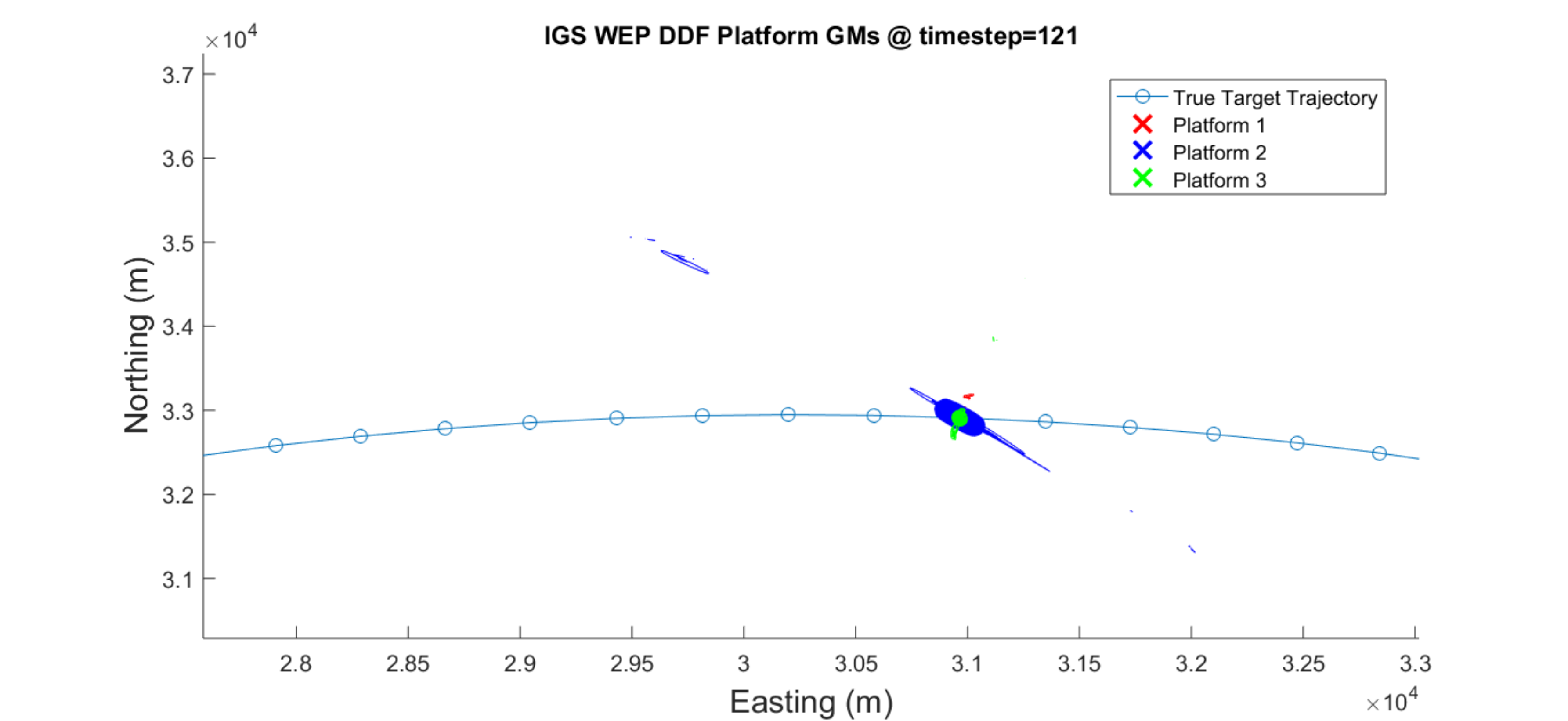} &
\hspace*{-0.2 in}
\includegraphics[width=\figsize]{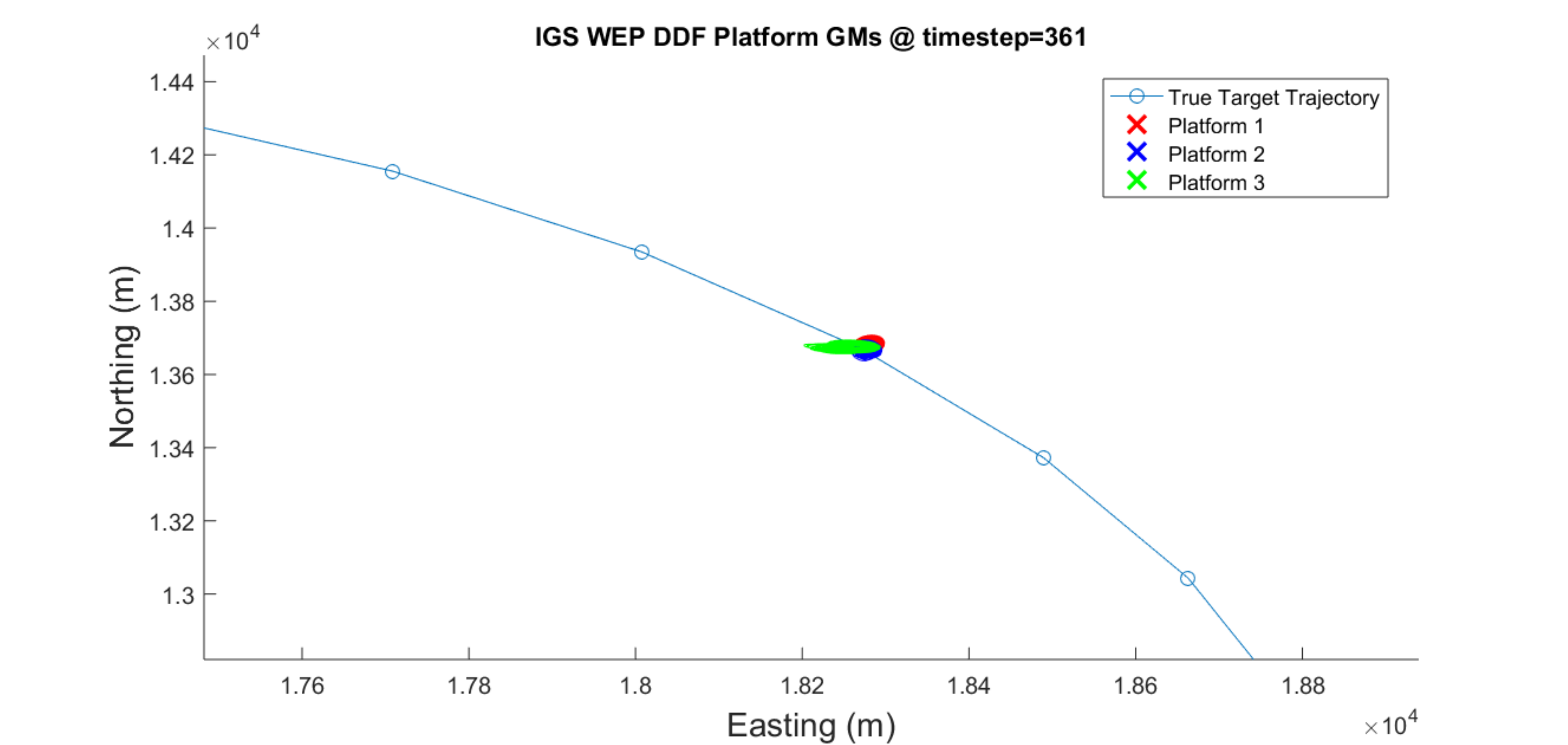} \\
\scriptsize (d) & \scriptsize (e) & \scriptsize (f)
\end{tabular}
\caption{\scriptsize Typical platform marginal GMs for IGS WEP DDF across all maneuvering modes for target's estimated E-N position: (a)-(c) prior to DDF updates; (d)-(f) following DDF updates (mixture component 2$\sigma$ ellipses shown, with colors corresponding to platforms).  }
\label{fig:immtracking_IGSGMresults}
\end{figure}

Figures \ref{fig:immtracking_FOCIGMresults} and \ref{fig:immtracking_IGSGMresults} show snapshots of the marginal East-North target position pdfs under DDF with FOCI and IGS, just before and just after 3 of the 7 fusion instances. These plots show that IGS leads to GM fusion results that are much closer to what is expected for periodic Bayesian combination of the information collected at each platform, as the GM components for each platform remain tightly clustered near the true target trajectory. The shapes of these pdfs also generally agree with the centralized fusion result, which effectively finds the intersection of each platform's uncertainty ring while avoiding double counting of common prior information from the shared switching process dynamics. While FOCI tends to also produce some GM components that are clustered near the true target trajectory, it also produces many other extraneous components at each platform that are far from the true target trajectory. Note that the fusion pdfs produced by each platform under IGS or FOCI do not match across platforms. This is expected, due to the delayed spreading of information imposed by the asymmetric circular communication topology and due to information loss from WEP DDF.

Since centralized fusion produces a tightly clustered pdf for the true target state, the minimum mean squared error estimate (MMSE) of $x^t$ derived from the platform GMs provides a sensible basis by which to assess and compare the effective amount of information gained by each platform at each fusion instance under either of the GM WEP DDF methods. In this case, the MMSE estimator $\hat{x}^{t,i}_{k}$ for platform $i$ at time $k$ corresponds to the mean of the fused mixture pdf $p^{i,f}(x^t_{k}|y_{1:k})$, which is given by the marginal of the LHS of \pareqref{IMMStateDDF} with respect to modes $m_k$. Likewise, the local marginal mixture covariance provides the estimation error covariance matrix. 


\begin{figure}[h!]
\centering
\newcommand{\figsize}{10.5cm}
\begin{tabular}{@{}c@{}}
\includegraphics[width=\figsize]{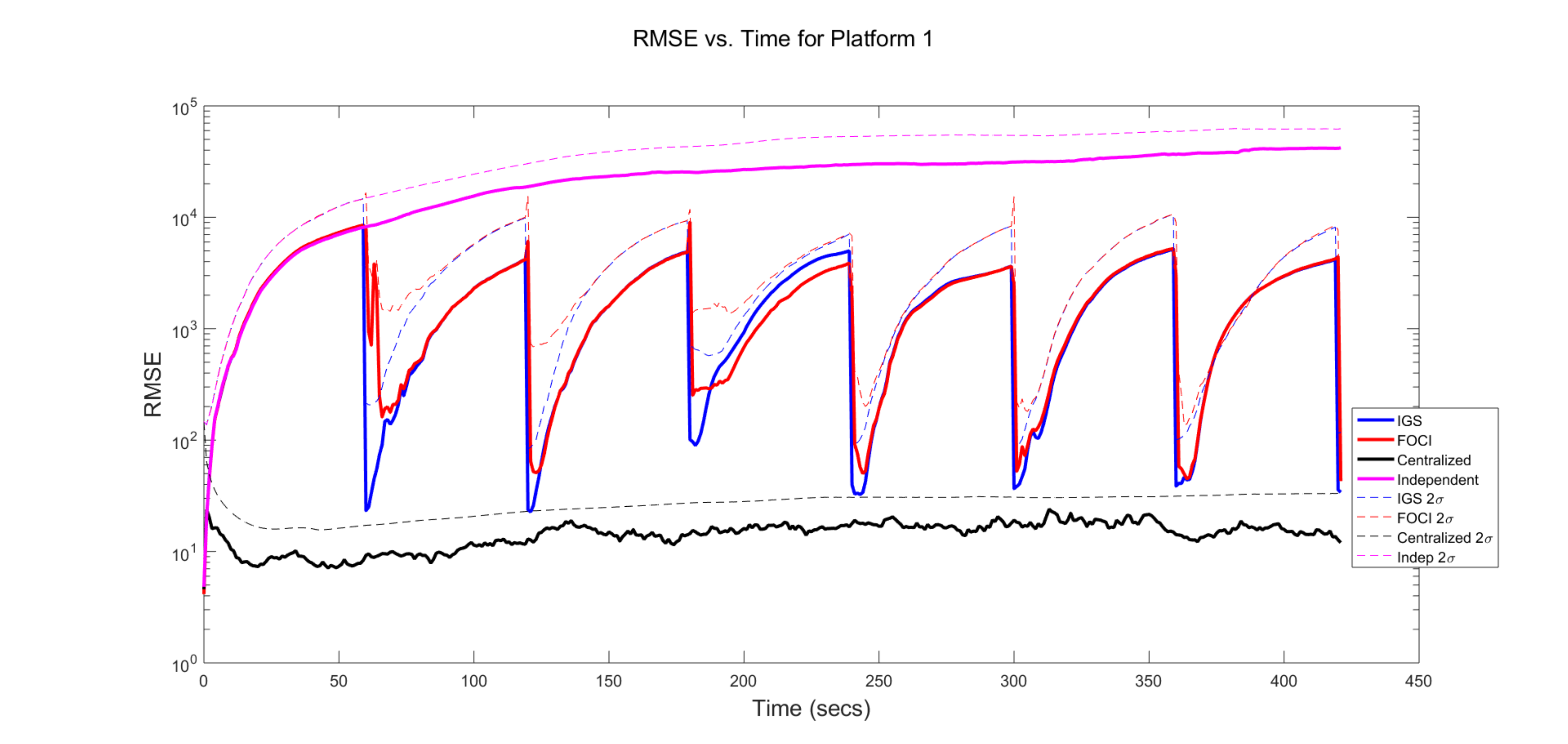} \\
\scriptsize (a) \\
\includegraphics[width=\figsize]{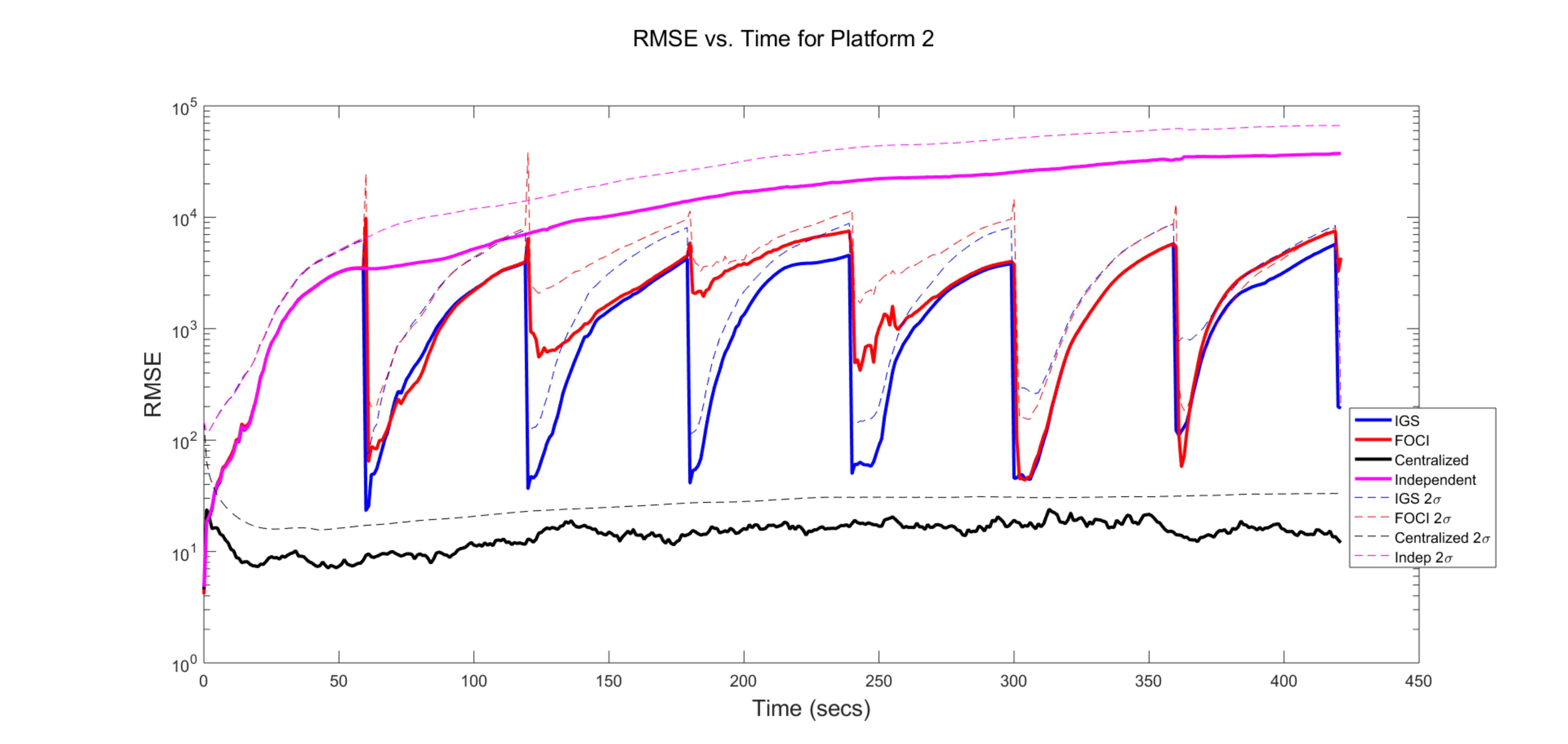} \\
\scriptsize (b) \\
\includegraphics[width=\figsize]{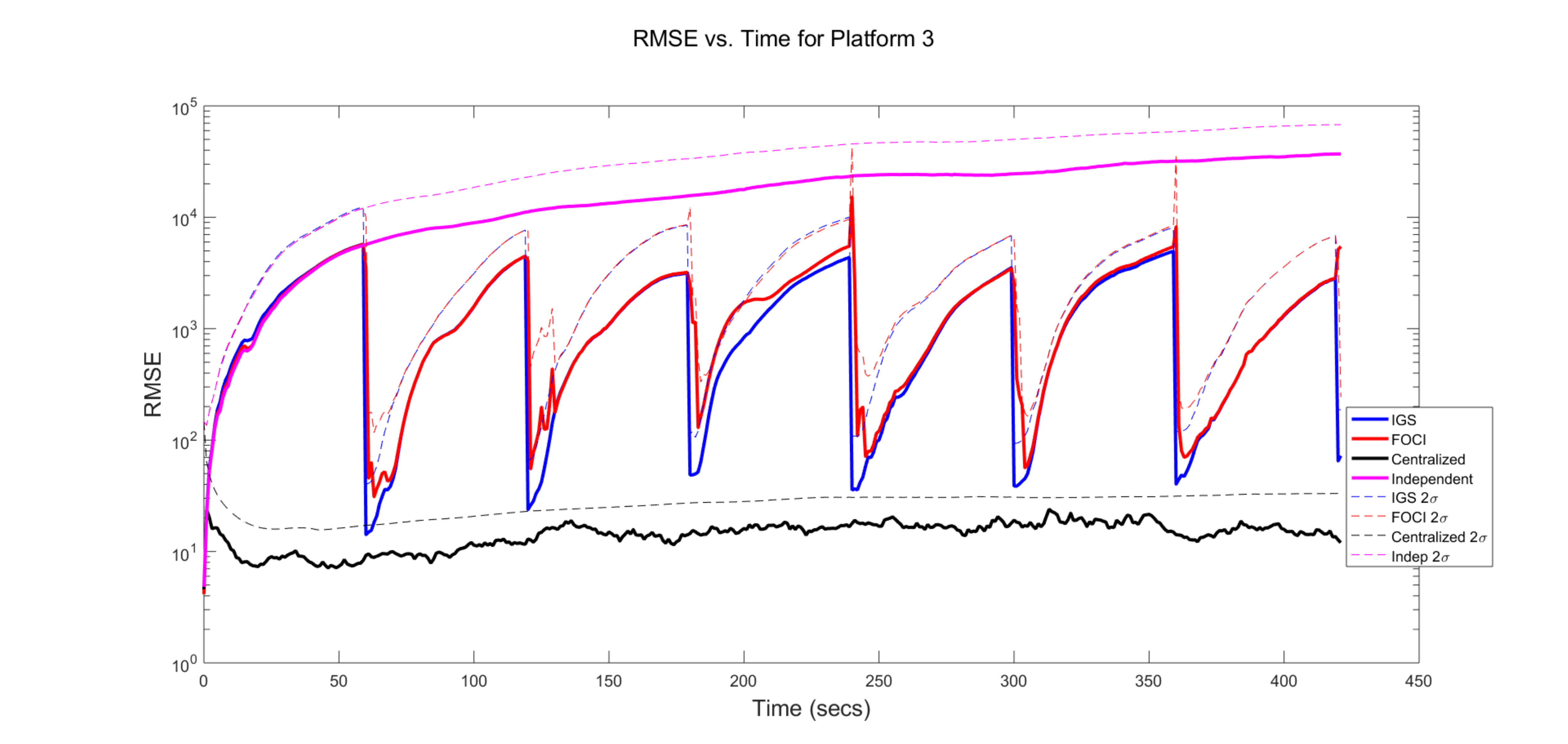} \\
\scriptsize (c) 
\end{tabular}
\caption{\scriptsize Platform tracking RMSEs and 2$\sigma$ bounds vs. time for different fusion methodologies, averaged over 50 Monte Carlo trials (errors shown on log scale). }
\label{fig:immtracking_RMSEresults}
\end{figure}

Figure \ref{fig:immtracking_RMSEresults} shows the root mean square error (RMSE) and $2\sigma$ bounds (derived from the square root of the trace of estimation error covariance) for the MMSE target state estimate of each platform under each fusion method vs. time, averaged over all 50 Monte Carlo runs. The FOCI results tend to exhibit large spikes in the $2\sigma$ estimation uncertainty, reflecting the contribution of extraneous mixture components that show up after WEP DDF. In several fusion instance, the FOCI estimate displays poor tracking behavior following fusion. In particular, for the early part of platform 2's tracks and for the middle portion of platform 3's tracks, the error does not drop significantly following FOCI fusion, indicating that severe biases enter via the FOCI GM fusion pdf. In contrast, IGS fusion shows much better and more consistent performance overall, with position errors generally in range of 10s of meters and velocity errors generally in range of single digits. Furthermore, the $2\sigma$ estimation uncertainty for IGS drops consistently and significantly after each fusion instance, as expected. It can also be seen that in all instances, both IGS and FOCI remain `conservative' in the MSE sense relative to the centralized optimal fusion result (and hence statistically consistent), although IGS is less conservative overall, especially in the time windows immediately following DDF updates. 

\begin{figure}[t]
\hspace{-0.5in}
\newcommand{\figsize}{16cm}
\includegraphics[width=\figsize]{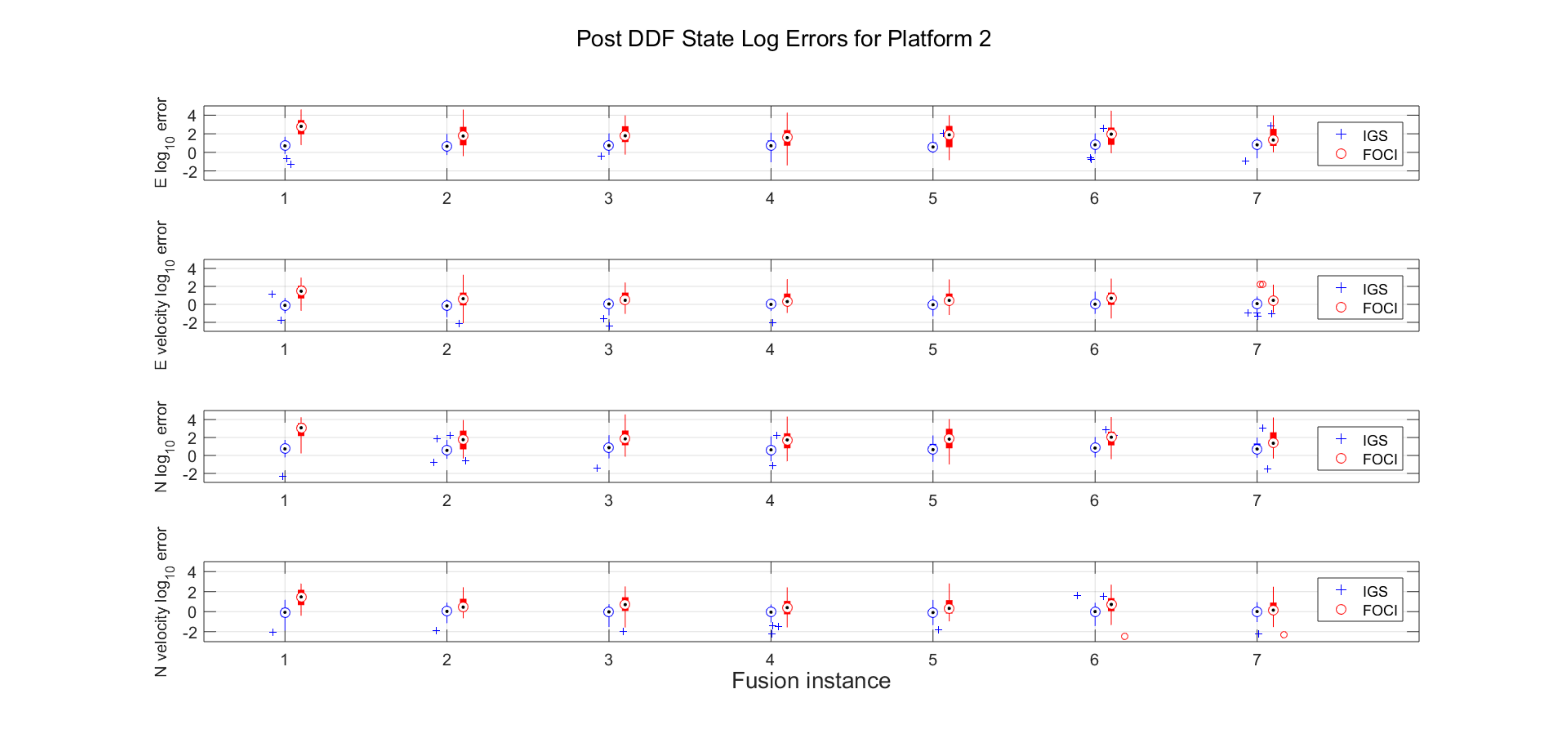} 
\caption{\scriptsize Platform 2 distributions for log of absolute state error immediately following DDF events. }
\label{fig:immtracking_Platform2logstateerrors}
\end{figure}

To examine the state estimation error discrepancies between IGS and FOCI more closely, Figure \ref{fig:immtracking_Platform2logstateerrors} shows the distributions of the base 10 logarithm of the absolute MMSE state errors for each estimated target state following DDF updates at Platform 2 across all 50 Monte Carlo runs (the results for Platforms 1 and 3 are similar and not shown here). The FOCI E-N position estimation errors are typically 1 or 2 orders of magnitude larger than for IGS, and the FOCI estimates overall show much higher variability compared to IGS. This underscores the ability of IGS to reliably produce high fidelity GM WEP DDF approximations at each platform. This in turn allows each platform to maintain statistically correct pdfs between DDF updates via numerous multimodal hybrid dynamic state predictions and partially observable nonlinear local measurement updates.  

Finally, given the more complex nature of this 4D hybrid dynamical tracking problem relative to the previous 2D static toy problems and quasi-static target search application, it is worth commenting on the computational performance of IGS and FOCI. On each DDF update for a given pair of platforms, IGS required $92.4 \pm 57.8$ msecs to jointly optimize $\omega$ and approximate the fusion GM for each maneuvering mode, whereas FOCI required $87.5 \pm 34.9$ msecs. 
The Matlab code for these fusion approximations again did not leverage parallelization or other optimization strategies. The run times for IGS and FOCI could thus be significantly improved for online applications. 
\section{Conclusions}
This paper presented novel approximation strategies for Bayesian decentralized data fusion with Gaussian mixture models. 
These methods exploit the fact that the fusion posterior for a general Gaussian mixture DDF problem is a mixture of non-Gaussian component pdfs, each of which can be approximated by Gaussian pdfs to obtain a high-fidelity GM approximation for recursive fusion. This leads to a parallelizable decomposition of the fusion posterior that is equally applicable to both exact and approximate Bayesian DDF updates. Two classes of Monte Carlo importance sampling algorithms -- indirect global sampling (IGS) and direct local sampling (DLS) -- were developed to exploit the natural structure of mixture fusion problems and obtain the required fusion mixand approximations. IGS and DLS were demonstrated on several simulated synthetic and practical application examples, including multi-platform target search and range-based maneuvering target tracking. The simulation results showed that the approximations developed here provide significant improvements over existing Gaussian mixture fusion approximations in terms of computational efficiency, reliability, and scalability to large mixture models.  

The insights underlying the IGS and DLS methods developed here are theoretically applicable to state space models of any size. 
However, since these methods are based on Monte Carlo importance sampling techniques in practice, some care must still be exercised when dealing with high dimensional problems. Techniques such as the Laplace approximation and heavy-tailed importance pdf sampling are useful to optimize sample efficiency in such cases, but other numerical sensitivities may still be present. For example, first-hand experience for exact DDF problems in 6 dimensions and higher has shown that saddle points can be problematic for gradient-based and quasi-Newton numerical optimization techniques for Laplace approximation, especially when the number of mixture terms in the common information pdf becomes large. Algorithm \ref{alg:ISOptAlg} can also fail to provide reliable results with small sample sizes in cases where there is little or no overlap between the GM pdfs being fused (e.g. if mixand components are too far apart). Related to this, there are no formal guarantees that the FOCI approximation provides the best importance sampling distribution for Algorithm \ref{alg:ISOptAlg}, though it has been empirically observed to work well in practice. Additional mitigation strategies can be introduced to handle these and other problem-dependent issues, but are left as avenues for future work. 

Finally, the methods developed here could also be extended to other applications of Bayesian inference that generally involve Gaussian mixture model division, most notably forward-backward mixture-based smoothing in nonlinear Markovian dynamical systems \cite{Vo-TSP-2012, Lee-JGCD-2015} \footnote{interestingly, the closed-form solution derived by \cite{Vo-TSP-2012} bypasses the GM division problem but does not yield a `true' forward-backward algorithm as a result, since the sizes of the backward state messages grow over time rather than remaining constant, as in the classic forward-backward algorithm} and mixture-based algorithms for multi-target tracking via finite set statistics filters \cite{vo2006gaussian, vo2007analytic, uney2013distributed}. 
The theoretical connections and potential applications of the IGS and DLS methods to such problems provide yet another interesting avenue for future research.

\bibliographystyle{elsarticle-num}
\bibliography{gmf_bib}
\end{document}